\documentclass[12pt]{article}

\newlength{\abstractwidth}
\usepackage{amsmath, nccmath}
\usepackage{amsfonts}
\usepackage{amssymb}
\usepackage{latexsym}

\usepackage{color}
\usepackage{graphicx}
\usepackage{tikz}
\usepackage{dsfont}
\usepackage{pgfplots}

\usetikzlibrary{arrows,shapes,positioning}
\usetikzlibrary{decorations.markings}
\usepackage[rightcaption]{sidecap}
\tikzstyle arrowstyle=[scale=1]
\tikzstyle directed=[postaction={decorate,decoration={markings,
    mark=at position .65 with {\arrow[arrowstyle]{stealth}}}}]
\tikzstyle reverse directed=[postaction={decorate,decoration={markings,
    mark=at position .65 with {\arrowreversed[arrowstyle]{stealth};}}}]
\usetikzlibrary{positioning}

\usepackage{hyperref}
\definecolor{darkred}{rgb}{0.8,0.1,0.1}
\hypersetup{colorlinks=true, linkcolor=darkred, citecolor=blue, linktoc=page}

\usepackage{subcaption}

\usepackage[nosort]{cite}

\renewcommand{\thefootnote}{\fnsymbol{footnote}}
\renewcommand{\thanks}[1]{\footnote{#1}}
\newcommand{\starttext}{
\setcounter{footnote}{0}
\renewcommand{\thefootnote}{\arabic{footnote}}}

\newcommand{\bea}{\begin{eqnarray}}
\newcommand{\eea}{\end{eqnarray}}
\newcommand{\be}{\begin{eqnarray}}
\newcommand{\ee}{\end{eqnarray}}
\newcommand{\bma}{\begin{matrix}}
\newcommand{\ema}{\cr\end{matrix}}

\newtheorem{thm}{Theorem}[section]
\newtheorem{lem}[thm]{Lemma}

\newtheorem{cor}[thm]{Corollary}

\def\cC{{\cal C}}
\def\cD{{\cal D}}

\def\cF{{\cal F}}

\def\cN{{\cal N}}
\def\cO{{\cal O}}

\def\mA{\mathfrak{A}}
\def\mB{\mathfrak{B}}
\def\mC{\mathfrak{C}}

\def\ZZ{{\mathbb Z}}
\def\RR{{\mathbb R}}

\def\CC{{\mathbb C}}

\def\Re{{\rm Re \,}}
\def\Im{{\rm Im \,}}

\def\half{{1\over 2}}
\def\thalf{{\tfrac{1}{2}}}
\def\p{\partial}

\def\a{\alpha}
\def\b{\beta}
\def\g{\gamma}

\def\f{\varphi}

\def\ep{\varepsilon}
\def\om{\omega}

\def\no{\nonumber}
\def\sm{\smallskip}

\newcommand{\x}[1]{\textcolor{cyan}{[\bf #1 ]}}

\begin{document}
\starttext
\setcounter{footnote}{0}

\begin{flushright}
August 2022
\end{flushright}

\vskip 0.3in

\begin{center}

{\Huge \bf Exploring the Strong-Coupling Region \\[5pt] of $\mathbf{ SU(N)}$ Seiberg-Witten Theory }

\vskip 0.3in

{\large \bf Eric D'Hoker$,{ }^{1}$ Thomas T. Dumitrescu,${ }^{1}$ and Emily Nardoni${ \phantom ,}^{2}$} 

\vskip 0.1in

 { \sl ${}^{1}$ Mani L. Bhaumik Institute for Theoretical Physics}\\
{\sl Department of Physics and Astronomy }\\
{\sl University of California, Los Angeles, CA 90095, USA} \\

\vskip 0.1in

 { \sl ${}^{2}$ Kavli Institute for the Physics and Mathematics of the Universe (WPI)}\\
{\sl The University of Tokyo, Kashiwa, Chiba 277-8583, Japan}

\bigskip

{\tt \small dhoker@physics.ucla.edu, tdumitrescu@physics.ucla.edu, \\ emily.nardoni@ipmu.jp}

\end{center}

\begin{abstract}

\noindent We consider the Seiberg-Witten solution of pure~$\mathcal{N} =2$ gauge theory in four dimensions, with gauge group~$SU(N)$. A simple exact series expansion for the dependence of the~$2 (N-1)$ Seiberg-Witten periods~$a_I(u), a_{DI}(u)$ on the~$N-1$ Coulomb-branch moduli~$u_n$ is obtained around the~$\mathbb{Z}_{2N}$-symmetric point of the Coulomb branch, where all~$u_n$ vanish. This generalizes earlier results for $N=2$ in terms of hypergeometric functions, and for $N=3$ in terms of Appell functions. Using these and other analytical results, combined with numerical computations, we explore the global structure of the K\"ahler potential~$K = \frac{1}{2\pi} \sum_I \text{Im}(\bar a_I  a_{DI})$, which is single valued on the Coulomb branch.  Evidence is presented that~$K$ is a convex function, with a unique minimum at the $\mathbb{Z}_{2N}$-symmetric point. Finally, we explore candidate walls of marginal stability in the vicinity of this point, and their relation to the surface of vanishing K\"ahler potential. 

\end{abstract}

\newpage

\baselineskip=15pt
\setcounter{equation}{0}
\setcounter{footnote}{0}

\newpage
\setcounter{tocdepth}{2}
\tableofcontents
\newpage


\section{Introduction}
\setcounter{equation}{0}
\label{sec:intro}

\subsection{An Expansion for the $SU(N)$ Seiberg-Witten Periods}

%

The Seiberg-Witten (SW) solution of $\mathcal{N}=2$ $SU(2)$ gauge theory in four dimensions~\cite{Seiberg:1994rs,Seiberg_1994} and its generalizations have led to many new insights into the dynamics of gauge theories and string theories at strong coupling. In the original work \cite{Seiberg:1994rs}, the exact low-energy effective Lagrangian on the Coulomb branch of the pure~$SU(2)$ gauge theory was determined by computing the periods of a suitable SW one-form over the homology cycles of an auxiliary genus-one Riemann surface -- the SW curve. 
 
 \sm
 
In this paper, we revisit the generalization of the SW solution to pure $SU(N)$ gauge theory~\cite{Klemm:1994qs,Argyres:1994xh,Klemm:1995wp}, beginning with a brief sketch of its salient features. These theories have a Coulomb branch that is parameterized by $N-1$ complex coordinates $u_n$, $n=0,\ldots,N-2$, in correspondence with the gauge-invariant products of the $SU(N)$ vector multiplet scalars. At generic points on the Coulomb branch, the low-energy theory on the Coulomb branch is a~$U(1)^{N-1}$ gauge theory described by $N-1$ abelian $\cN=2$ vector multiplets $A_I$, $I=1,\cdots,N-1$, whose complex scalar bottom components we denote by $a_I$. Note that the~$a_I(u)$ are locally (but not globally) holomorphic functions of the Coulomb branch moduli.

\sm

The leading long-distance interactions of these vector multiplets are completely determined if we also specify~$N-1$ locally holomorphic functions~$a_{DI}(u)$, which can be thought of as vector multiplet scalars in a magnetic dual description. Then the K\"ahler potential describing the sigma model for the scalars is given by
\bea
\label{k1}
K = { i \over 4 \pi} \sum_{I=1} ^{N-1} \Big ( a_I \, \bar a_{DI} - \bar a_I \, a_{DI} \Big )\,,
\eea
while the symmetric matrix of complexified~$U(1)^{N-1}$ gauge couplings is given by~$\tau_{IJ} = {\p a_{DI} / \p a_J}$. It follows from~${\cal N}=2$ supersymmetry that the positive-definite K\"ahler metric~$g_{I {J}} = {\partial^2 K} / \partial a^I \partial \bar{a}^J$ is (up to a positive constant) the same as the imaginary part of~$\tau_{IJ}$. 

\sm

The SW solution can be expressed through the~$2(N-1)$  special coordinates $a_I(u),a_{DI}(u)$ (also called SW periods), which in turn are identified with the periods of a meromorphic  one-form~$\lambda$ on a canonical basis of homology cycles $(\mA_I,\mB_I)$ of a genus $N-1$ hyperelliptic  curve~${\cal C}(u)$ that depends on the moduli,
\bea
\label{per1}
2 \pi i \, a_I = \oint _{\mA_I} \lambda \, ,
\qquad 
2 \pi i \, a_{DI} = \oint _{\mB_I} \lambda \, .
\eea
The derivatives of the SW differential with respect to the moduli $u_n$ are holomorphic one-forms, which is sufficient to ensure the positivity of the K{\"a}hler metric.  Different choices of  homology basis $(\mA_I,\mB_I)$ act on the special coordinates as electric-magnetic duality transformations: a change of homology basis by $M\in \text{Sp}(2(N-1),\mathbb{Z})$ preserves the canonical intersection pairing, while the period vector $v=(a_{DI}, a_I)$ transforms under such a duality transformation in the doublet representation. 

\sm

The presentation of the~$SU(N)$ SW solution sketched above is deceptively simple: obtaining explicit, tractable expressions for the periods that are valid in varied regions of moduli space is in general a formidable challenge.  For gauge group $SU(2)$, the periods are elliptic integrals and can be expressed in terms of Gauss hypergeometric functions  \cite{Seiberg:1994rs}. The latter have well-known analytic continuations in the entire $u$-plane. For general $N$,  efficient methods for obtaining the period integrals in certain limits have been developed in~\cite{DHoker:1996pva,DHoker:1996yyu,DHoker:1997mlo,Masuda:1996xk,Edelstein:1999fz,Edelstein:1999tb,DHoker:2020qlp}.   The $SU(N)$ periods are known to satisfy Picard-Fuchs differential equations as functions of the moduli~$u_n$ \cite{Klemm:1995wp,Ito:1995ga} (see also~\cite{Isidro:1996hj,Alishahiha:1996rb}). For~$SU(3)$ gauge group, the solutions to these Picard-Fuchs equations were shown in~\cite{Klemm:1995wp} to be given by Appell $F_4$ functions. However, the complexity of the Picard-Fuchs equations increases rapidly with~$N$.
 
 \sm
 
A principal result of this paper is the derivation of a simple, exact series expansion for the $a_I, a_{DI}$ periods around the origin of moduli space, where all~$u_n$ vanish, generalizing the earlier results for $N=2$ and $N=3$. This result is expressed as Theorem \ref{thm:1} below; it is then refined in several Corollaries and exploited in various applications. For $SU(2)$ and $SU(3)$ gauge groups, our series expansion reduces to the known hypergeometric and Appell function representations, respectively. For higher $N$, it takes the form of a series expansion in the moduli $u_n$, which is optimal in the sense that the coefficient of each monomial $u_0^{\ell_0}\cdots u_{N-2}^{\ell_{N-2}}$ consists of a single factorized term.

\subsection{Exploring the~$SU(N)$ K\"ahler Potential}

In addition to the SW periods~$a_{I}, a_{DI}$ themselves, another object whose properties we wish to illuminate is the K{\"a}hler potential~\eqref{k1}. Besides being an important ingredient in the low-energy Lagrangian, our interest in the K\"ahler potential~$K$ derives from the observation \cite{Cordova:2018acb} that $K$ plays a critical role in a certain supersymmetry-breaking scenario that connects four-dimensional $\cN=2$ Yang-Mills theory with gauge group $G$ to a non-supersymmetric $G$ gauge theory with two Weyl fermions transforming in the adjoint representation of $G$ -- in short, $N_f=2$ adjoint QCD.  The renormalization group flow from the supersymmetric to the non-supersymmetric theory is triggered by the soft supersymmetry-breaking deformation $\mathcal{T}_\text{UV} \sim \text{Tr} (\bar{\phi}\phi)$, which gives mass to the complex $\cN=2$ vector multiplet scalars.  Crucially, the operator $\mathcal{T}_\text{UV}$ is the bottom component of the protected~${\cal N} = 2$ stress-tensor supermultiplet; it can therefore be reliably tracked to the low-energy description on the Coulomb branch, where it is identified with the K{\"a}hler potential, i.e.~$\mathcal{T}_\text{UV} \rightarrow \mathcal{T}_\text{IR} \sim K$. Several comments are in order:

\begin{itemize}
\item[(i)] Since~${\cal T}_\text{UV}$ is a well-defined operator and it flows to~$K$ in the IR, it follows that the K\"ahler potential given by~\eqref{k1bis} is a well-defined function on the Coulomb branch, i.e.~it does not suffer from the usual K\"ahler ambiguities, and it is invariant under $\text{Sp}(2(N-1), \ZZ)$ electric-magnetic duality transformations.\footnote{~The former statement is similar to constraints on K\"ahler potentials in theories with four supercharges~\cite{Komargodski:2010rb,Dumitrescu:2011iu}. The latter statement is manifest from the definition~\eqref{k1bis}.}

\item[(ii)] Even though~$\mathcal{T}_\text{UV} \sim \text{Tr} (\bar{\phi}\phi)$ is classically positive-definite, quantum effects can render its expectation value negative. Indeed we show below that there is a region of the Coulomb branch -- which we term the strong-coupling region -- where~$K < 0$. Note that this region is well defined, because~$K$ is well defined (see (i) above).\footnote{~This region should not be confused with the strong-coupling chamber for massive BPS states, which will also make an appearance below.}  

\item[(iii)] The deformation by~${\cal T}_\text{UV}$ leads to a supersymmetry-breaking scalar potential proportional to~$K$ on the Coulomb branch. In this way, the properties of $K$ lead to a prediction for the vacuum structure of  non-supersymmetric~$N_f = 2$ adjoint QCD, which is reliable if the supersymmetry-breaking mass scale is small compared to the strong coupling scale $\Lambda$ of the~${\cal N} = 2$ theory.  In upcoming work \cite{DDGNtemp}, we utilize this perspective to explore the phases of $N_f=2$ adjoint QCD with gauge group $SU(N)$.
\end{itemize}

\begin{figure}[t!]
\centering
\includegraphics[width=0.55\textwidth]{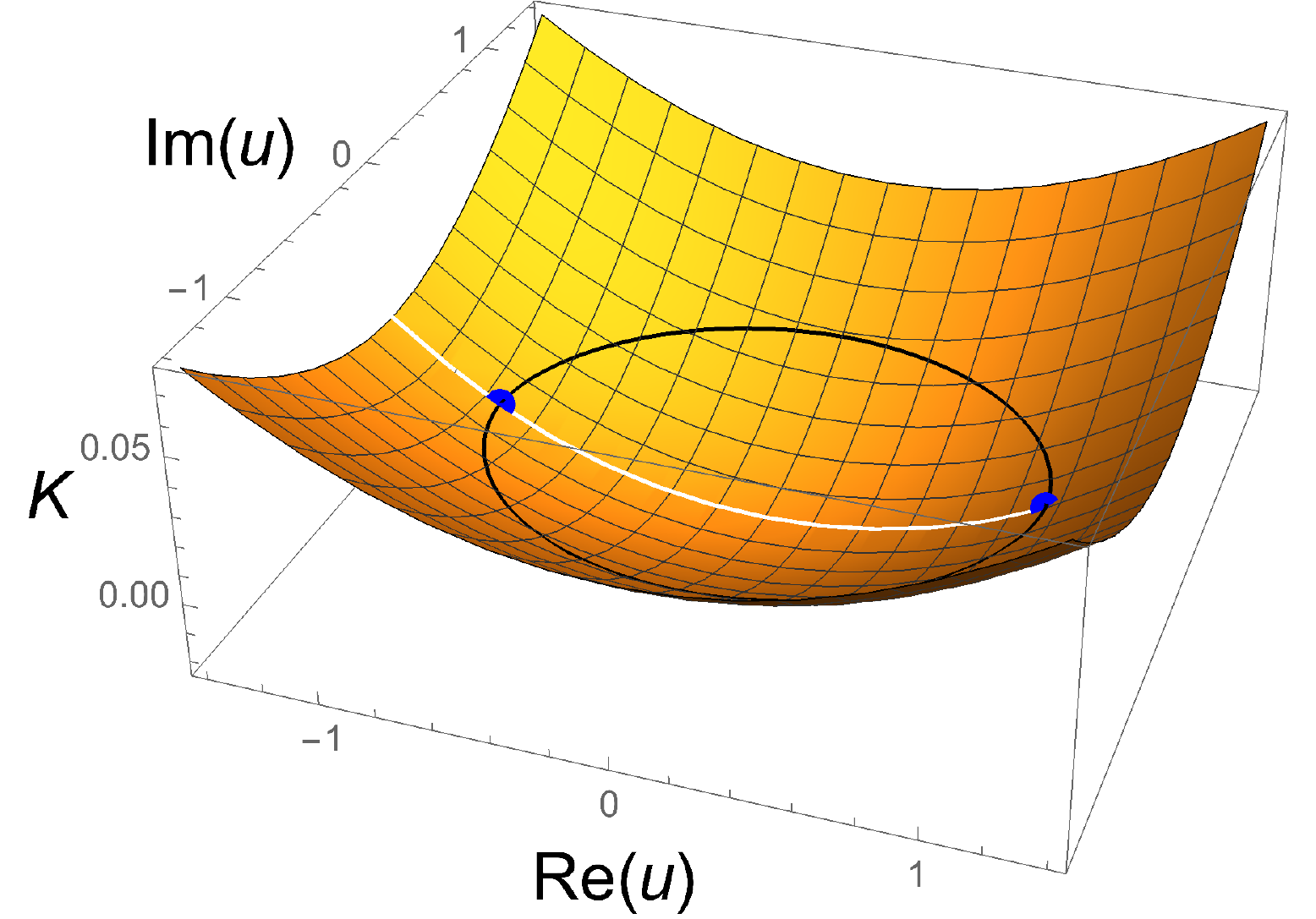}
\caption{The K{\"a}hler potential for gauge group $SU(2)$ in the complex $u$-plane (with $\Lambda=1$), made using the hypergeometric function representation of the periods (see section~\ref{sec:su2check}).  The black curve indicates the $K=0$ contour; it is the boundary of the strong-coupling region where~$K < 0$. Note that the monopole and dyon points (indicated by the blue dots) lie on this~$K = 0$ boundary.  \label{fig:Ksu2better}}
\end{figure}

With the preceding motivation in mind, we here apply our explicit and (relatively) simple expressions for the SW periods, along with other analytic and numeric methods, to study the global structure of the SW K{\"a}hler potential~$K$, focusing on its stationary points and convexity properties. For the case of gauge group $SU(2)$, $K(u)$ is a convex function of the single complex modulus $u$, with a single minimum at the origin~$u = 0$ of moduli space. This minimum is depicted in figure \ref{fig:Ksu2better}, and was also previously discussed in \cite{Luty:1999qc,Cordova:2018acb} in the context of supersymmetry-breaking.

\sm

For general $N>2$, $K$ is a function of $N-1$ complex variables, and the extraction of its properties is considerably more involved. We prove that for general $N$, the origin of moduli space where all $u_n=0$ is a stationary point (see section \ref{sec:stat}). Since the SW curve is invariant under a $\mathbb{Z}_{2N}$ symmetry when all the $u_n=0$  -- corresponding to the action of the unbroken $\mathbb{Z}_{4N}$ $R$-symmetry on the moduli space --  we refer to this point as the $\mathbb{Z}_{2N}$-symmetric point (or simply the~$\ZZ_{2N}$ point). We compute the value $K(0)$ at the  $\mathbb{Z}_{2N}$-symmetric point and show that it is negative, scaling with $N$ as $K(0) \sim - N^2$ (and given explicitly in \eqref{symm}). Then, much as in $SU(2)$, there is a region around the origin for which $K$ is negative, $K < 0$, which we refer to as the strong coupling region. The boundary of this region for $SU(2)$ is depicted by the black curve in figure \ref{fig:Ksu2better}, which contains the singular monopole and dyon points. For general $N$, the $N$ multi-monopole points -- the $N>2$ generalization of the monopole and dyon points studied in \cite{Douglas:1995nw} --  also lie on the $K=0$ boundary of this region. In section \ref{sec:num} we numerically study this $K=0$ surface for the case of gauge group $SU(3)$, slices of which are depicted in figures \ref{fig:K0contoursu} and \ref{fig:K0contoursv}. 

\sm

It is natural to conjecture that the origin is the unique global minimum of the K{\"a}hler potential, and that $K$ is everywhere convex, for all $N$. While this conjecture is as of yet proven only for gauge group $SU(2)$,  it is supported by the following evidence that we present throughout the paper:

\begin{itemize}

\item  If~$K$ has another stationary point, it occurs at a negative value of $K$, i.e.~within the strong coupling region. The proof of this statement is given in section \ref{sec:kneg}.  Thus one may restrict the search for another minimum to this region.

\item For gauge group $SU(3)$, we have thoroughly explored the K{\"a}hler potential numerically. This analysis is presented in section \ref{sec:num}. Our numerical studies have shown $K$ to be convex on every slice upon which we have evaluated it, with no evidence for a minimum away from the~$\ZZ_{2N}$-symmetric point at the origin.

\item We obtain some partial analytic results that are valid for all~$N$ and consistent with the convexity conjecture. For instance, on the slice parameterized by a single modulus $u_0$ (with all other~$u_n = 0$) there is no stationary point for $\text{Im}(u_0)\neq 0$ (see section \ref{sec:pp}).

\end{itemize}

\noindent It would be interesting to find a definitive proof (or disproof) of this conjecture, but we leave this problem for the future.

\subsection{Charting Candidate Curves of Marginal Stability}

It has been known since~\cite{Seiberg:1994rs} that there are walls of marginal stability on the Coulomb branch. In the simplest cases, such walls are loci where a massive BPS particle becomes marginally unstable to decay into two (or more) other BPS particles -- or conversely, loci where two (or more) BPS particles can form threshold bound states.\footnote{~In general more complicated phenomena are possible when a wall is crossed.} This occurs when the complex central charges of the BPS particles in question, which are determined by the SW periods, have aligned phases. The~$SU(2)$ case was completely understood in~\cite{Seiberg:1994rs,Ferrari:1996sv}: a wall of marginal stability separates the moduli space into a strong-coupling chamber around the origin, and a weak-coupling chamber extending out to infinity. This wall precisely coincides with the black $K=0$ contour depicted in figure \ref{fig:Ksu2better}, i.e.~the~$SU(2)$ strong-coupling chamber for massive BPS states is the same as our strong-coupling region defined by~$K < 0$. 

\sm

Already the case of~$SU(3)$ gauge group is much richer, see for instance~\cite{Taylor:2001hg,Taylor:2002sg,Galakhov:2013oja} for discussions specifically focusing on this case, with many additional references to the large literature on BPS states and wall crossing in 4d~${\cal N} = 2$ theories. 

\sm

As another application of our formulas for the SW periods of~$SU(N)$ gauge theory, in section~\ref{sec:MS} we map out some candidate walls of marginal stability within the strong coupling~$K < 0$ region near the origin of the Coulomb branch. We emphasize that the scope of our analysis is narrow: we do not claim to find all walls of marginal stability, nor do we analyze the much more delicate question of which BPS states actually decay or form bound states as we cross these walls. The results in this section should be viewed as motivation for a more detailed study of this problem.  


\subsection*{Outline}

The remainder of this paper is organized as follows:

In section \ref{sec:periods} we present a simple series expansion for the $SU(N)$ SW periods around the $\mathbb{Z}_{2N}$-symmetric point at the origin of moduli space, as well as various simplifications of this expression for different special values of the moduli, and for low values of $N$.  
The proof of the main theorem is relegated to appendix  \ref{sec:A}, and the proof that the expansion reproduces the $SU(3)$ Appell functions appears in appendix \ref{sec:B}.

\sm

In section \ref{sec:Kahler1} we build on these results to express the K{\"a}hler potential in a simple diagonal form. We evaluate $K$ at the~$\ZZ_{2N}$-symmetric point, showing that it behaves (up to an~${\cal O}(1)$ positive coefficient) as $-N^2$ for $N\gg 1$, and additionally discuss the structure of the K{\"a}hler potential for restricted values of the moduli. 

\sm

In section \ref{sec:Kahler2} we collect a number of general results regarding the structure of the K{\"a}hler potential, including a proof that the~$\ZZ_{2N}$-symmetric point is a stationary point (as expected from the~$\ZZ_{2N}$-symmetry), and that  $K$ is negative at an arbitrary stationary point. We also re-express the derivatives of $K$ with respect to the moduli as two-real-dimensional integrals, which is useful both for proving some of the results in this section, and for numerical computations in later sections. 

\sm

In section \ref{sec:KP3} we restrict to the case of gauge group $SU(3)$. By mapping the SW curve and differential to an elliptic problem, we compute the SW periods on the cusp slice of moduli space, where the discriminant of the curve vanishes -- and which includes both the multi-monopole and the Argyres-Douglas~\cite{Argyres:1995jj} points of~$SU(3)$. We use these results to evaluate $K$ at special points of the moduli space. Next, we  describe the results of a numerical exploration of $K$, presenting evidence that the K\"ahler potential is convex and that the~$\ZZ_{2N}$-symmetric point at the origin is the only minimum. Appendix \ref{sec:C} includes a brief review of elliptic functions and modular forms as needed for the analysis of this section, and numerical methods for evaluating~$K$ are discussed in appendix~\ref{sec:appnum}. 

\sm

In section \ref{sec:MS} we investigate candidate walls of marginal stability within the~$K < 0$ strong-coupling region. We mostly focus on special slices of $SU(3)$ moduli space, but also present some results for general~$N$. A summary of the BPS particles that are stable in the strong-coupling chamber of the~$SU(N)$ theory (with emphasis on the case~$N = 3$) appears in appendix \ref{sec:BPSapp}.

\subsection*{Acknowledgements}

The research of ED is supported in part by the National Science Foundation under grants PHY-19-14412 and  PHY-22-09700. The research of EN is supported by World Premier International Research Center Initiative (WPI), MEXT, Japan. EN also acknowledges the Aspen Center for Physics where part of this work was performed, which is supported by National Science Foundation grant PHY-1607611. TD is supported by a DOE Early Career Award under DE-SC0020421, by the Simons Collaboration on Global Categorical Symmetries, and by the Mani L. Bhaumik Presidential Chair in Theoretical
Physics at UCLA.  We are grateful to P.~Dumitrescu, A.~Neitzke, and~F.~Yan for useful discussions. We especially thank E.~Gerchkovitz for collaboration and discussion on closely related work~\cite{DHoker:2020qlp,DDGNtemp}.

\newpage

\section{Expanding the Periods around the $\ZZ_{2N}$ Point}
\setcounter{equation}{0}
\label{sec:periods}

In this section we analyze the SW periods in the vicinity of the~$\ZZ_{2N}$-symmetric SW curve, or equivalently the~$\ZZ_{2N}$-symmetric origin, where all~$u_n = 0$, of the Coulomb branch. 

\subsection{Seiberg-Witten Review} 

We consider pure $\cN=2$ supersymmetric Yang-Mills theory with gauge group $SU(N)$ (for any~$N \geq 2$) and no hypermultiplets in four space-time dimensions. We begin by briefly reviewing the Seiberg-Witten~(SW) solution for this class of theories, which was found in~\cite{Seiberg:1994rs,Klemm:1994qs,Argyres:1994xh,Klemm:1995wp}. This solution determines the vector multiplet scalars~$a_I(u)$ and their magnetic duals~$a_{DI}(u)$ as locally holomorphic functions of the gauge invariant Coulomb branch moduli~$u_n~(n = 0, 1, \ldots, N-2)$. This is accomplished by expressing them as periods of a meromorphic SW one-form (or differential)~$\lambda$ over a canonical basis of homology one-cycles $\mA_I$ and $\mB_I$,
\bea
\label{2.SWP}
2 \pi i \, a_I = \oint _{\mA_I} \lambda \, ,
\hskip0.91in 
2 \pi i \, a_{DI} = \oint _{\mB_I} \lambda \, .
\eea
on a family of  curves $\cC(u)$ (known as SW curves) that depend holomorphically on the moduli~$u_n$. Recall that, given a compact, oriented surface a canonical homology basis of oriented one-cycles $(\mA_I, \mB_I)$ is defined by the following intersection pairings,
\bea
\label{2.CHB}
 \#(\mA_I, \mB_J) = \delta_{IJ}~, \qquad \#(\mA_I, \mA_J) = 0~, \qquad \#(\mB_I,  \mB_J) = 0~,
\eea
where~$\#(X, Y) = - \#(Y, X)$ denotes the antisymmetric intersection pairing of the oriented one-cycles~$X$ and~$Y$.

\sm

It is standard to introduce a locally defined and holomorphic prepotential~$\cF(a)$, which captures the relationship between the electric and magnetic periods~$a_I(u)$ and~$a_{DI}(u)$, as well as the symmetric matrix~$\tau_{IJ}(u)$ of effective holomorphic~$U(1)^{N-1}$ gauge couplings on the Coulomb branch, 
\bea
\label{2.SWF}
a_{DI} = { \p \cF \over \p a_I}~, \qquad \tau_{IJ} = { \p a_{DI} \over \p a_J} = {\p^2 \cF \over \p a _I \p a_J} = \tau_{JI}~, \qquad (I,J=1,\cdots, N-1)~.
\eea

The SW curve $\cC(u)$ is parametrized by the $N-1$ complex moduli $u_0, \cdots, u_{N-2}$. For each value of the~$u_n$, the curve is hyper-elliptic and can be chosen to take the following form,
\bea
\label{2.SWC}
y^2 = A(x)^2 -\Lambda ^{2N} \, ,
\hskip 0.8in 
A(x) = x^N - \sum _{n=0}^{N-2} u_n x^n\, ,
\eea
where $\Lambda$ is the strong-coupling scale of the non-Abelian gauge theory.\footnote{~\label{fn:Lrel}For completeness, we note that our conventions for the curve differ from those in \cite{DHoker:1997mlo,DHoker:2020qlp} by an $N$-dependent redefinition of the strong-coupling scale $4 \Lambda_\text{there}^{2N} = \Lambda_\text{here}^{2N}~.
$} For the remainder of this paper we set~$\Lambda = 1$, so that all quantities are dimensionless. In terms of the data in~\eqref{2.SWC}, the SW differential~$\lambda$ is given by
\bea
\label{2.SWD}
\lambda = { x A'(x) dx \over y}\, .
\eea
By construction, the derivatives of the SW differential $\lambda$ with respect to the moduli $u_n$ produce   holomorphic Abelian differentials, modulo exact differentials, 
\bea
\label{2.holodif}
{ \p \lambda \over \p u_n} =\om_n - dx { \p \over \p x } \left ( { x^{n+1} \over \sqrt{A(x)^2-1}} \right )\, ,
\hskip 0.5in  \om_n = { x^n dx \over \sqrt{A(x)^2-1}}~.
\eea
Here the~$\om_n$ (with~$n = 0, \ldots, N-2$) are holomorphic Abelian differentials of the first kind, which furnish a basis for the Dolbeault cohomology $H^{(1,0)}(\cC(u), \CC)$.  It follows that the matrix $\tau_{IJ}(u)$ is the period matrix of the curve $\cC(u)$ for a given set of $u_n$. By the Riemann bilinear relations, $\tau_{IJ}(u)$ is symmetric and has positive definite imaginary part, as required for a matrix of complexified~$U(1)^{N-1}$ gauge couplings in a unitary theory.

\subsection{The $\ZZ_{2N}$-Symmetric Curve}

The Seiberg-Witten curve $\cC(0)$ at the origin of the Coulomb branch, obtained by setting all~$u_n=0$ in~\eqref{2.SWC}, is given by 
\bea
y^2 = x^{2N}-1~.
\eea
Note that $\cC(0)$ is manifestly invariant under the following~$\ZZ_{2N}$ transformation,
\bea
(x,y) \to (\ep x, \pm y)~, \qquad \ep=e^{\frac{2 \pi i}{2N}}~,
\eea
with~$\ep$ a~$2N$-th root of unity. This symmetry descends from a physical $\mathbb{Z}_{4N}$ discrete $R$-symmetry of the~$\cN = 2$ gauge theory, whose quotient by fermion parity~$(-1)^F$ acts on the bosonic moduli space coordinates~$u_n$ via the following~$\ZZ_{2N}$ action,
\bea
u_n \; \to \; e^{\frac{2\pi i (N-n)}{2N}} u_n~.
\eea
Thus the origin, where all~$u_n = 0$, is the unique point on the Coulomb branch where this~$\ZZ_{2N}$ symmetry is unbroken. For this reason we refer to $\cC(0)$ as the $\ZZ_{2N}$-symmetric curve.

\begin{figure}[htb]
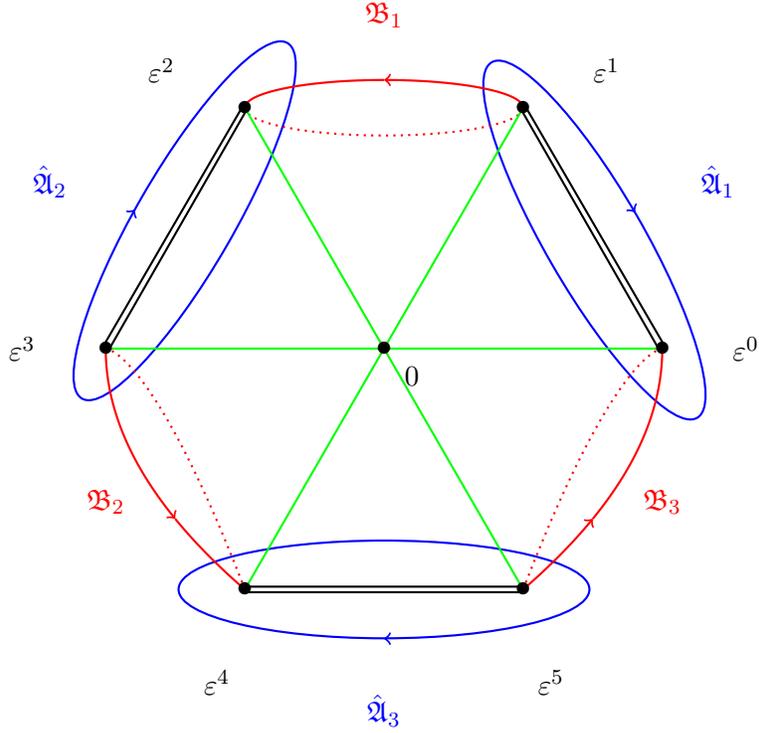

\begin{center}
\tikzpicture[scale=3.7]
\scope[xshift=0cm,yshift=0cm]
\draw [thick] (-0.5,-0.856) -- (0.5,-0.856);
\draw [thick] (-0.5,-0.876) -- (0.5,-0.876);
\draw [thick] (-0.495,0.856) -- (-0.995,-0.015);
\draw [thick] (-0.505,0.876) -- (-1.005,0.015);
\draw [thick] (0.495,0.856) -- (0.995,-0.015);
\draw [thick] (0.505,0.876) -- (1.005,0.015);
\draw[thick, color=blue, rotate=0] (0,-0.866)  ellipse (21pt and 5pt);
\draw[thick, color=blue, rotate=60] (0.04,0.85)  ellipse (21pt and 5pt);
\draw[thick, color=blue, rotate=-60] (0.04,0.85)  ellipse (21pt and 5pt);
\draw  [red, thick, domain=0:90,->] plot ({0.5*cos(\x)},{0.866+0.1*sin(\x)});
\draw  [red, thick, domain=90:180] plot ({0.5*cos(\x)},{0.866+0.1*sin(\x)});
\draw  [red, thick, dotted, domain=180:360] plot ({0.5*cos(\x)},{0.866+0.1*sin(\x)});
\draw  [red, thick, domain=0:90,->] plot ({-0.75-0.25*cos(\x)},{-0.866*sin(\x/2)});
\draw  [red, thick, domain=90:180] plot ({-0.75-0.25*cos(\x)},{-0.866*sin(\x/2)});
\draw  [red, thick, dotted, domain=180:360] plot ({-0.75-0.25*cos(\x)},{-0.866*sin(\x/2)^3});
\draw  [red, thick, domain=0:90] plot ({0.75+0.25*cos(\x)},{-0.866*sin(\x/2)});
\draw  [red, thick, domain=90:180,<-] plot ({0.75+0.25*cos(\x)},{-0.866*sin(\x/2)});
\draw  [red, thick, dotted, domain=180:360] plot ({0.75+0.25*cos(\x)},{-0.866*sin(\x/2)^3});
\draw [thick, blue,->] (0.895,0.5)--(0.902,0.49);
\draw [thick, blue,->] (0.01,-1.043)--(0,-1.043);
\draw [thick, blue,->] (-0.902,0.49) -- (-0.895,0.5);
\draw [thick, green] (0,0) -- (1,0);
\draw [thick, green] (0,0) -- (0.5,0.866);
\draw [thick, green] (0,0) -- (-0.5,0.866);
\draw [thick, green] (0,0) -- (-1,0);
\draw [thick, green] (0,0) -- (-0.5,-0.866);
\draw [thick, green] (0,0) -- (0.5,-0.866);
\draw (0,0) node{$\bullet$};
\draw (1,0) node{$\bullet$};
\draw (-1,0) node{$\bullet$};
\draw (0.5,0.866) node{$\bullet$};
\draw (-0.5,0.866) node{$\bullet$};
\draw (0.5,-0.866) node{$\bullet$};
\draw (-0.5,-0.866) node{$\bullet$};
\draw (0.1,-0.1) node{\small $0$};
\draw (1.3,0) node{\small $\ep^0$};
\draw (0.8,1) node{\small $\ep^1$};
\draw (-0.8,1) node{\small $\ep^2$};
\draw (-1.3,0) node{\small $\ep^3$};
\draw (-0.6,-1.2) node{\small $\ep^4$};
\draw (0.6,-1.2) node{\small $\ep^5$};
\draw [blue] (1.2,0.6) node{\small $\hat \mA_1$};
\draw [blue] (-1.2,0.6) node{\small $\hat \mA_2$};
\draw [blue] (0,-1.3) node{\small $\hat \mA_3$};
\draw [red] (0,1.2) node{\small $\mB_1$};
\draw [red] (-1,-0.55) node{\small $\mB_2$};
\draw [red] (1,-0.55) node{\small $\mB_3$};
\endscope
\endtikzpicture
\caption{Branch points $x = \ep^n$ for the $\ZZ_{2N}$-symmetric curve $y^2=x^{2N}-1$ for the special case~$N = 3$, for which $\ep=e^{2\pi i /6}$. The branch cuts are denoted by black double lines, while the~$\hat \mA_J$ and~$\mB_I$ cycles defined above are shown in blue and red, respectively. The integration paths used to define the function $Q(\xi)$ (see~\eqref{2.Q} below) are shown in green.  \label{fig:1}}
\end{center}
\end{figure}

The branch points of $\cC(0)$ are the $2N$ roots of unity $\ep^n$, with  $n=0, \cdots, 2N-1$, and the branch cuts in the hyper-elliptic representation may be chosen to lie along the intervals~$[\ep^{2I-2}, \ep^{2I-1}]$ with $I=1, \cdots, N$. We define the cycles~$\mA_I$ and~$\mB_I$ (with $I=1, \cdots, N$) as follows,
\bea
\mA_I =\bigcup _{J=1}^I \hat \mA_J \,,
\hskip 0.7in
\hat \mA_I = (\ep^{2I-2}, \ep^{2I-1})\,,
\hskip 0.7in
\mB_I = (\ep^{2I-1}, \ep^{2I})~.
\eea
Here the intervals indicating the cycles are somewhat schematic. In truth, the cycles are closed curves surrounding the branch points, as indicated in detail in figure~\ref{fig:1} for the case~$N=3$. Crucially, we choose the orientations of the cycles as indicated in the figure: clockwise for~$\hat \mA_J$,  $\mA_I$ and counter-clockwise for $\mB_I$. This ensures that~$\mA_I$ and~$\mB_I$ comprise a canonical homology basis in the sense of~\eqref{2.CHB}.    

\sm

{\it Warning:} The cycles $\mA_I$ and~$\mB_I$ defined above (which are very convenient near the origin) do not coincide with other common duality frames used in SW theory, e.g.~the standard electric frame that is simply related to the UV theory by Higgsing (and that, in a suitable sense, becomes weakly coupled at infinity in the moduli space), or the standard magnetic frame that becomes weakly-coupled at the multi-monopole point where the maximal number of mutually local monopoles become massless.  Rather, our $\mA_I$ and~$\mB_I$ cycles are related to these bases by an electric-magnetic duality transformation. We will see this explicitly below.

\sm

Note that the $N$ pairs of $\mA$- and $\mB$-cycles defined above are not independent: the union of all~$\mA$-cycles is homologically trivial, and so is the union of all~$\mB$-cycles. Via~\eqref{2.SWP}, this translates into 
\bea
\sum_{I=1}^N a_I = \sum _{I=1}^N a_{DI} =0\,,
\eea
as required in~$SU(N)$ gauge theory.

\subsection{Expanding Around the $\ZZ_{2N}$-Symmetric Curve}

Given the conventions spelled out above, the periods~$a_I(u)$, $a_{DI}(u)$ (with~$I=1,\cdots, N-1$) can be expressed as follows, 
\bea
\label{2.aQ}
a_I =   \sum_{J=1}^I \big \{ Q(\ep^{2J-1} ) - Q(\ep^{2J-2}) \big \}~, \qquad
a_{DI} = Q(\ep^{2I} ) - Q(\ep^{2I-1}) ~, \qquad \ep=e^{\frac{2 \pi i }{2N}}~.
\eea
Here~$Q(\xi)$ is a function on the~$2N$-th roots of unity~$\xi$ (i.e.~$\xi^{2N} = 1$), which is defined as the integral of the SW differential~$\lambda$ in~\eqref{2.SWD} along a path from~$x = 0$ to~$x = \xi$ (i.e.~one of the green paths in figure~\ref{fig:1}), 
\bea
\label{2.Q}
 \pi i \, Q(\xi) = \int _0 ^\xi \lambda~.
\eea
As usual, the factors of 2 in (\ref{2.aQ})  account for the fact that the integral over a full cycle is twice the integral over the corresponding  interval on a single sheet of the curve.

\sm

The function $Q(\xi)$ is a hyper-elliptic integral whose series expansion in powers of the moduli $u_n$ is given by the following theorem:
{\thm 
\label{thm:1} The function $Q(\xi)$ has the following series expansion around~$u_n = 0$, 
\bea
Q(\xi) =
 \sum _{{\{\ell_n\}= 0 \atop n=0, \ldots, N-2}}^\infty V_{L,M} ( \xi )  \, 
{ u_0^{\ell_0} \cdots u_{N-2}^{\ell_{N-2}} \over \ell_0! \, \cdots \ell_{N-2}! }~,
\qquad
L = \sum_{j=0}^{N-2} j \ell_j ~,
\qquad
M=\sum_{j=0}^{N-2} \ell_j~,
\eea
where the coefficients $V_{L,M}(\xi)$ are  given by,
\bea
V_{L,M}(\xi) = 
{  2^{M-(L +1)/N}   \over 2 \pi^2 N} \, \xi ^{NM+L+N+1}  
\Gamma ( \tfrac{L+1}{N}) \Gamma( \tfrac{NM-L-1}{2N}  )^2 \sin^2 \big ( \pi  \tfrac{NM-L-1}{2N}  \big ) \,.
\eea}

\noindent The proof of this theorem is essentially calculational; we defer the details to appendix~\ref{sec:A}. Note that the resulting series expansion is optimal, i.e.~the coefficient of each monomial $u_0^{\ell_0} \cdots u_{N-2}^{\ell_{N-2}}$ consists of a single factorized term.
 The following corollaries are  direct consequences of Theorem~\ref{thm:1}.
{\cor 
\label{cor:1}
The summation over $\ell_0$ may be carried out to express $Q(\xi)$ in terms of an infinite series of Gauss hypergeometric functions $F(a, b; c; z)={}_2F_1(a, b; c; z)$,
\bea
Q(\xi) = \sum _{\ell_1, \ldots , \ell_{N-2}=0 }^\infty  
{ 2^{M_0 -(L +1)/N} \,  \over 2 \pi^2 N} \, \xi ^{NM_0 +L+N+1}  \, \Gamma ( \tfrac{L+1}{N})  \, Y_{M_0} (\xi^N, L)
\, { u_1^{\ell_1} \cdots u_{N-2}^{\ell_{N-2}} \over \ell_1! \, \cdots \ell_{N-2}! }~,
\eea
where $L= \sum _{j=0}^{N-2} j \ell_j$ and $M_0 = \sum _{j=1}^{N-2} \ell_j$, 
while the coefficients $Y_{M_0} (\xi^N, L)$ are given by 
\bea
Y_{M_0} (\xi^N, L) & = & 
2  u_0 \, \xi^N  \cos^2 \big ( \pi  \tfrac{NM_0  -L-1}{2N}  \big )  \Gamma(\tfrac{NM_0+N  -L-1}{2N}  )^2 
 F\left (\tfrac{NM_0 +N-L-1}{2N}, \, \tfrac{NM_0+N-L-1}{2N} ; \, \tfrac{3}{2}; \, u_0^2 \right )
\no \\ && + 
\sin^2 \big ( \pi  \tfrac{NM_0  -L-1}{2N}  \big )  \Gamma(\tfrac{NM_0 -L-1}{2N}  )^2 \, F \left (\tfrac{NM_0 -L-1}{2N}, \, \tfrac{NM_0 -L-1}{2N}; \, \thalf; \, u_0^2 \right )\,.
 \eea}

 {\cor
 \label{cor:4}
In the special case where $u_n=0$ for all $n \not=0$, the function $Q(\xi)$ is given by a linear combination of Gauss hypergeometric functions $Q(\xi) = \xi Q_1+ \xi^{N+1} Q_{N+1}$, with
\bea
Q_1 & = &  {  u_0 \, \Gamma ( 1+ \tfrac{1}{2N}) \over 2 \, \Gamma (\tfrac{3}{2}) \, \Gamma( \half + \tfrac{1}{2N}) } \, 
F \big (\tfrac{N-1}{2N}, \tfrac{N-1}{2N}; \tfrac{3}{2}; u_0^2 \big )\,,
\no \\
Q_{N+1} & = &  { \Gamma ( \thalf+ \tfrac{1}{2N}) \over 2 \, \Gamma(\tfrac{1}{2}) \, \Gamma(1+\tfrac{1}{2N})}
\,   F \big (-\tfrac{1}{2N}, - \tfrac{1}{2N}; \thalf ; u_0^2 \big )\,.
\eea}

\noindent The proof of these corollaries, using the results of Theorem \ref{thm:1} as well as the standard series representation of~$F(a, b; c; z)$, is essentially straightforward and left to the reader. Note that while the hypergeometric functions appearing in these corollaries may be analytically continued to all values of~$u_0$, it is in general not clear how this affects the convergence of the resulting series expansions.

\subsection{Comparison to known~$SU(2)$ results} 
\label{sec:su2check}

When~$N = 2$, the only modulus is~$u_0$. Then~\eqref{2.aQ} (with~$\ep = i$) together with Corollary~\ref{cor:4} implies that
\bea
\label{2.su2us}
a & =&  Q(i) -Q(1) =  (i-1) Q_1 - (i + 1) Q_3~, \no \\
 a_D &=& Q(-1) - Q(i) = -  (1+i)Q_1 -  (1 - i) Q_3~,
\eea
where the functions $Q_1$ and $Q_3$  are defined in Corollary \ref{cor:4} by, 
\bea
Q_1(u_0) & = & {u_0 \Gamma({5 \over 4}) \over 2 \Gamma({3 \over 2}) \Gamma({3 \over 4}) } F(\tfrac{1}{4}, \tfrac{1}{4}; \tfrac{3}{2}; u_0^2)~, 
\no \\
Q_3(u_0) & = & { \Gamma({3 \over 4}) \over 2 \Gamma({1 \over 2}) \Gamma({5 \over 4}) } F(-\tfrac{1}{4}, -\tfrac{1}{4}; \tfrac{1}{2}; u_0^2)~.
\eea
Note that these functions have two symmetric branch cuts running from~$u_0 = \pm1$ to~$u_0 = \pm \infty$ along the real axis. 

\sm

Let us compare this to the~$SU(2)$ periods determined by Seiberg and Witten (SW) in \cite{Seiberg:1994rs,Seiberg_1994}; we will follow the conventions of their~\cite{Seiberg_1994}. Taking into account an overall factor of~$\sqrt 2$ that results from differently normalized strong coupling scales ($\Lambda_\text{us} = \sqrt{2} \Lambda_\text{SW}$, as in footnote~\ref{fn:Lrel}), we obtain, 
\bea
a_\text{SW}(u_0) & = & \sqrt{1 + u_0} \, F(-\tfrac{1}{2}, \tfrac{1}{2} ; 1 ; \tfrac{2}{ 1 + u_0})~, 
\no \\ 
a_{D, \text{SW}}(u_0) & = & {i(u_0 - 1) \over \sqrt 2} \, F(\tfrac{1}{2}, \tfrac{1}{2}; 2; \tfrac{1-u_0}{2})~. 
\eea
Here we are using a representation in terms of hypergeometric functions that was spelled out in~\cite{Alvarez-Gaume:1996ohl} (see also section 4.1 of~\cite{Cordova:2018acb}). The conventions are such that~$a_\text{SW}(u_0)$ has a branch cut running from the monopole point~$u_0 = 1$ to~$- \infty$ along the real axis, while~$a_{D, \text{SW}}(u_0)$ has a branch cut running from the dyon point~$u_0 = -1$ to~$-\infty$ along the real axis. Note that~$a_{D,\text{SW}} = 0$ at the monopole point~$u_0 = 1$. By contrast, we find that~$a = a_D \neq 0$ at~$u_0 = 1$.  

\sm

We claim that our periods~$a, a_D$ and the SW periods~$a_\text{SW}, a_{D, \text{SW}}$ are related by an electric-magnetic duality transformation. This transformation can be determined by analytically continuing~$a_\text{SW}(u_0)$ to the origin~$u_0 = 0$ by going above the monopole point and through the upper half plane.\footnote{~Due to the monodromy around the monopole point, other continuation paths will lead to different duality transformations.} We can then verify that\footnote{~This is straightforward for the second relation~$a_{D, \text{SW}} = a - a_D$~by explicitly expanding all hypergeometric functions around~$u_0 = 0$, where all of these expansions converge. In order to verify that~$a_\text{SW} = -a$ we can use {\tt Mathematica}, whose conventions for analytically continued hypergeometric functions agree with ours.} 
\begin{equation}
a_\text{SW} = - a~, \qquad a_{D, \text{SW}} = a - a_D~, \qquad \begin{pmatrix} - 1 & 0 \\ 1 & - 1 \end{pmatrix} \in SL(2, \ZZ)~.
\end{equation}

\subsection{Comparison to known~$SU(3)$ results} 

Explicit formulas for the periods in the case~$N = 3$ were obtained in~\cite{Klemm:1995wp} using Picard-Fuchs equations. The authors expressed their results in terms of Appell $F_4$ functions, which can be defined by the following series expansion,
\bea
F_4(a,b,c_1,c_2; x,y) = \sum _{m,n=0}^\infty {\Gamma (m+n+a) \Gamma (m+n+b) \Gamma (c_1) \Gamma (c_2) 
\over \Gamma (a) \Gamma (b) \Gamma (m+c_1) \Gamma (n+c_2) {\, m! \, n!}} \, x^m y^n\,.
\eea
We recover their results in the following corollary: 

{\cor 
\label{cor:2.3}
The periods for gauge group $SU(3)$ are given  by the relations (\ref{2.aQ}) in terms of the function $Q(\xi)$, which for $SU(3)$ simplify as follows, 
\begin{align}
a_1 & =  Q(\ep^1) - Q(\ep^0)~,& 
a_{D1} & =  Q(\ep^2) -  Q(\ep^1)~,
\no \\
a_2 & =  Q(\ep^1) - Q(\ep^0) +  Q(\ep^3) -  Q(\ep^2)~,&
 a_{D2} & = Q(\ep^4) -  Q(\ep^3)~.
\end{align}
The function~$Q(\xi)$ can be expanded in inequivalent representations of $\ZZ_{6}$,
\bea
Q(\xi) = \xi^4 Q_{0,0} + \xi^2 Q_{1,0} + \xi^0 Q_{2,0} + \xi^1 Q_{0,1} + \xi^5 Q_{1,1} + \xi^3 Q_{2,1}\,.
\eea
The formula for $Q(\xi)$ given in Theorem \ref{thm:1} may be recast in terms of Appell functions $Q_{s,t}$  expressed as follows in terms of the variables $x=4u^3/27$ and $y=v^2$ with $u=u_1$ and $v=u_0$, 
\bea
Q_{0,0} & = & { 2^{{1 \over 3}} \, 3^{{3 \over 2}} \, \Gamma(\tfrac{2}{3})^3 \over 4 \pi^2 } \,
F_4(-\tfrac{1}{6}, -\tfrac{1}{6}, \tfrac{2}{3}, \tfrac{1}{2}; x,y) \,,
\no \\
Q_{0,1} & = & {  2 \pi \over 2^{{1 \over 3}} \, 3^{{3 \over 2}} \, \Gamma(\tfrac{2}{3})^3  } \, 
v \, F_4(\tfrac{1}{3}, \tfrac{1}{3}, \tfrac{2}{3}, \tfrac{3}{2}; x,y) \,,
\no \\
Q_{1,0} & = &  { 2 \pi \over 2^{{1 \over 3}} \, 3^2 \, \Gamma(\tfrac{2}{3})^3  } \, 
u \, F_4(\tfrac{1}{6}, \tfrac{1}{6}, \tfrac{4}{3}, \tfrac{1}{2}; x,y) \,,
\no \\
Q_{1,1} & = & { 2^{{1 \over 3}} \, \Gamma(\tfrac{2}{3})^3 \over 4 \pi^2 } \, 
u v \, F_4(\tfrac{2}{3}, \tfrac{2}{3}, \tfrac{4}{3}, \tfrac{3}{2}; x,y) \,.
\eea
Additionally, we have $Q_{2,1}=0$, while $Q_{2,0}$ cancels out of all periods.
These expressions, including their normalizations, agree with~\cite{Klemm:1995wp}.}

\sm

The proof of this corollary is given in appendix \ref{sec:B}. We note that the double infinite series for the Appell function is absolutely convergent for $\sqrt{|x|} + \sqrt{|y|} <1$ which gives the following region of absolute convergence in terms of $u$ and $v$,
\bea
\label{2.b1}
\tfrac{ 2}{\sqrt{27}} \,  |u|^{{3 \over 2}} + |v| <1\,.
\eea
Beyond this region, partial analytic continuation formulas are known for the Appell functions,\footnote{~These are obtained by expressing $F_4$ as an infinite sum of  hypergeometric functions, such as
\bea
F_4(a,b,c_1,c_2; x,y) = \sum _{n=0}^\infty {\Gamma (n+a) \Gamma (n+b) \Gamma (c_2) 
\over \Gamma (a) \Gamma (b) \Gamma (n+c_2) {\, n!}} \, y^n\, F(n+a,n+b;c_1;x) \,,
\eea
and applying inversion formulas for the hypergeometric functions.
}
 \bea
 \label{analF4}
 F_4(a,b,c_1,c_2;x,y) &  & =
{ \Gamma (c_1) \Gamma (b-a) \over \Gamma (b) \Gamma (c_1-a)} \, 
(-x)^{-a} F_4 (a,a+1-c_1,a+1-b,c_2; \tfrac{1}{x},\tfrac{y}{x})
\no \\ && \ +
{ \Gamma (c_1) \Gamma (a-b) \over \Gamma (a) \Gamma (c_1-b)} \, 
(-x)^{-b} F_4 (b,b+1-c_1, b+1-a, c_2;\tfrac{1}{x},\tfrac{y}{x})\,,
\qquad
\eea
which  gives the following region in terms of $u$ and $v$,
\bea
\label{2.b2}
1+|v|< \tfrac{ 2}{\sqrt{27}} \,  |u|^{{3 \over 2}} \, ,
\eea
allowing us to explore the region of large $|u|$ and small $|v|$. 
Recent progress on the analytic continuation of $F_4$ may be found in \cite{Ananthanarayan:2020xut}.

\subsection{Expanding Periods of Holomorphic Abelian Differentials}

To evaluate the series expansion of the holomorphic Abelian differentials $\omega_n$ for the family of SW curves $\cC(u)$, we use their relation with the SW differential given in (\ref{2.holodif}).  The second term on the right side in (\ref{2.holodif}) is an exact differential of a single-valued holomorphic function for $0\leq n \leq N-2$, and thus integrates to zero on all closed homology cycles.  We can thus write the periods of the holomorphic Abelian differentials as follows, 
\bea
2 \pi i \, \p_n a_{I} & = & \oint _{\mA_I} { x^n \, dx \over \sqrt{A(x)^2-1}}= \oint_{\mA_I} \omega_n \,,
\no \\
2 \pi i \, \p_n a_{DI} & = & \oint _{\mB_I} { x^n \, dx \over \sqrt{A(x)^2-1}}= \oint_{\mB_I} \omega_n \,,
\eea
which shows that they are simply derivatives of the SW periods~$a_I, a_{DI}$ with respect to $u_n$.\footnote{~Here we use the shorthand $ \p f / \p u_n = \p_nf$. Note that the matrix of complexified gauge couplings $\tau_{IJ}$ is identified with the period matrix of the curve as $\tau_{IJ} = \p_n a_{DI} (\p_n a_{J})^{-1}$.} Using \eqref{2.aQ}, these in turn may be expressed in terms of $u_n$-derivatives of the function $Q(\xi)$,
\bea
\label{2.aQ2}
\p_n a_{I} & = &   \sum_{J=1}^I \big \{  \p_nQ(\ep^{2J-1} ) - \p_nQ(\ep^{2J-2}) \big \} \,,
\no \\
\p_n a_{DI} & = &  \p_nQ(\ep^{2I} ) - \, \p_nQ(\ep^{2I-1})~.
\eea
The derivatives $\partial_n Q(\xi)$ are given by Theorem \ref{thm:1} as follows,
\bea
\p_nQ(\xi) =
 \sum _{{\ell_m=0 \atop m=0, \ldots, N-2}}^\infty 
{ \ell_n \over u_n} \, { u_0^{\ell_0} \cdots u_{N-2}^{\ell_{N-2}} \over \ell_0! \, \cdots \ell_{N-2}! } 
\, V_{L,M} ( \xi ) \,,
\eea
where $L, M$ and $V_{L,M}(\xi)$ are the same as in Theorem \ref{thm:1}. In the above sum, it is understood that whenever $u_n=0$ one also has $\ell_n=0$ in the first factor in the summand.


\section{Expanding the K\"ahler Potential around the $\ZZ_{2N}$ Point}
\setcounter{equation}{0}
\label{sec:Kahler1}


In this section we express the K\"ahler potential for pure $\cN=2$ Seiberg-Witten theory with gauge group $SU(N)$, defined in terms of the periods $a_I$ and $a_{DI}$ by (\ref{k1}), which we repeat here for convenience,
\bea
\label{k1bis}
K = { i \over 4 \pi} \sum_{I=1} ^{N-1} \Big ( a_I \, \bar a_{DI} - \bar a_I \, a_{DI} \Big )\,,
\eea
 in terms of the functions $Q(\xi)$, defined in (\ref{2.Q}) as (hyper-) elliptic integrals. The series expansion of the K\"ahler potential is then readily obtained from the series expansion of the functions $Q(\xi)$, derived in Theorem \ref{thm:1}. 
We thus have the following theorem:

{\thm
\label{thm:2}
In terms of the Fourier coefficients $Q_j$ in the decomposition of the function $Q(\xi)$ into inequivalent representations of $\ZZ_{2N}$, 
\bea
\label{2.Qxi}
Q(\xi) = \sum _{j=0}^{2N-1} Q_j \, \xi ^j\,,
\eea
the K\"ahler potential takes the following diagonal form,
\bea
\label{2.Kthm}
K={ N \over 2 \pi  } \sum_{{j=1 \atop j \not= N}}^{2N-1}  |Q_j |^2  \tan \left ( {  \pi j \over  2N} \right )\,.
\eea
The function $Q_0$ does not enter the expression for the K\"ahler potential, and $Q_N=0$.}

\sm 

To prove Theorem \ref{thm:2}, we use the expression (\ref{2.Q}) for the periods in terms of the function $Q(\xi)$, as well as Theorem \ref{thm:1} giving a decomposition of $Q(\xi)$ in powers of $\xi$. 

\sm

To show that $Q_N=0$ we observe that the coefficient of $\xi^N$ in $V_{L,M}(\xi)$ of Theorem \ref{thm:1}  can arise only when $NM-L-1 \equiv 0 $ (mod $2N$). When this relation holds, the sine function that appears in~$V_{L,M}(\xi)$ vanishes. Moreover, the $\Gamma$-function of the same argument is non-singular since we have $NM-L \geq 2$  whenever at least one $\ell_n \not=0$ and $NM-L-1=-1$ when all $\ell_n=0$.  As a result, $Q_N=0$. 

\sm

To evaluate the K\"ahler potential in terms of the functions $Q_j$, we substitute the expansion~(\ref{2.Qxi}) into the expressions for the periods in (\ref{2.aQ}) and carry out the sum over $J$, 
\bea
a_I = -  \sum_{j=1}^{2N-1} Q_j { 1-\ep^{2Ij} \over 1+\ep^j}~,
\qquad
a_{DI} =  \sum_{j=1}^{2N-1} Q_j (1- \ep^{-j}) \ep^{2Ij}\,.
\eea 
The K\"ahler potential is then given by
\bea
- 4 \pi i K & = & \sum_{j,k=1}^{2N-1} {  Q_j \bar Q_k \over (1+\ep^j) (1+\bar \ep^k)} 
\sum_{I=1}^{N-1} \Big [
(\ep^k-\ep^{-k}) (\ep^{-2Ik} -\ep^{2I(j-k)} ) 
\no \\ && \hskip 1.9in 
+(\ep^j - \ep^{-j}) (\ep^{2Ij} - \ep^{2I(j-k)}) \Big ]\,.
\eea
The summation over $I$ gives the following,
\bea
\sum_{I=1}^{N-1} \ep^{2Ik} = -1 + \sum_{I=0}^{N-1} \ep^{2Ik} =  N \delta_{k\equiv 0} -1~.
\eea
Here we use $\delta_{k \equiv 0} = \delta_{k\equiv 0 \, ({\rm mod} \, N)}$ to represent the Kronecker symbol mod $N$. 
In carrying out the summations over $I$ the contribution of the additive term $-1$ cancels out. The contributions $j=N$ and $k=0$ cancel in view of $Q_N=0$, a fact that was established above. This leaves only the contributions $\delta_{j-k\equiv 0}$,
\bea
K & = & {N \over 4 \pi i} \sum_{j,k=1}^{2N-1} { Q_j \bar Q_k \over (1+\ep^j) (1+\bar \ep^k)} 
 \Big [ \ep^k-\ep^{-k}+\ep^j - \ep^{-j} \Big ]  \delta_{j-k\equiv 0}\,.
\eea
Next, we solve $\delta_{j-k\equiv 0}$ which gives $k=j+\alpha N$ for some integer $\alpha$. 
The solutions for the ranges of $k$ and $j$ involved in the sums  are as follows,
\bea
k & = & j \hskip 1.35in 1 \leq j \leq N-1 ~ \hbox{ and } ~ N+1 \leq j \leq 2N-1
\no \\
k & = & j+N \hskip 1in 1 \leq j \leq N-1
\no \\
k & = & j-N \hskip 1in N+1 \leq j \leq 2N-1
\eea
The fact that $Q_N=0$ implies that the contributions $j=N$ and $k=N$ are absent. For the solutions $k=j \pm N$, we have $\ep^j+\ep^k=0$, so that their contributions to $K$ vanish. This leaves only the contribution from $k=j$,
\bea
K & = & {N \over 2 \pi i} \sum_{{j=1 \atop j \not= N}}^{2N-1} { |Q_j |^2 \over |1+\ep^j|^2} 
(\ep^j - \ep^{-j}) \,.
\eea
Expressing $\ep= e^{ 2 \pi i/2N}$ in terms of real variables we readily obtain (\ref{2.Kthm}), thereby completing the proof of Theorem \ref{thm:2}.

\subsection{The Value of~$K$ at the~$\ZZ_{2N}$-Symmetric Point}

At the symmetric point, we have $u_n=0$ for all $n=0,\cdots, N-1$. Using the results of Corollary~\ref{cor:1} for $u_0=0$, we find that $Q(\xi)= \xi^{N+1} Q_{N+1}$ where $Q_{N+1}$ is given by the corollary. Using the expression (\ref{2.Kthm})  for the K\"ahler potential, with the only non-vanishing contribution from $j=N+1$, we readily obtain
\bea
\label{symm}
K(u_n = 0) = - {N \over 8 \pi^2} { \Gamma ( \thalf +\tfrac{1}{2N})^2 \over \Gamma (1+\tfrac{1}{2N})^2} \,
\cot \left ({ \pi \over 2N} \right )~.
\eea
This formula has two noteworthy features:
\begin{itemize}
\itemsep=0in
\item $K(u_n = 0)$ is negative.
\item $K(u_n = 0)$ scales as $-N^2/4 \pi^2$ for $N \gg 1$. 
\end{itemize}

\subsection{Structure of the K\"ahler Potential on Restricted Moduli}

An immediate consequence of Theorem  \ref{thm:2} is that the number $N_Q$ of independent functions~$Q_j$ that can contribute to the K\"ahler potential is bounded from above, $N_Q \leq 2N-2$, with equality being attained for generic moduli. Setting some of the moduli to zero may decrease the number $N_Q$ to values smaller than $2N-2$.  In this subsection, we shall give a formula for $N_Q$ as a function of the choice of non-vanishing  moduli. 

\sm

At the $\ZZ_{2N}$-symmetric point, all moduli $u_n$ vanish so that only $Q_{N+1}$ is non-zero and we have $N_Q=1$ for any value of $N$. On the slice $u_n=0$ for all $n=1,\cdots, N-2$ and $u_0\not=0$ we have $N_Q=2$ as shown in 
Theorem~\ref{thm:1}. More generally, the number of independent $Q_j$ functions contributing to the K\"ahler potential equals the number of distinct values, other than $\ep^0$ and $\ep^N$,  taken by the roots of unity function $\ep^{NM+N+1+L}$ in the expression for $Q(\xi)$ given in Theorem~\ref{thm:1}.  

\sm

Let $u_{j_1} , \ldots~, u_{j_p}$ be the set of distinct moduli that differ from zero, while all other moduli vanish, and define the set $S=\{ j_1, j_2, \cdots, j_p \} \subset \{ 0, 1, 2, \cdots, N-2 \}$. In terms of these data, the roots of unity function takes the following values,
\bea
N_Q = \# \bigg ( \Big \{  \ep^{N+1} \prod_{j \in S} \ep^{(N+j)\ell_j} \Big \}_{0 \leq \ell_j \leq 2N-1} \setminus \{ \ep^0, \ep^N \} \bigg )\,.
\eea
Here the $\ell_j$ are allowed to take all possible values in the range $0 \leq \ell_j \leq 2N-1$. 
A significant simplification occurs when $N+j$ is even for every $j \in S$. In this case $\ep^0$ and $\ep^N$ never belong to the range of the root of unity function and we simply have, 
\bea
N_Q = \#  \Big \{  \ep^{N+1} \prod_{j \in S} \ep^{(N+j)\ell_j} \Big \} _{0 \leq \ell_j \leq 2N-1}\,.
\eea
As an example, consider the case $N=4k$, with only the modulus $u_{2k}$  turned on,
\bea
N_Q = \#  \Big \{  \ep^{N+1} (-i)^{\ell_{2k}} \Big \} _{0 \leq \ell_{2k} \leq 3} = | \ZZ_4| = 4\,.
\eea
Note that the counting procedure outlined above is correlated with the breaking pattern of the~$\ZZ_{2N}$ symmetry on the moduli space.

\section{Some Exact Properties of the K\"ahler Potential}
\setcounter{equation}{0}
\label{sec:Kahler2}

In this section we present a number of general results about the K\"ahler potential, which offer evidence for its overall structure advocated in this paper. The derivations of these results are direct and exact, i.e.~they do not rely on the series expansion of the periods around the  $\mathbb{Z}_{2N}$ symmetric point presented in Theorem~\ref{thm:1}.

\subsection{$K$ is Negative at an Arbitrary Stationary Point}
\label{sec:kneg}

At an arbitrary stationary point of $K$, the value of $K$ is negative. This may be established by using the partial derivative of $K$ with respect to an arbitrary modulus $u_n$ and using the fact that $a_I$ and $a_{DI}$ are holomorphic in $u_n$,
\bea
{\p K \over \p u_n} = {i \over 4 \pi} 
\sum_I { \p a_I \over \p u_n} \left ( \bar a_{DI} - \sum_J \tau_{IJ} \bar a_J \right ) =0\,.
\eea
Since the matrix $\p a_I / \p u_n$ is invertible, the vanishing condition required at an arbitrary stationary point of $K$ simplifies and we have, 
\bea
\bar a_{DI} - \sum_J \tau_{IJ} \, \bar a_J =0\,,
\eea
for $I=1,\ldots, N-1$. Using this relation to eliminate $a_{DI}$ and $\bar a_{DI}$ from $K$, we obtain, 
\bea
K \big |_{\text{stationary}}
= -{ 1 \over 2 \pi} \sum _{I,J} a_I \bar a_J \, \Im (\tau_{IJ}) 
= -{ 1 \over 2 \pi} \sum _{I,J} a_{DI} \bar a_{DJ} \, (\Im \tau)^{-1} _{IJ} \quad \leq \quad 0~.
\eea
This inequality is strict as long as the vectors $a_I$ and $a_{DI}$ are not both identically zero.

 \subsection{$K$ as a Two-Dimensional Integral}
 
In this subsection, we shall prove that derivatives of the K\"ahler potential~$K$ can be written as real, two-dimensional integrals over the SW curve~${\cal C}(u)$, via the following formulas,
\bea
\label{5.Ka}
{ \p K \over \p \bar u_n}  
= { i \over 16 \pi^3} \int _{{\cal C}(u)} \lambda \wedge { \p \bar \lambda \over \p \bar u_n} 
=
{1 \over 8 \pi^3} \lim _{R \to \infty} \int _{|x|<R} d^2x  \, {  xA'(x) \,  \bar x^{n} \over |A(x)^2 - 1|}~,
\eea
 where $d^2x = { i \over 2} dx \wedge d \bar x$.  The SW differential $\lambda$ is meromorphic in all its ingredients and has only double poles at $P_\pm=\pm \infty$.  The formula is established using calculations similar to those used to prove the Riemann bilinear relations on the integrals involving Abelian differentials.  Indeed, the starting point is the relation, 
\bea
\label{5.Riem}
{ \p K \over \p \bar u_n}  
=  {i \over 16 \pi^3} 
\sum_I \left ( \oint _{\mA_I} \lambda  \oint _{\mB_I} \bar \om_n
- \oint _{\mB_I} \lambda  \oint _{\mA_I} \bar \om_n \right )\,,
\eea
where the holomorphic differentials $\om_n$ were given in (\ref{2.holodif}) by
\bea
{\p \lambda \over \p u_n} = \om_n + (\hbox{exact differential})~, \qquad \om_n = { x^n dx \over \sqrt{A(x)^2-1}}~.
\eea
To recast (\ref{5.Riem}) in terms of a two-dimensional integral, we introduce  a simply-connected domain $M_\ep$  in $\CC$, where $M_\ep$ is obtained from the SW curve~${\cal C}(u)$ by cutting the latter open along the $\mA_I$ and $\mB_I$ cycles of a canonical homology basis and removing coordinate discs, of coordinate radius $\ep >0$, centered at $P_\pm$ with boundaries $\gamma_\pm$. In the simply-connected domain $M_\ep$, we may write the closed form $\om_n$ as the exact differential $\om_n = d f_n$ of a function $f_n$ that is single-valued in $M_\ep$. With this set-up, we 
 evaluate the following integral,
 \bea
 \int _{M_\ep} \lambda \wedge \bar \om_n = \int _{M_\ep} \lambda \wedge d \bar f_n
 = - \int _{M_\ep} d (\bar f_n \lambda) = - \oint _{\p M_\ep} \bar f_n \lambda\,.
 \eea
Decomposing the integral over $\p M_\ep$ into integrals over cycles, we obtain (see e.g.~\cite{Farkas})
\bea
- \oint _{\p M_\ep} \bar f_n \lambda =
 \sum_I \left ( \oint _{A_I} \lambda \oint _{B_I} \bar \om_n - \oint _{B_I} \lambda \oint _{A_I} \bar \om_n  \right )
 + \oint _{\gamma_+} \bar f_n \lambda + \oint _{\gamma_-} \bar f_n \lambda\,.
 \eea
 The integrals over $\mA_I$ and $\mB_I$ cycles may be read off from the properties of the SW differential and the period matrix, while the integrals over $\gamma_\pm$ vanish, as may be seen by expanding $\bar f_n(z)$ in a series near the points $P_\pm$. Taking the limit $\ep \to 0$ gives a prescription for regularizing the double pole in the surface integral, and we obtain,
 \bea
 \int_{{\cal C}(u)} \lambda \wedge \bar \om_n = 2 \pi i  \sum_I \left ( -a_{DI} + \sum _J \bar \tau_{IJ} a_J \right ){ \p \bar a_I \over \p \bar u_n}\,.
 \eea
 From the definition of $K$ we obtain,
 \bea
 { \p K \over \p \bar u_n} = { i \over 4 \pi} \sum_I \left ( -  a_{DI} +  \sum_J \bar \tau_{IJ} a_J  \right ){ \p \bar a_I \over \p \bar u_n}\,.
 \eea
 Re-expressing this relation in terms of the variables $u_n$ we obtain the first equality in~(\ref{5.Ka}); substituting the values for the SW differential and making the regularization explicit 
establishes the second equality in~(\ref{5.Ka}).
  
 As an aside, the convexity of $K$ for gauge group $SU(2)$ can be proven by differentiating~\eqref{5.Ka}, and showing that the determinant of the Hessian is strictly positive. For general $N>2$ such arguments can be used to demonstrate positivity of the derivatives of $K$ on certain sub-slices of moduli space, but we have not succeeded in generalizing them to prove the conjecture that the origin is the only minimum of $K$.

\subsection{The~$\ZZ_{2N}$-Symmetric Point is a Stationary Point of~$K$}
\label{sec:stat}

Here we verify that the symmetric point, defined by $u_n=0$ for all $n=0,1, \ldots, N-2$, is a stationary point of~$K$, as expected from the fact that~$K$ is invariant under~$\ZZ_{2N}$ rotations. To establish this directly, we use formula~\eqref{5.Ka} above for the gradient of~$K$ as a surface integral, 
\bea
{ \p K \over \p \bar u_n}  
= {1 \over 8 \pi^3}  \int _\CC d^2 x { x A'(x) \, \bar x^{n}  \over |A(x)^2-1|}\,.
\eea
At the symmetric point, we have $A(x) = x^N$, so that,
\bea
{ \p K \over \p \bar u_n}  \bigg |_{u=0}
=  {N \over 8 \pi^3} \int _\CC d^2 x \, { x^N \, \bar x^{n}  \over |x^{2N}-1|}\,.
\eea
Under a rotation on $x$ by angle $2\pi/2N$,
\bea
x= \ep y~, \qquad \ep = e^{2\pi i/2N}~,
\eea
the denominator and the measure $d^2x$  are invariant, but the remaining part of the numerator is not invariant, and we have,
\bea
\int _\CC d^2 x \, { x^N \, \bar x^n  \over |x^{2N}-1|} = 
\ep ^{N-n} \int _\CC d^2 y \, { y^N \, \bar y^n  \over |y^{2N}-1|}\,.
\eea
Since $\ep^{N-n} \not= 1$ for all $n=0,\cdots, N-2$, the integral must vanish for all $n$, which shows that the symmetric point is a stationary point. Note that the periods $a_I$ are manifestly not zero at that point, so that the value of $K$ at the symmetric point must be negative (as was already shown in \eqref{symm}).

\subsection{$\p K / \p \bar u_0 \not=0 $ whenever $\Im (u_0) \not=0$}
\label{sec:pp}

We shall now prove a stronger statement that implies the result of section~\ref{sec:stat} above: $\Im (\p K / \p \bar u_0)>0$ for $\Im(u_0)>0$, where we stress that both inequalities are strict. This result will imply that $\p K / \p \bar u_0 \not=0 $ whenever $\Im (u_0) \not=0$. Our starting point is (\ref{5.Ka}) for $A(x) = x^N - u_0$ and we change integration variable from $x$ to $z=x^N$ to obtain
\bea
{ \p K \over \p \bar u_0} = 
{ 1 \over 8 \pi^3 N} \int _\CC { d^2 z \over |z|^{2-\frac{2}{N}} } \, { z \over |z-u_0+1| \,  |z-u_0-1|}\,.
\eea
We decompose $z$ and $u_0$ into real coordinates $z=x+iy$ and $u_0=v_1+iv_2$ with $x,y,v_1, v_2 \in \RR$, and take the imaginary part of the above equation, 
\bea
\Im { \p K \over \p \bar u_0}  =  
{ 1 \over 8 \pi^3 N} \int_{-\infty} ^\infty dx \, \int _{-\infty}^\infty dy \, { \f(x,y,u_0) \over |x^2+y^2|^{1-\frac{1}{N}} } \,,
\eea
where the function $\f$ is given by,
\bea
\f(x,y,u_0) = { y \over \big ( (x-v_1+1)^2 + (y-v_2)^2 \big )^\half \big ( (x-v_1-1)^2 + (y-v_2)^2 \big )^\half}\,.
\eea
Decomposing the integration over $y$ into positive and negative parts, and reducing both integrations to $y>0$, we obtain, 
\bea
\Im { \p K \over \p \bar u_0}  & = &  
{ 1 \over 8 \pi^3 N} \int_{-\infty} ^\infty dx \, \int _0^\infty dy \, { \f(x,y,u_0) + \f(x,-y,u_0) \over |x^2+y^2|^{1-\frac{1}{N}} } \,.
\eea
Note that~$(x-v_1\pm 1)^2 +(y+v_2)^2 > (x-v_1\pm 1)^2 +(y-v_2)^2$ as long as $v_2>0$. Thus, for $\Im(u_0)=v_2>0$ we have
\bea
 \f(x,y,u_0) - \f(x,-y,u_0) >0~,
 \eea
uniformly throughout the domain of integration, and thus $\Im ( \p K / \p \bar u_0)  >0$, strictly.

\subsection{$\p K / \p u_0 \not=0 $ for real $u_0 \not=0$}
\label{sec:pp1}

A subtle analysis, which is much more involved than the one given above for $\Im (u_0)\not=0$ and  that we shall not reproduce here, allows one to show that $\p K / \p u_0$ is non-zero also when $u_0$ is purely real and non-zero. The result is obtained by a detailed bound on various combinations that appear in the integrand.


\section{Exploring the K\"ahler Potential for $SU(3)$}
\setcounter{equation}{0}
\label{sec:KP3}

The goal of this section is to explore the behavior of the K\"ahler potential for the case of gauge group~$SU(3)$ using a variety of complementary analytic and numerical techniques.

\subsection{Expansion Around the $\ZZ_6$-Symmetric Point}

The exact series expansion around the $\ZZ_6$-symmetric curve given by Theorem \ref{thm:1}, or via Appell functions in Corollary \ref{cor:2.3}, is absolutely convergent in a finite neighborhood  of moduli space surrounding the origin~$u = v = 0$.  The boundaries of this region are set by the singularities that arise when branch points collide. 
For $SU(3)$ gauge group the K\"ahler potential takes the following form (see Theorem \ref{thm:2}),
\bea
K(u,v) = {  \sqrt{3} \over 2 \pi} \Big [ |Q_{0,1}|^2 + 3 |Q_{1,0}|^2 -3 |Q_{0,0}|^2 - |Q_{1,1}|^2 \Big ]\,.
\eea
The special slice $u=0$ reduces to Gauss hypergeometric functions, 
\bea
K(0,v) = {  \sqrt{3} \over 2 \pi} \Big [ |Q_{0,1}|^2  -3 |Q_{0,0}|^2 \Big ]\,,
& \hskip 0.4in & 
Q_{0,0} = { 2^{{1 \over 3}} \, 3^{{3 \over 2}} \, \Gamma(\tfrac{2}{3})^3 \over 4 \pi^2 } \,
F(-\tfrac{1}{6}, -\tfrac{1}{6}; \tfrac{1}{2}; v^2) \,,
\no \\ &&
\label{u0}
Q_{0,1} = { 2 \pi \, v \over 2^{{1 \over 3}} \, 3^{{ 3 \over 2}} \, \Gamma(\tfrac{2}{3})^3  } \, 
F(\tfrac{1}{3}, \tfrac{1}{3}; \tfrac{3}{2}; v^2) \,,
\eea 
as does the special slice $v=0$,
\bea
K(u,0) = {  \sqrt{3} \over 2 \pi} \Big [ 3 |Q_{1,0}|^2 -3 |Q_{0,0}|^2  \Big ]\,,
& \hskip 0.4in &
Q_{0,0} = { 2^{{1 \over 3}} \, 3^{{3 \over 2}} \, \Gamma(\tfrac{2}{3})^3 \over 4 \pi^2 } \,
F(-\tfrac{1}{6}, -\tfrac{1}{6}; \tfrac{2}{3}; \tfrac{4 u^3}{27}) \,,
\no \\ &&
\label{v0}
Q_{1,0} =  { 2 \pi \, u \over 2^{{1 \over 3}} \, 3^2 \, \Gamma(\tfrac{2}{3})^3  } \, 
F(\tfrac{1}{6}, \tfrac{1}{6}; \tfrac{4}{3}; \tfrac{4 u^3}{27}) \,.
\eea
  The full control over the analytic continuation of the hypergeometric function allows for a complete picture of $K$ in either the $u=0$ or $v=0$ slices through moduli space.

\subsection{The Cusp Slice}

In this subsection, we explore the behavior of the SW periods and of the K\"ahler potential on the cusp slice for gauge group $SU(3)$, namely the section of moduli space along which the discriminant of the SW curve vanishes.

\subsubsection{Definition}

The SW curve may be parametrized by $u=u_1$ and $v=u_0$, following the notation of \cite{Klemm:1995wp}, 
\bea
y^2 = (x^3-ux-v-1)(x^3-ux-v+1)\,, \qquad \lambda = { ( 3x^3-ux) dx \over y}\,.
\eea
Both factors on the right side of $y^2$ cannot vanish simultaneously, so that the discriminant of the SW curve factors into the product $\Delta _+ \Delta _-$ of the discriminants of each factor (up to an overall constant), with
\bea
\label{3.Deltapm}
\Delta _\pm = 4 u^3 - 27 {(v \mp 1)^2}\,.
\eea
The cusp divisor is the union of the vanishing sets of $\Delta_+$ and $\Delta_-$. The two sets are related by $(u,v) \to (-u,-v)$, and the SW curve and differential are invariant provided we also let $(x,y) \to (-x,-y)$. Thus, we concentrate on $\Delta_+$ for which $v$ is given as a function of $u$ by,
\bea
\label{3.uv}
v = 1 +  \left ( { 4 u^3 \over 27} \right )^{{ 1 \over 2}}\,.
\eea 
Note that the $N=3$ Argyres-Douglas points \cite{Argyres:1995jj}, as well as the multi-monopole points (i.e.~the $N=3$ generalization of the $SU(2)$ monopole and dyon points, studied for instance in~\cite{Douglas:1995nw,DHoker:1997mlo,DHoker:2020qlp}), lie on the cusp slice, as they satisfy
\bea
\label{spp}
(u^3,v^2)_{\text{AD}}=(0,  1) \,, \hskip 0.8in (u^3,v^2)_{\text{mon}}=({27}/{4}, 0)\,.
\eea
Inspection of the boundary of absolute convergence of the Appell functions in~(\ref{2.b1}) and~(\ref{2.b2}) reveals that the above cusp relation sweeps out a divisor that intersects with the boundary of convergence. For this reason the Appell function solution, even if analytically continued with the help of (\ref{analF4}), is expected to be of limited use  for evaluating the periods and the K\"ahler potential on the cusp slice.

\subsubsection{Mapping to an Elliptic Problem}

Instead, we shall use the special properties of the cusp slice to solve it with the help of elliptic functions and modular forms. We begin by substituting the cusp relation (\ref{3.uv}) into the expression for the SW curve, 
\bea
y^2 = \left ( x^3-ux -2 - \left ( \tfrac{ 4 u^3}{27} \right )^{{ 1 \over 2}} \right )
\left ( x^3-ux  - \left ( \tfrac{ 4 u^3}{27} \right )^{{ 1 \over 2}} \right )\,.
\eea
It will be convenient to scale out the modulus $u$ by using the new variables $\xi,\kappa$,
\bea
\label{4.newvar}
\xi = \left ( { 3 \over u} \right )^\thalf x \,,
\hskip 0.9in 
\kappa = \left ( { 27 \over 4u^3} \right )^{{1 \over 2}}\,,
\eea
in terms of which the SW curve and differential become, 
\bea
y^2 & = & \left ( { u \over 3} \right )^3 (\xi+1)^2 (\xi-2) (\xi^3 -3 \xi -2 -4\kappa)\,,
\no \\
\lambda & = & { \sqrt{3u} \, \xi (\xi-1) d\xi \over \sqrt{(\xi-2)(\xi^3 - 3 \xi -2 -4 \kappa)}}\,.
\eea
Note that the factor $(\xi+1)^2$ in the equation for the SW curve cancels out from the SW differential. This cancellation is the crucial ingredient in reducing the SW differential to an elliptic differential whose denominator is the square root of a quartic polynomial. By contrast the denominator of the original SW differential involved the square root of a polynomial of degree 6. The explicit knowledge of one of the roots of the quartic polynomial, namely $\xi=2$, allows us to send that point to infinity using a M\"obius transformation from the variable $\xi$ to a new variable $\chi$. The resulting polynomial is now cubic. Choosing the remaining freedom in the M\"obius transformation to cancel the term quadratic in $\chi$ in the cubic polynomial determines the appropriate change of variables uniquely, 
\bea
\xi = 2 - { 4 \kappa \over \chi -3}\,.
\eea
In terms of $\chi$ the SW differential becomes,
\bea
\label{4.SWa}
\lambda = { 4 \sqrt{3u} \, (\chi-3 -2\kappa)(\chi-3 - 4 \kappa) d \chi 
\over (\chi-3)^2 \sqrt{4 \chi^3 - 12 (9+8\kappa) \chi +8(8 \kappa^2 + 36 \kappa + 27)}}~.
\eea
The square root in the denominator is now over a polynomial of degree three in $\chi$ whose quadratic term vanishes.

\subsubsection{Uniformization}

The SW differential obtained in (\ref{4.SWa}) may be uniformized in terms of the Weierstrass elliptic function $\wp(z)$ with periods $1$ and $\tau$, using the differential equation it satisfies,
\bea
\wp'(z)^2 = 4 \wp(z)^3 - g_2 \wp(z) - g_3 \,,
\eea
where $g_2$ and $g_3$ are the standard modular forms of weight 4 and 6 respectively, with respect to the periods $1$ and $\tau$.\footnote{~A summary of  elliptic functions and modular forms needed here  is provided in appendix \ref{sec:C}.} However, in mapping the SW differential to the elliptic problem, we need to leave the periods $2\om$ and $2\om '$ with $\om'/\om =\tau$ to be determined by the SW problem. Restoring arbitrary periods may be carried out by using the degrees of homogeneity in the periods, which are 2, 4 and 6 for $\wp(z)$, $g_2$ and $g_4$ respectively. Thus, we uniformize the Seiberg-Witten curve and differential by making the following change of variables, 
\bea
\label{pg2g3}
\wp(z) = \chi \, (2 \om)^2 \,,
& \hskip 0.8in & 
g_2 = 12(8\kappa+9) \, (2 \om)^4\,,
\no \\ &&
g_3 = - 8 (8 \kappa^2 + 36 \kappa +27) \, (2 \om)^6\,.
\eea
The complex coordinate $z$ takes values in the fundamental parallelogram with vertices $\{ 0, 1, \tau, \tau+1 \}$. 
The uniformized SW differential is then given by, 
\bea
\lambda = 8 \, \om \, \sqrt{3u} \,  f(z) \, dz\,,
\eea
where the elliptic function $f(z)$ is given by,
\bea
f(z) = 1 - { 24 \, \om^2 \kappa \over \wp(z) - 12 \om^2} + { 128 \, \om^4 \kappa ^2 \over ( \wp (z) - 12 \om^2)^2 } \,.
\eea
The reduced discriminant $\Delta = g_2^3 - 27 g_3^2$ and the $j$-function evaluate as follows,
\bea
\label{Deltaj}
\Delta (\tau) = - 2^{12} \times 27 \, (2 \om)^{12} \kappa ^3 (\kappa+1)\,,
\hskip 0.8in
j (\tau) = - 27 \, {(8 \kappa +9)^3 \over \kappa ^3 (\kappa +1)}\,.
\eea
Given $\kappa$ in terms of $u$ as in (\ref{4.newvar}), the modulus $\tau$ may be obtained by the standard expression in terms of hypergeometric functions $F= {}_2F_1$ \cite{BatemanII, Wiki},
\bea
\tau  = {i \over \sqrt{3}} { F\left (\tfrac{1}{3} , \tfrac{2}{3}; 1; -\kappa \right ) \over F \left (\tfrac{1}{3} , \tfrac{2}{3}; 1; 1+\kappa \right ) }\,.
\eea
A final rearrangement of $\lambda$ is made to obtain an expression that may be easily integrated to obtain the periods. 
To do so, we define a point $z_0$ such that $12\om^2 = \wp(z_0)$. By matching zeros and poles we obtain the following alternative expression for $f(z)$,
\bea
f(z) = { 1 \over 4} + { 3 \over 8 \wp(z_0)} \Big ( \wp(z-z_0) + \wp (z+z_0) \Big )\,.
\eea
This formula may be checked directly by using the addition formula for Weierstrass functions.

\subsubsection{Periods on the Cusp Slice}

In summary we obtain the following formula for the SW differential on the cusp slice,
\bea
\lambda =   {\sqrt{3u}  \over 4 \om} \, dz \Big ( 8 \om^2 + \wp(z-z_0) + \wp (z+z_0)  \Big )\,.
\eea
The SW curve for the cusp slice is a genus-one curve with two punctures at $z= \pm z_0$,  resulting from a non-separating degenerating of the genus-two SW curve for $SU(3)$. Correspondingly, the SW differential has double poles at $z=\pm z_0$. The homology generators of the underlying compact genus-one Riemann surface may be chosen as follows,
\bea
\mA &:& z \to z+1\, ,
\hskip 0.8in
\mB : z \to z+\tau\,.
\eea
Of the remaining two homology cycles of the genus-two curve, one cycle tends to infinity under the non-separating degeneration and corresponds to the curves from $-z_0 $ to $+z_0$, while the other cycle tends to zero. 

\sm 

The periods of the SW differential $\lambda$ on the cycles $\mA$ and $\mB$ are readily evaluated using the Weierstrass $\zeta$-function, which satisfies $\wp(z) = - \zeta '(z)$, and the monodromy relations given in~(\ref{C.zeta}) of appendix \ref{sec:C}, 
\begin{align}
2 \pi i a & =  \sqrt{3u} \,  \big ( 2 \om - 2 \eta \big ) \,, & 2 \om \, \eta & = \zeta (\tfrac{1}{2}) \,,
\no \\
2 \pi i a_D & =  \sqrt{3u} \, \big ( 2 \om' - 2 \eta ' \big ) \,,& 2 \om \, \eta ' & = \zeta (\tfrac{\tau}{2})\,.
\end{align}
The modular transformation properties of the periods are manifest in view of (\ref{omzetamod}).\footnote{~Actually,  the periods $a$ and $a_D$ are subject to the larger Fourier-Jacobi group $\ZZ^2 \ltimes SL(2,\ZZ)$, which acts on $\omega$ and $\eta$ by a common shift, and similarly for $\om'$ and $\eta'$. A systematic investigation of the non-separating degeneration of  genus-two Riemann surfaces was presented in \cite{DHoker:2018mys}.} The K\"ahler potential,
\bea
K(u,v) = { 3 |u| i \over  4\pi^3} \Big ( (\om -\eta) (\bar \om' - \bar \eta ') - (\bar \om - \bar \eta) ( \om' -  \eta ') \Big )\,,
\eea
is manifestly modular invariant. For later use, it will be convenient to recast $K$ in the following way,
\bea
\label{ks}
K(u,v) = { 3 |u| i \over  16 \pi^3 |2 \om|^2 } \bigg [ \Big ( 4\om^2 -2 \zeta(\tfrac{1}{2}) \Big ) 
\Big ( 4 \bar \om^2 \bar \tau  - 2 \overline{ \zeta ( \tfrac{\tau}{2} )} \Big ) - {\rm c.c.} \bigg ]\,.
\eea
Manifestly, only the single-valued combination $\om^2$ appears (recall that $\om$ is double valued).

\subsection{Values of~$K$ at Special Points}

As noted around \eqref{spp}, the Argyres-Douglas points and multi-monopole points lie on the cusp slice, and thus we may use \eqref{ks} to evaluate the K\"ahler potential at these points. 

\subsubsection{Argyres-Douglas Points}

The Argyres-Douglas points are located at $(u,v)=(0, \pm 1)$, which lie on the cusp slice since they satisfy $\Delta_+\Delta_-=0$ in (\ref{3.Deltapm}). Thus, we may obtain the behavior of the K\"ahler potential at the Argyres-Douglas points by taking the limit $u \to 0$ on the cusp slice.

\sm

As $ u \to 0$, we have $\kappa \to \infty$ in view of (\ref{4.newvar}), and thus $j (\tau) \to 0$ in view of (\ref{Deltaj}), which implies that $\tau =\rho= e^{2 \pi i/3}$, up to modular transformations. Since $g_3(\rho)$ is finite and $\kappa \to \infty$, the last equation in (\ref{pg2g3}) implies that we must have $\om \to 0$ as $u \to 0$. This result is consistent with the fact that $g_2(\rho)=0$ and the relation between $g_2$, $\kappa$ and $\omega$ in (\ref{pg2g3}). Using the result for $\eta$ and $\eta '$ for $\tau = \rho$ from (\ref{table1}), we readily find, 
\bea
2 \om \eta = { \pi \over \sqrt{3}} \,,
\hskip 0.9in
2 \om \eta ' = -{ \pi \over 2 \sqrt{3}} - i { \pi \over 2}\,.
\eea
The K\"ahler potential evaluates as follows,
\bea
K (0,\pm1)  = - \lim _{u \to 0} \, {  \sqrt{3}  |u|  \over   4 \pi |2 \om|^2}   \,.
\eea
We obtain $\om$ from its relation with $g_3(\rho)$, which in turn is derived from the value of $E_6(\rho)$ given in (\ref{table1}), and we find,  
\bea
\label{KADpt}
K (0,\pm 1) 
=  - { 3 \sqrt{3}  \over 4 \pi}  \left | { 16 \over   g_3(\rho)} \right |^{{1 \over 3}} 
 = - { 2^{{1 \over 3}} \, 3^{{9 \over 2}}  \, \Gamma (\tfrac{2}{3})^6 \over (2 \pi)^5} 
 = - 0.1112829388
\eea
In particular, the K\"ahler potential is negative at the Argyres-Douglas points.

\subsubsection{Multi-Monopole Points}

At the multi-monopole points we have $v=0$, and $u=u_*$ with $4 u_*^3 = 27 $ and $\kappa^2 =1$. The root $\kappa =-1$ corresponds to a singular curve where $j= \infty$ and $\tau= i \infty$ up to modular transformations, and is the proper value for the multi-monopole point (by contrast $\kappa=1$ corresponds to a regular curve). Using the values of (\ref{table1}), we obtain $\eta $ and $\eta'$  in the limit of large $\tau$,
\bea
\om = { \pi \over 2 \sqrt{3}}\,,
\hskip 0.6in 
\om' = { \pi \tau \over 2 \sqrt{3}}\,,
\hskip 0.6in 
\eta = { \pi \over 2 \sqrt{3}}\,,
\hskip 0.6in
\eta ' = { \pi \tau \over 2 \sqrt{3}} - i \sqrt{3}\,.
\eea
Hence $\om - \eta$ vanishes in this limit, while $\om'-\eta'$ remains finite. As a result, we have,
\bea
K(u_*,0) =0\,.
\eea
This is as expected, and also confirms the result obtained by substituting for $u=u_*$ on the $v=0$ slice in \eqref{v0}. 
We will have more to say about the $K=0$ hypersurface inside the $N=3$ moduli space in section \ref{sec:num} below.

\subsubsection{Behavior of $K$ for Large $u$}

Large $u \to \infty $ corresponds to $ \kappa \to 0$, which implies  $j(\tau) \to \infty$ and thus $\tau \to i \infty$. Using~(\ref{pg2g3}) and the values of $g_2$ and $g_3$ at infinity, we obtain
\bea
g_2(i \infty) = {4 \pi^4 \over 3} \,,
\hskip 0.6in 
g_3(i \infty)  ={8 \pi^6 \over 27} \,,
\hskip 0.6in
4 \om ^2 = - { \pi^2 \over 9}\,.
\eea
We also have the following asymptotics for the $\zeta$-values at half periods,
\bea
\zeta (\thalf |i \infty) = { \pi^2 \over 6} \,,
\hskip 0.9in
\zeta (\tfrac{\tau}{2} | \tau)  \Big |_{\tau \to i \infty} \approx { \pi^2 \over 6 } \tau - i \pi\,.
\eea
This allows us to recast $K$ in the following form,
\bea
K _{u \to \infty} = { |u| \over 2 \pi^2} \left ( { 4 \pi \over 3} \tau_2 - 6 \right )\,.
\eea
The limiting behavior of $K$ may be obtained form the expression for the reduced discriminant, 
\bea
\Delta(\tau) \approx (2 \pi)^{12} e^{2 \pi i \tau} \approx - 3^3 \times 2^{12}\,  (2 \om)^{12} \kappa^3 
= - 2^9  \pi^{12} (3u)^{-{9 \over 2}}\,,
\eea
so that,
\bea
K \big|_{\text{cusp }, u \to \infty} = { |u| \over 2 \pi^2} 
\left (  2 \ln 2 -6 + 3 \ln \left | { 3 u \over \Lambda^2 } \right | \right ) \rightarrow + \infty~.
\eea


\subsection{A Numerical Study of $K$}
\label{sec:num}

In this subsection we summarize the results of a numerical exploration of the $SU(3)$ K{\"a}hler potential, which are complementary to the analytic results obtained thus far. Let us briefly review what has already been learned: 

\begin{itemize}

\item $K$ has a minimum at the symmetric point $u=v=0$, where the value of $K$ is negative.

\item On the 2-real-dimensional slices of moduli space $u=0$ and $v=0$, one may straightforwardly use the hypergeometric function representations \eqref{u0} and \eqref{v0} to conclude that $K$ does not have stationary points away from the origin. (This has not been proven on the analogous slices for general $N>3$, although the result of section \ref{sec:pp} provides some evidence in this direction.)

\item An arbitrary stationary point of $K$ must occur at negative $K$, and so it is of interest to map the real-codimension-1 boundary of this region, where~$K = 0$.  The multi-monopole points lie on this $K=0$ boundary, while the Argyres-Douglas points lie within it.

\end{itemize}

We would like to numerically evaluate $K$ to further elucidate its features, e.g.~whether it has other stationary points and what can be said about the~$K = 0$ surface. On a generic slice in moduli space, $K$ is given in terms of Appell $F_4$ functions whose arguments are functions of the moduli $u$ and $v$, as per Corollary \ref{cor:2.3}. 
Unfortunately analytic studies of the Appell functions are limited, and software tools such as {\tt Mathematica} and {\tt Maple} have considerable difficulty evaluating them directly.\footnote{~{\tt Mathematica} does not have a built in Appell $F_4$ function; while {\tt Maple} does have such a function, its evaluation fails outside its strict region of convergence.} 
We have developed two complementary techniques to evaluate~$K$ numerically, which are more fully described in  appendix~\ref{sec:appnum}. The first method (see appendix \ref{sec:F4}) converts the system of second order differential equations defining $F_4$ into the integration of a first order ODE along a ray in moduli space; the second method (see appendix \ref{sec:numint}) uses \eqref{5.Ka} to numerically compute the derivatives of $K$ with respect to the moduli, which are then numerically integrated to obtain $K$.

\begin{figure}
\centering
\begin{subfigure}[b]{0.95 \textwidth}
\centering
\includegraphics[width=0.65\textwidth]{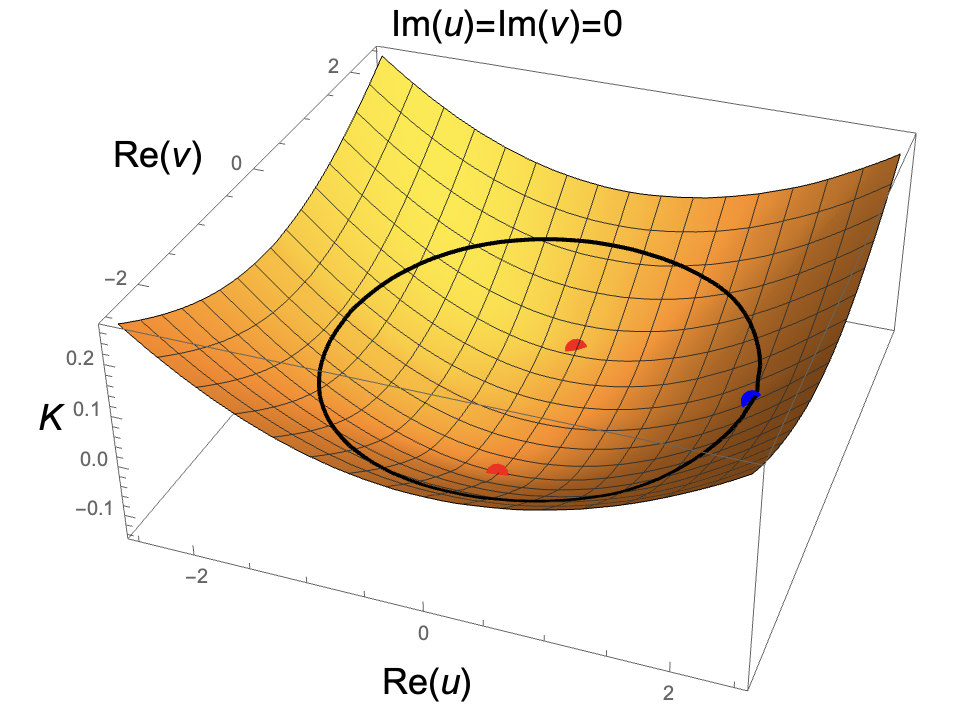}
\caption{$K$ plotted in the~$\Re u$-$\Re v$ plane, with~$\Im u = \Im v = 0$, evaluated  on a grid with spacing $\delta \text{Re} u=\delta \text{Re} v=0.25$. This slice includes the two Argyres-Douglas points, indicated by the red dots, and one of the multi-monopole points, indicated by the blue dot.  \label{fig:Krealurealv}}
\end{subfigure}
\hfill
\begin{subfigure}[b]{0.95 \textwidth}
\centering
\includegraphics[width=0.58\textwidth]{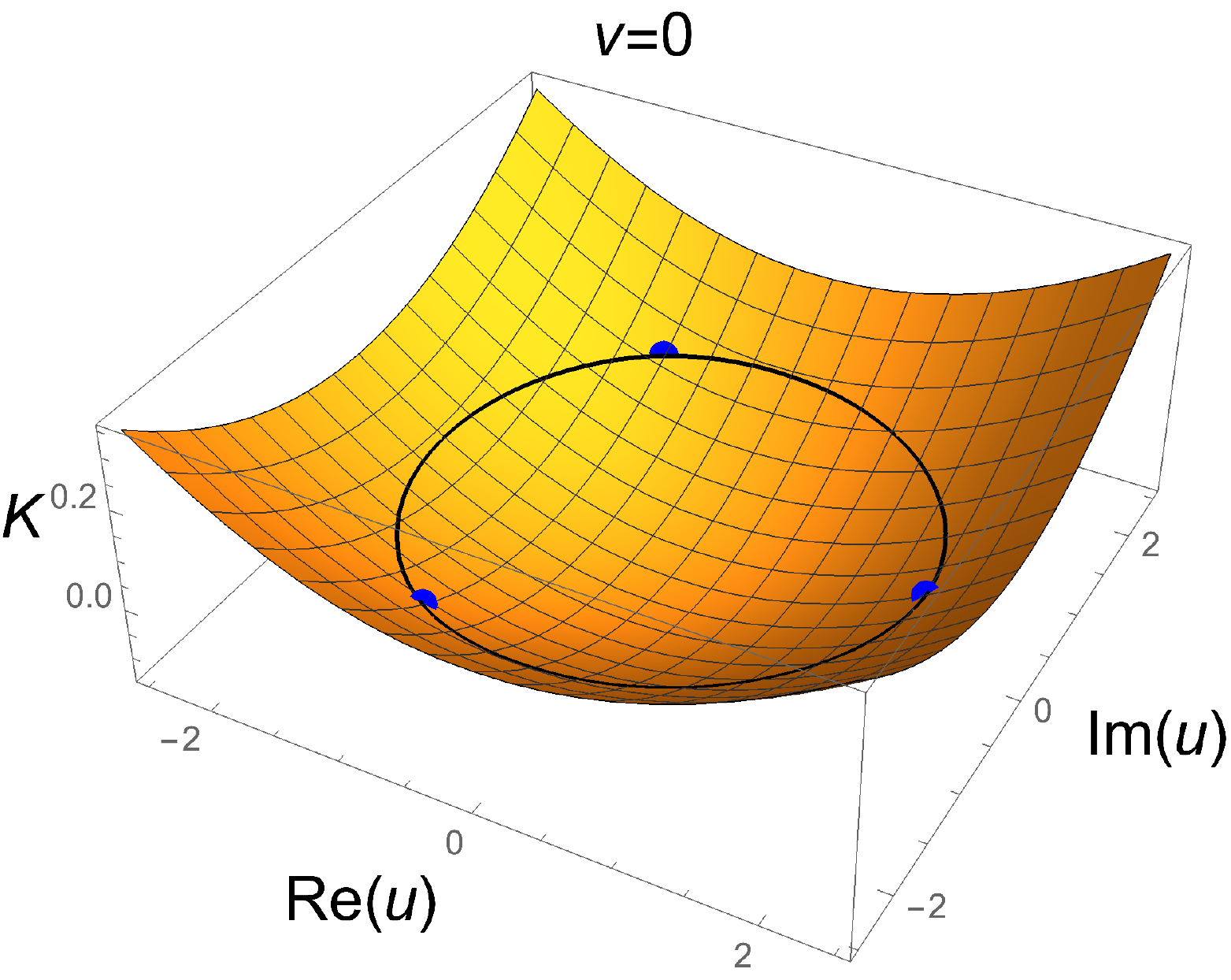}
\caption{$K$ plotted in the complex $u$-plane, with $v=0$, evaluated  on a grid with spacing $\delta \text{Re} u= \delta \text{Im} u=0.25$. The three multi-monopole points are indicated by blue dots. \label{fig:Kuplanev0}}
\end{subfigure}
\caption{
The $SU(3)$ K{\"a}hler potential, plotted numerically on two-real-dimensional slices of the moduli space, using the method of appendix \ref{sec:numint}. The $K=0$ contours are depicted in black.
\label{fig:Kplots}}
\end{figure}

\sm

\begin{figure}
\centering
\begin{subfigure}[b]{0.95 \textwidth}
\centering
\includegraphics[width=0.565\textwidth]{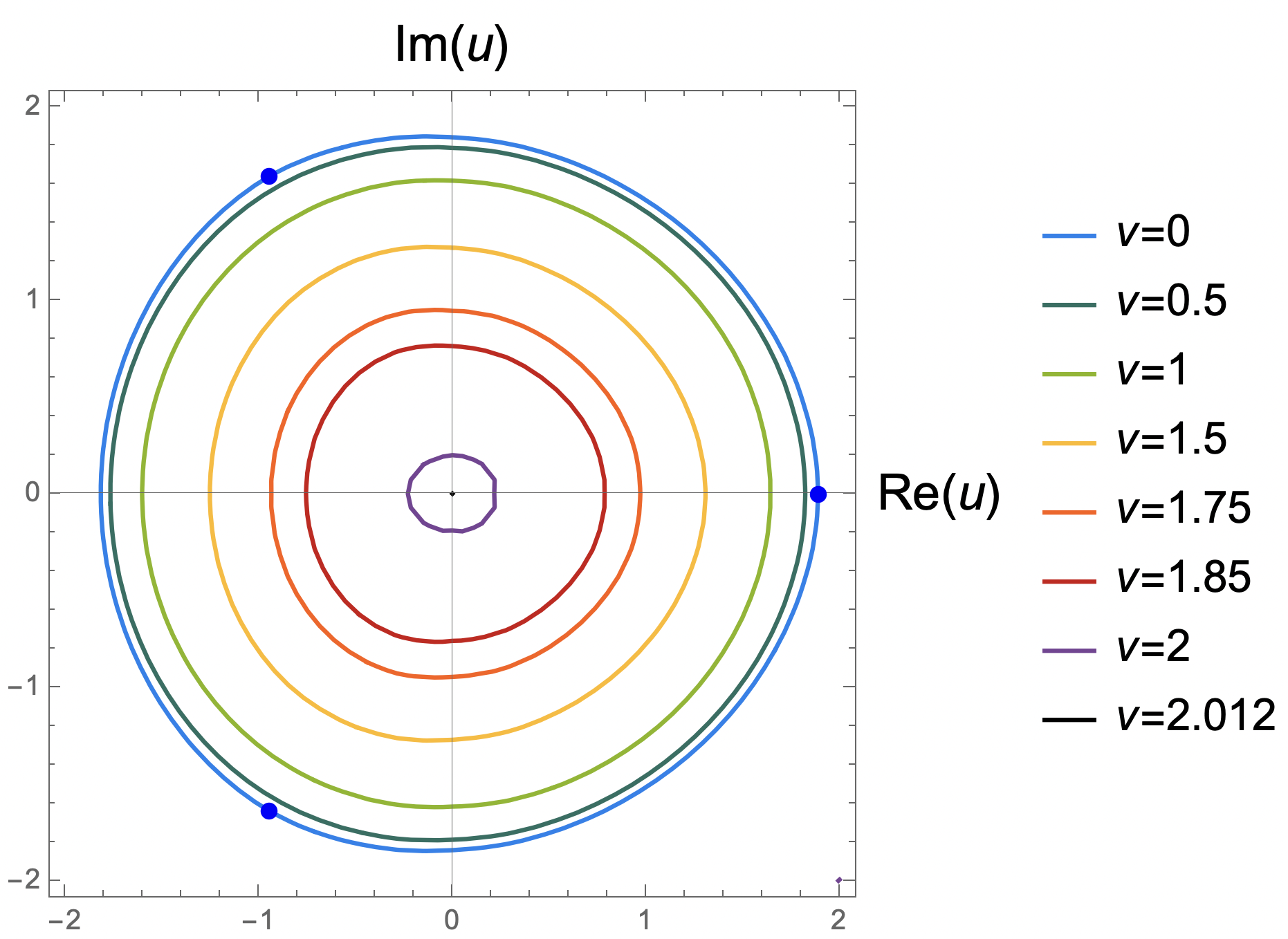}
\includegraphics[width=0.425\textwidth]{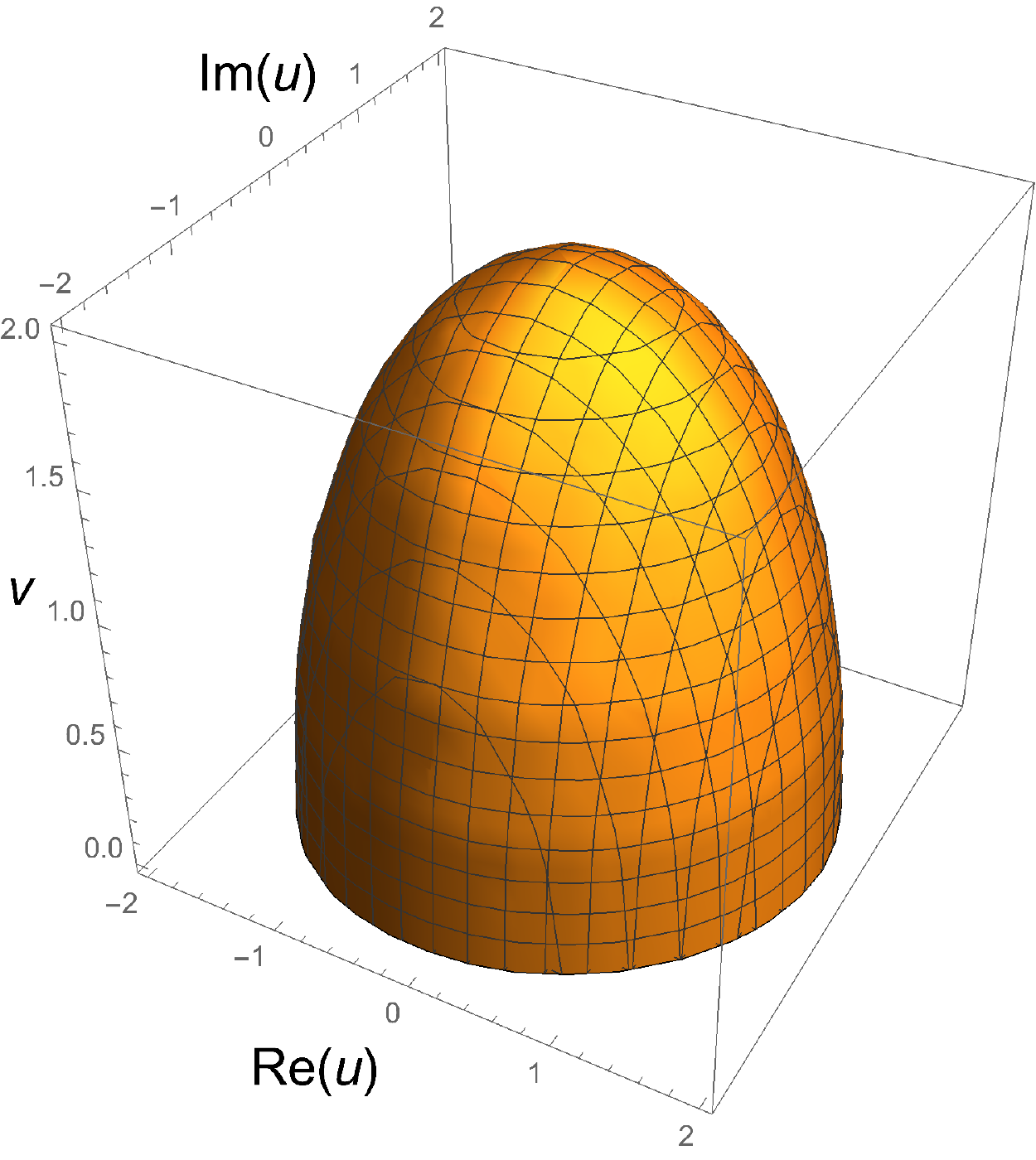}
\caption{ $K=0$ contours in the complex $u$-plane, for real $v \geq 0$. $K$ becomes positive on this slice when $v\gtrsim 2.012$. The blue $v=0$ contour in the left panel includes the three multi-monopole points, indicated by the blue dots. The $K=0$ surface plotted in the right panel is the 3D visualization of the contours plotted in the left panel, and is symmetric under $v\to -v$. \label{fig:K0contoursu}}
\end{subfigure}
\begin{subfigure}[b]{0.95\textwidth}
\centering
\includegraphics[width=0.565\textwidth]{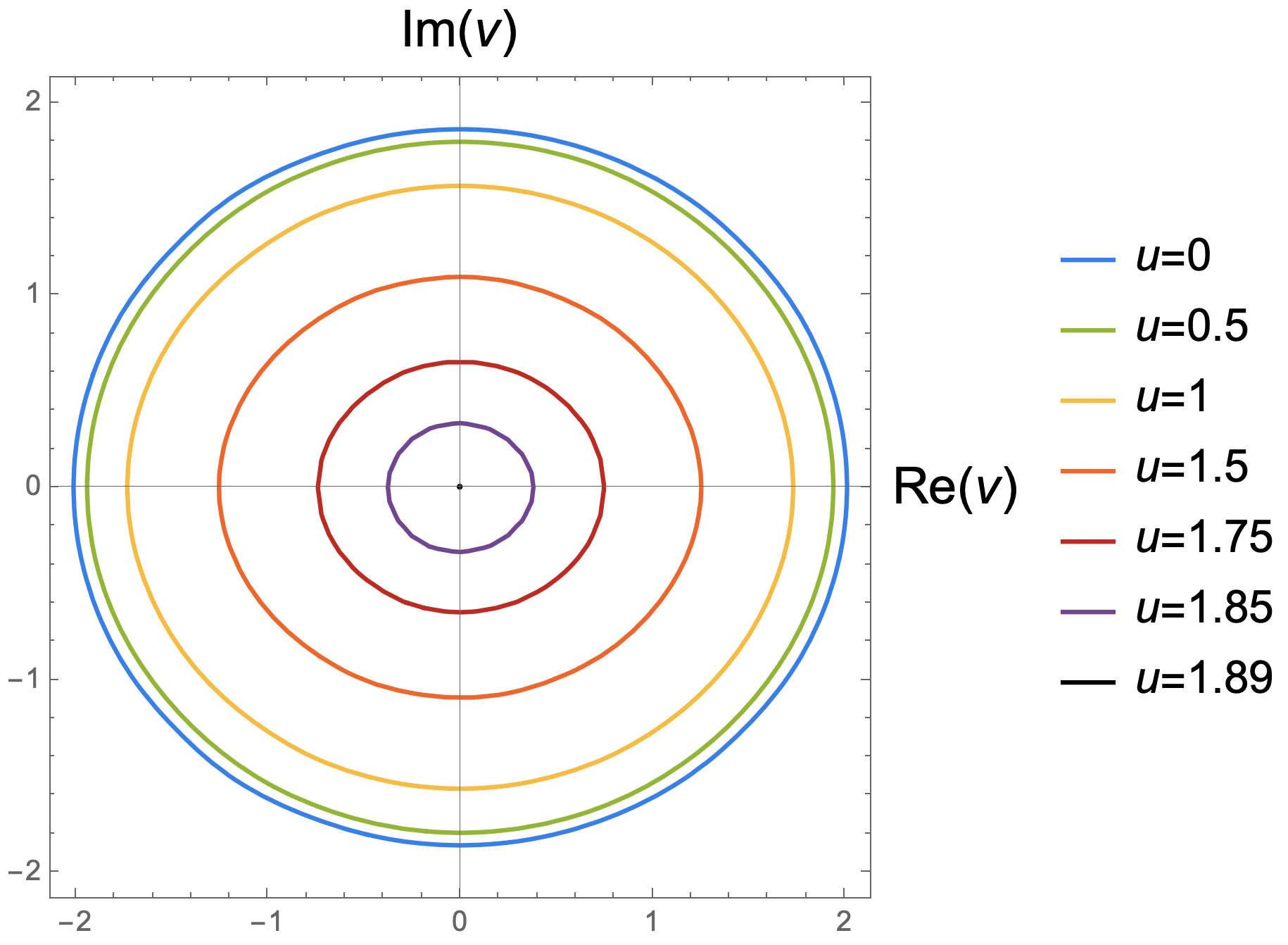}
\includegraphics[width=0.425\textwidth]{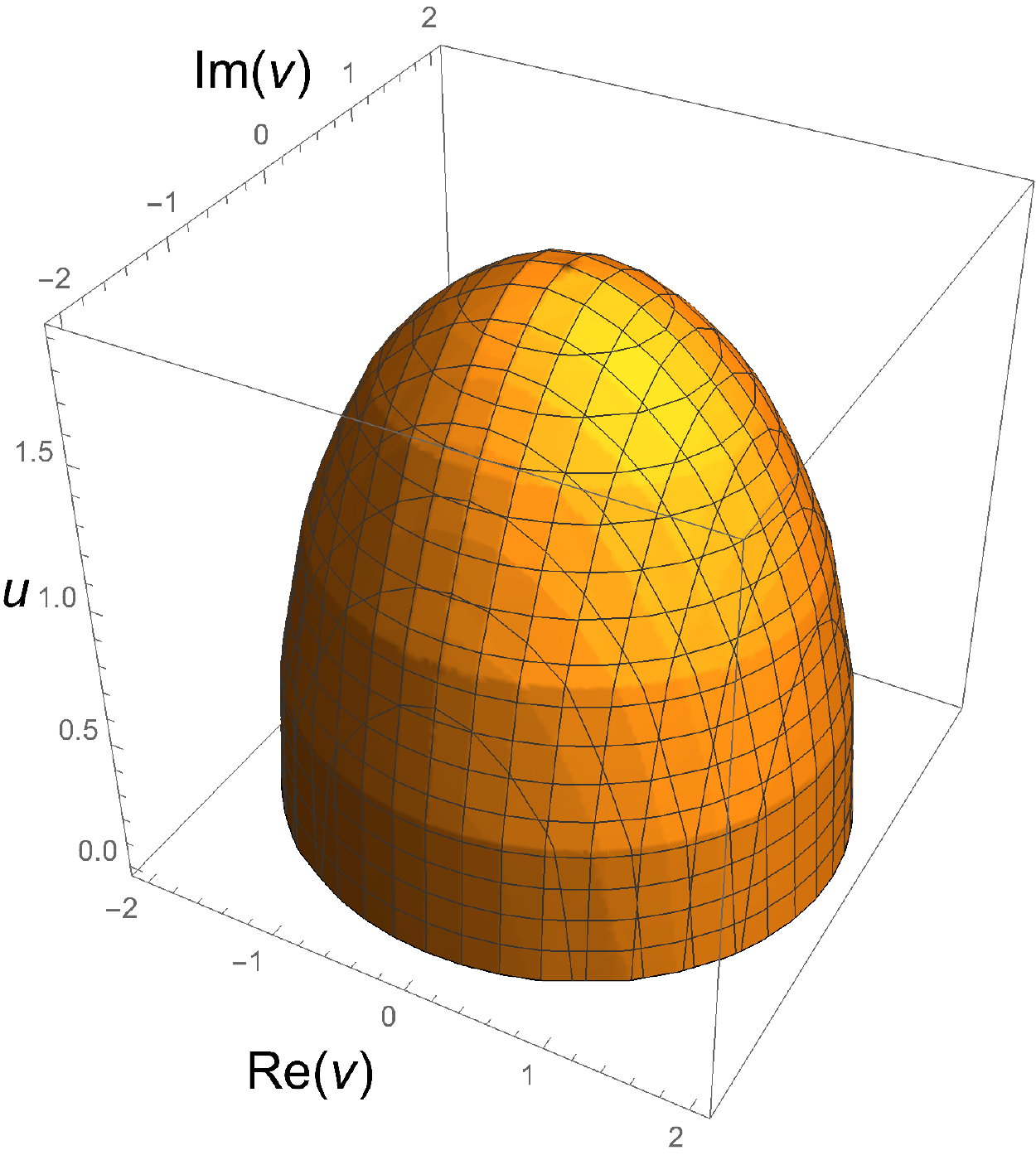}
\caption{ $K=0$ contours in the complex $v$-plane, for real $u \geq0$. $K$ becomes positive on this slice when $u\gtrsim 1.89$. The $K=0$ surface plotted in the right panel is the 3D visualization of the contours plotted in the left panel.\label{fig:K0contoursv}}
\end{subfigure}
\caption{$K=0$ contours for gauge group $SU(3)$, plotted via the method of appendix~\ref{sec:numint}. \label{fig:K0region}}
\end{figure}

\begin{figure}
\centering
\begin{subfigure}[b]{0.95 \textwidth}
\centering
\includegraphics[width=0.67\textwidth]{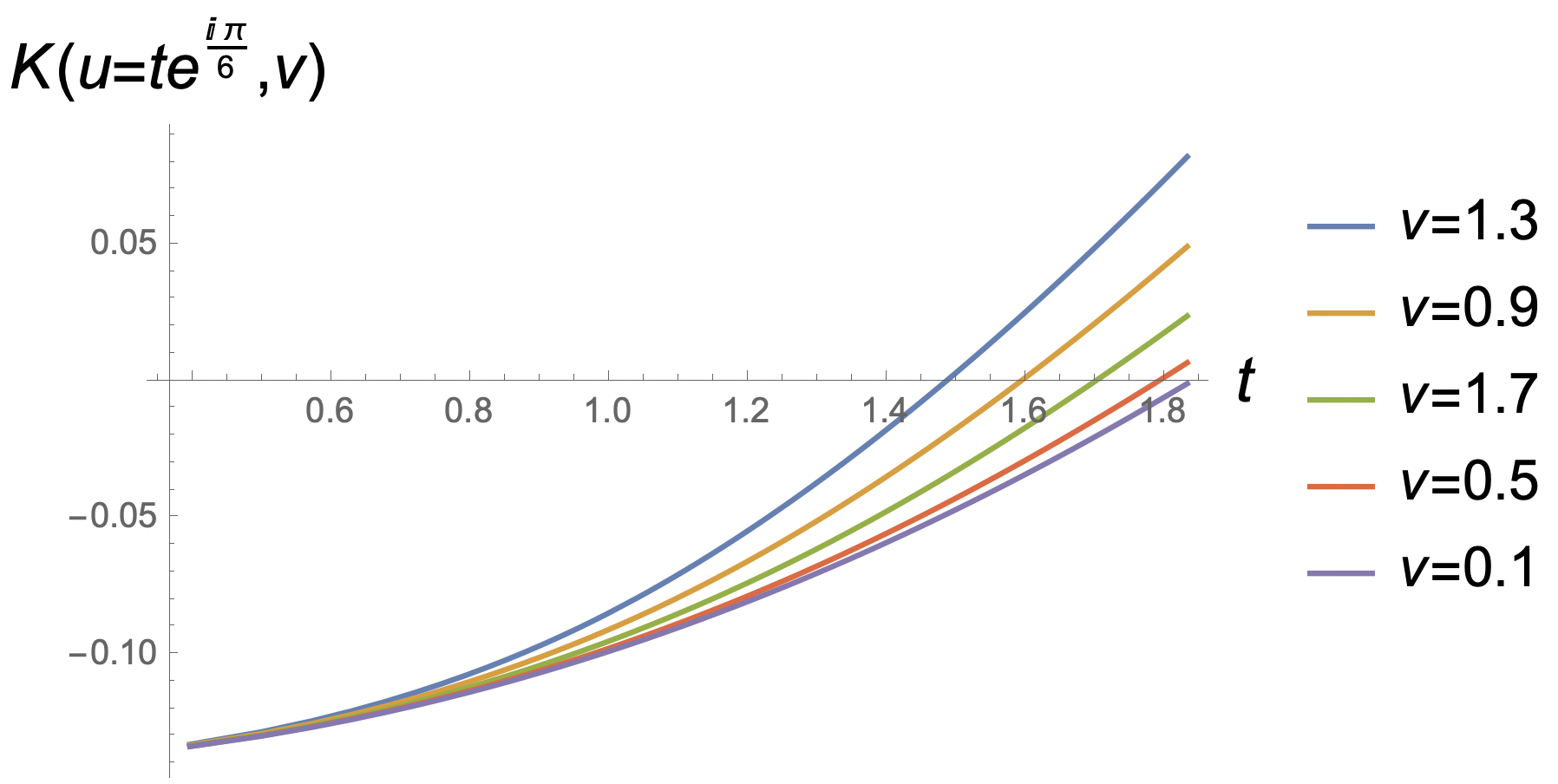}
\caption{ 
$K$ plotted along rays parameterized by $u=te^{\frac{i\pi}{6}}$, for various real values of $v$. Each ray consists of 100 data points. As $v$ is increased, the region of negative $K$ shrinks.
\label{fig:KericA}}
\end{subfigure}
\hfill
\begin{subfigure}[b]{0.95 \textwidth}
\centering
\includegraphics[width=0.45\textwidth]{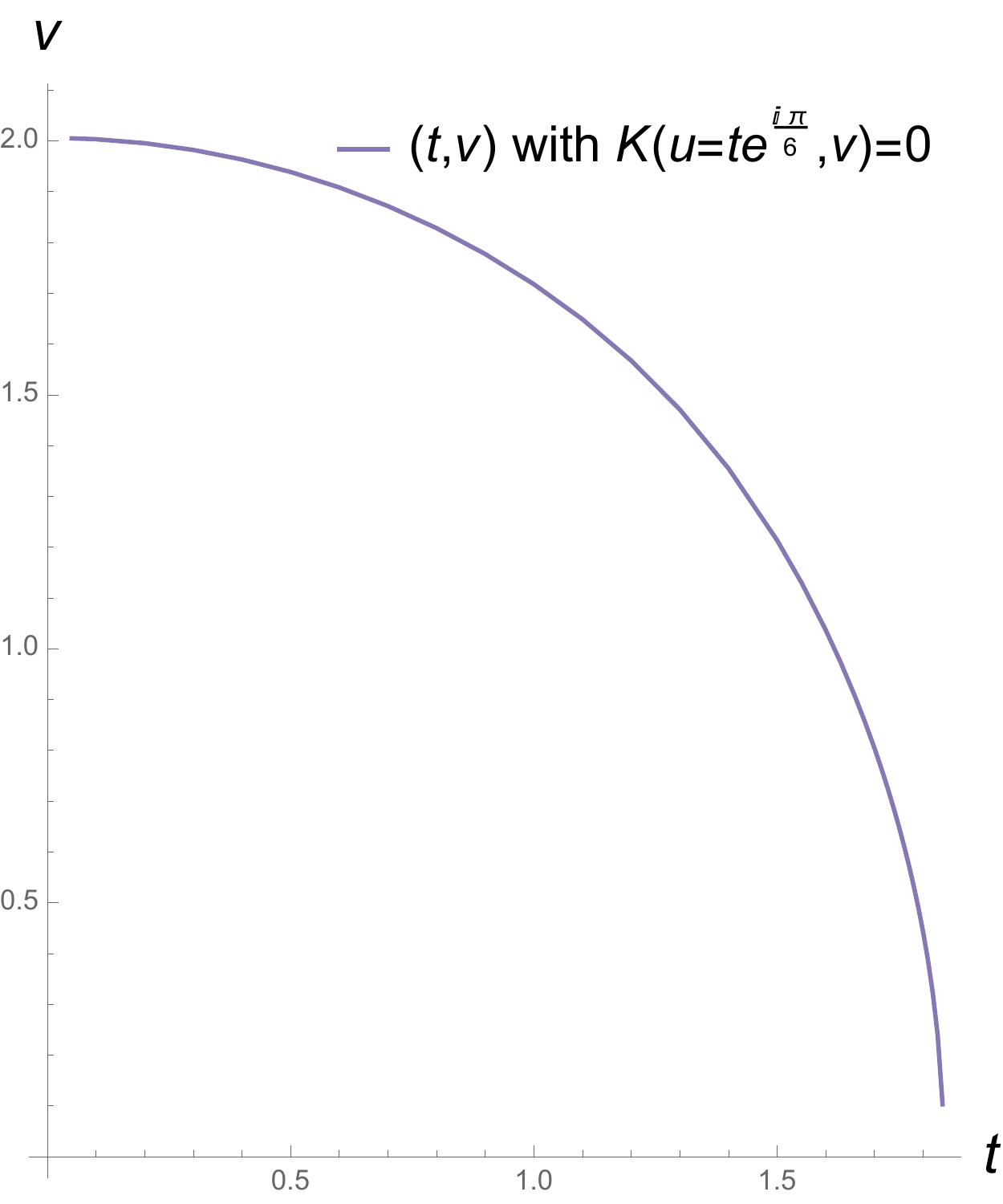}
\caption{
Plot of the (approximately, but not exactly ellipse-shaped)~$K = 0$ contour in the~$(t, v)$ plane, for various real values of~$t$ and~$v$, with~$u=t e^{\frac{i \pi}{6}}$. The curve is made up of 36 data points, and is symmetric under $v\to -v$. 
\label{fig:KericB}}
\end{subfigure}
\caption{Plots of $K$ made using the numerical techniques of  appendix \ref{sec:F4}.
\label{fig:Kplots2}}
\end{figure}

Using both of these numerical techniques, we have evaluated $K$ and its derivatives on  many slices through the four-real-dimensional moduli space, with representative plots appearing in figures \ref{fig:Kplots}-\ref{fig:Kplots2}.
In all cases we observe that $K$ is apparently convex, with no evidence for an extremum away from the origin of moduli space.

\sm

Figure \ref{fig:Kplots} depicts two representative numerical plots of $K(u,v)$ on two-dimensional slices of moduli space. In figure \ref{fig:Krealurealv}, $K$ is plotted in the $\text{Re}(u)$ and $\text{Re}(v)$ plane with  $\text{Im}(u)=\text{Im}(v)=0$. This slice includes the two Argyres-Douglas points at $v=\pm 1$, and one of the three multi-monopole points on the $K=0$ contour. Figure \ref{fig:Kuplanev0} is the $SU(3)$ analogue of figure \ref{fig:Ksu2better} for $SU(2)$, depicting $K$ in the complex $u$-plane at $v=0$ that includes all three multi-monopole points on the $K=0$ contour. Another slicing is shown in \ref{fig:KericA}, which depicts $K$ along rays in the complex $u$-plane for various fixed real values of $v$. 

\sm

All these plots indicate that~$K$ is negative around the $\mathbb{Z}_6$-symmetric point~$u = v = 0$ and goes to positive infinity as $u,v\to \infty$.
A visualization of the the~$K = 0$ hypersurface bounding the region of negative~$K$ surrounding the origin is depicted in figure \ref{fig:K0region}. In detail, figure~\ref{fig:K0contoursu} depicts the complex $u$-plane as a function of real $v$, and figure \ref{fig:K0contoursv} depicts the complex $v$-plane as a function of real $u$. In these slices, the~$K = 0$ surface is roughly shaped like a cigar, which caps off at approximately $|v| \sim 2.01$ and $|u| \sim 1.89$. Another view of this boundary is depicted in figure \ref{fig:KericB}, which shows an almost-circular section of the $K=0$ surface along a ray in the $u$-direction.

\newpage

\newpage

\section{Some Candidate Walls of Marginal Stability}
\setcounter{equation}{0}
\label{sec:MS}

Consider the Coulomb branch of the pure~$SU(N)$ gauge theory, where the low-energy gauge group is $U(1)^{N-1}$. In addition to the massless fields on the Coulomb branch, there can be massive particles. These are characterized by their mass~$M$, as well as their electric and magnetic charges under the low-energy~$U(1)^{N-1}$ gauge group, collectively denoted by a charge vector~$\vec \mu$, 
\bea
\vec \mu = \left(q_1, \ldots, q_{N-1}; g_1, \ldots, g_{N-1}\right) \in \ZZ^N \times \ZZ^N~,
\eea
with~$q_1, \ldots , q_{N-1} \in \ZZ$ the electric charges and $g_1, \ldots , g_{N-1} \in \ZZ$ the magnetic ones. 

\sm

The central charge in the~${\cal N} = 2$ supersymmetry algebra is a complex linear function of the charges that depends on the SW periods~\cite{Seiberg:1994rs},
\be
\label{zdef}
Z[\vec \mu] = \sum _{I=1}^{N-1} \left( q_I a_I + g_I a_{DI}\right)~.
\ee
A single-particle state of mass~$M$ and electromagnetic charge vector~$\vec \mu$ satisfies the following BPS bound~\cite{Witten:1978mh},
\be
M \geq \left|Z[\vec \mu] \right|~.
\ee
The particle is a short, or BPS, multiplet of the~${\cal N} = 2$ super-Poincar\'e algebra if and only if this bound is saturated, 
\be
\label{bpsmass}
M_\text{BPS} = \left| Z[\vec \mu]\right|~.
\ee

\sm

Consider two BPS particles with charge vectors~$\vec \mu, \vec \mu'$ and masses given by~\eqref{bpsmass}. Henceforth we assume that both charge vectors are non-zero, and that they are not integer multiples of one another.\footnote{~If~$\vec \mu = p  \vec \mu'$ with~$p \in \ZZ$, it is possible that the BPS particle of charge~$\mu$ is a threshold bound state of~$p$ BPS particles of charge~$\vec \mu'$. This famously happens for D0 branes in type IIA string theory. The structure of such bound states is particularly delicate and we will not discuss it.} Let us recall that two such particles can in principle form a single-particle bound state (which necessarily has electromagnetic charge vector~$\vec \mu + \vec \mu'$) that is also BPS. This happens precisely for threshold bound states (with zero binding energy) whose masses satisfy the following condition,
\be
\left|Z[\vec\mu + \vec \mu']\right| = \left|Z[\vec \mu] \right| + \left| Z[\vec \mu'] \right|~. 
\ee 
Since~$Z$ is a linear function of the charges, this condition saturates the triangle inequality, which is only possible when the complex numbers~$Z[\vec \mu]$, $Z[\vec \mu']$ (and hence also~$Z[\vec \mu + \vec \mu']$) are related by a real proportionality factor, 
\bea
\label{cwms}
Z[\vec \mu'] = \zeta \, Z[\vec \mu]~, \qquad \zeta \in \RR\,.
\eea
For fixed charge vectors, this real condition carves out a real-codimension-one slice on the Coulomb branch. We refer to this slice as a candidate wall of marginal stability for the BPS particles with charge vectors~$\vec \mu, \vec \mu'$. Whether or not these particles actually form a bound state upon crossing the wall is a more interesting and delicate question that we do not analyze here.

\subsection{Review of Marginal Stability for~$SU(2)$}

\begin{figure}[t!]
\centering
\includegraphics[width=0.35\textwidth]{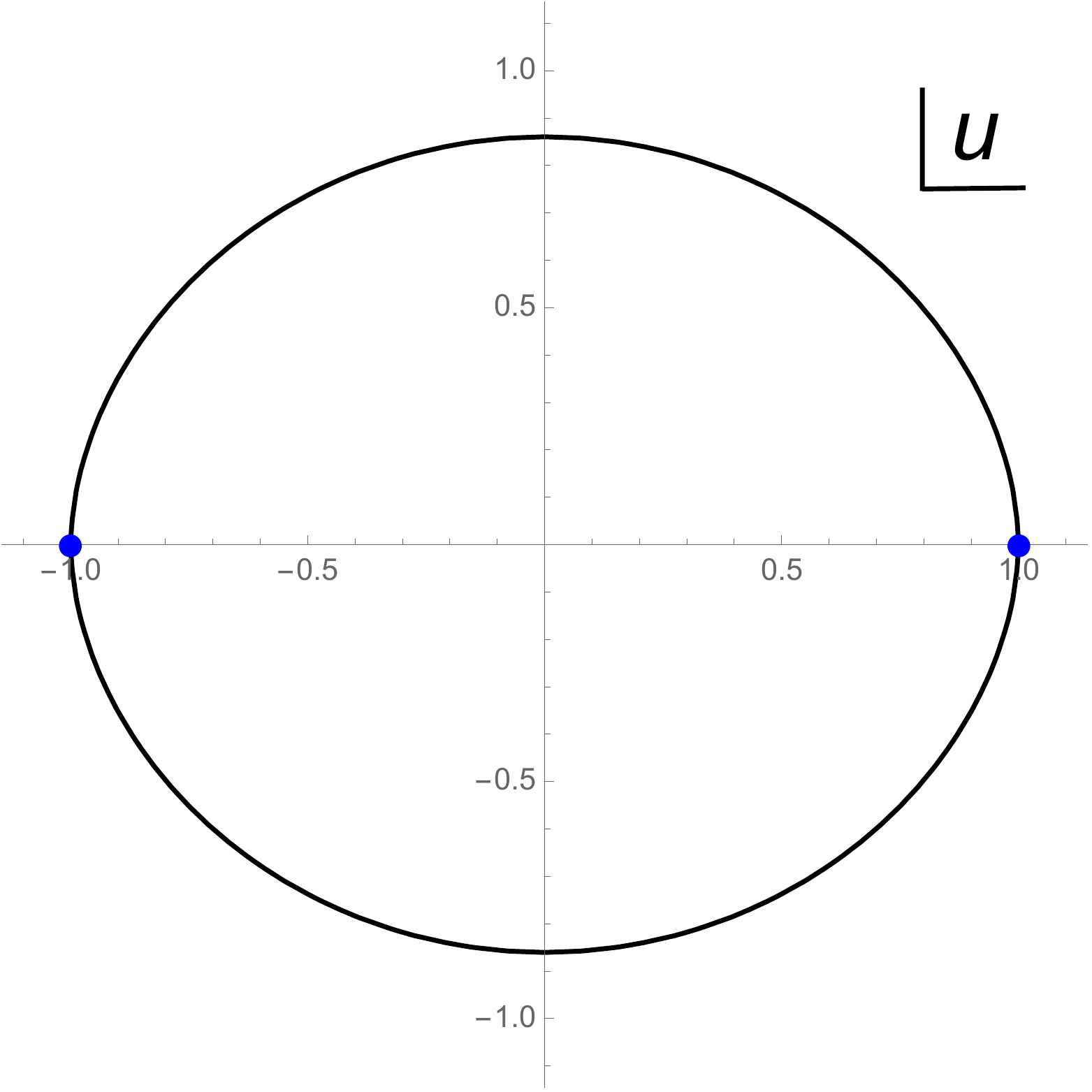}
\caption{The unique wall of marginal stability in~$SU(2)$ gauge theory. The monopole point at $u=1$ and the dyon point at $u=-1$ are indicated as blue dots.
\label{fig:su2ms}}
\end{figure}

\sm

For gauge group $SU(2)$, there is a single pair~$\vec \mu  = (q, g)$ of electromagnetic charges on the Coulomb branch, so that~$Z[\vec \mu] =  q a + g a_D$. Given two (non-vanishing and non-parallel) charge vectors~$\vec\mu=(q, g)$ and $\vec \mu' = (q', g')$, the condition~\eqref{cwms} for a candidate wall of marginal stability then reads
\bea
(q ' a + g' a_D) = \zeta (q a + g a_D )\,, \qquad \zeta \in \RR\,.
\eea
Since the charges are all real, this condition can be satisfied if and only if~$a$ and $a_D$ are themselves related by a real proportionality factor, or equivalently
\be
\label{su2wall}
\Im \left({a_D \over a}\right) = 0~.
\ee
This condition defines a (roughly elliptical) curve surrounding the origin~$u = 0$, which is plotted in figure~\ref{fig:su2ms}. As shown in~\cite{Seiberg:1994rs,Ferrari:1996sv}, inside the wall of marginal stability, there are are precisely two BPS states (together with their antiparticles): the monopole that becomes massless at~$u = 1$, and the dyon that becomes massless at~$u = -1$. Note that these points lie on the wall. The interior of the wall is known as the strong coupling chamber of the moduli space. As the wall is crossed towards the weakly coupled region at infinity, the monopole and the dyon form an infinite tower of BPS bound states comprising the~$SU(2)$ W-bosons and the infinite dyon towers that are visible at weak coupling.

\sm

Finally, note that the K\"ahler potential~$K \sim \Im a \bar a_D $ vanishes on the wall of marginal stability, because the condition~\eqref{su2wall} is equivalent to~$\Im a \bar a_D = 0$. Thus our notion of strong coupling region,  defined as the region where~$K < 0$, coincides with the standard strong coupling chamber for BPS particles in the case of~$SU(2)$ gauge group.

\subsection{Some Candidate Walls of Marginal Stability for~$SU(3)$} 

Here we use the expansion of the periods obtained in Corollary \ref{cor:2.3} for $SU(3)$ gauge group to determine candidate walls of marginal stability in special slices of the moduli space, and in a neighborhood that encompasses our strong-coupling region, where~$K < 0$. It is known that there is an open neighborhood of the origin -- termed the strong coupling chamber -- in which the BPS spectrum consists of exactly six stable particles (as well as their antiparticles), pairs of which become massless at the three multi-monopole points of~$SU(3)$ gauge theory~\cite{Lerche:2000uy,Alim:2011kw,Chuang:2013wt}. However, the precise extent of this chamber in moduli space is not known, and our results can serve as a starting point for a more detailed analysis of this question. 

\subsubsection{The $v=0$ Slice}

\begin{figure}[t!]
\centering
\includegraphics[width=0.45\textwidth]{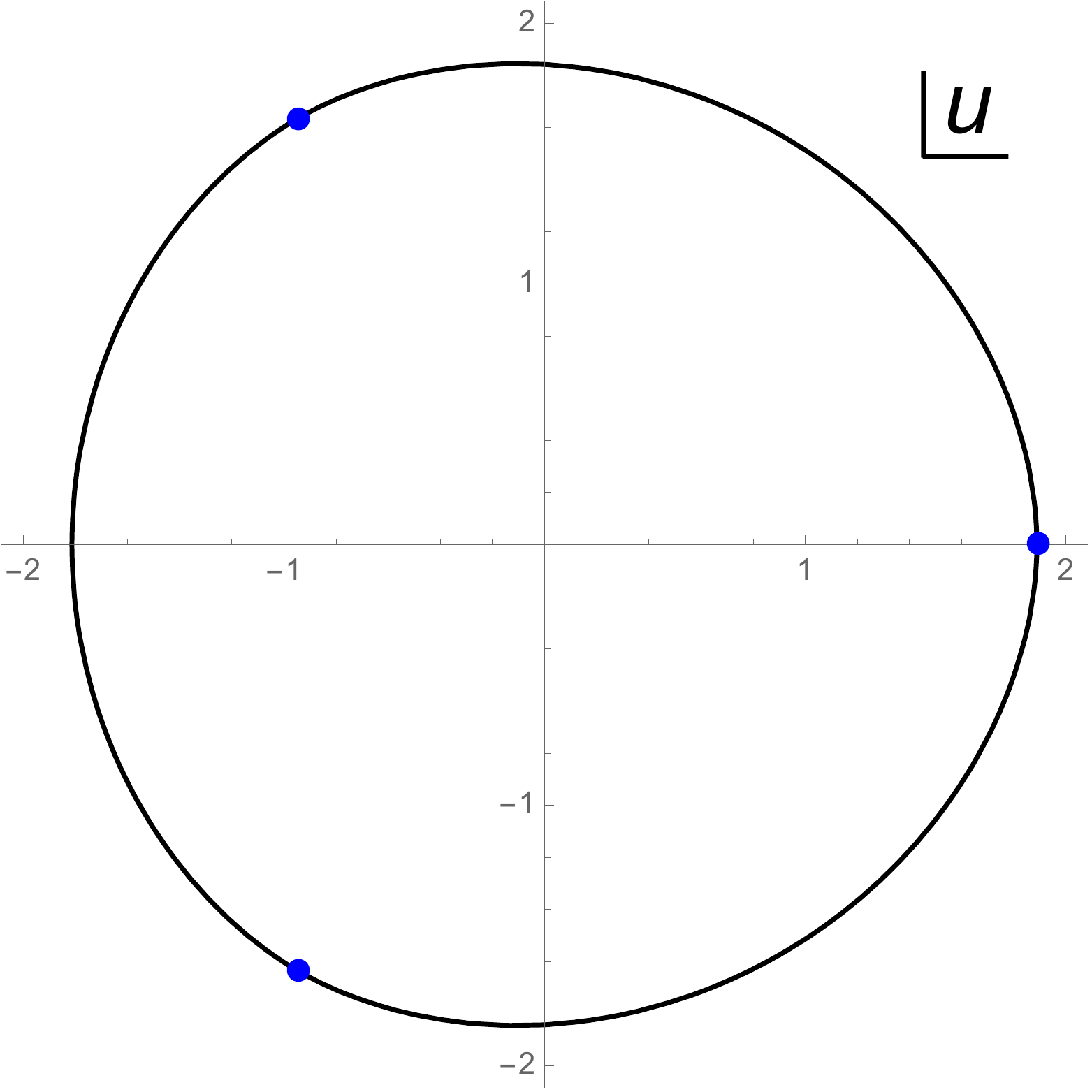}
\caption{Candidate wall of marginal stability in the $u$-plane (where $v=0$) for $SU(3)$ gauge group. The three multi-monopole points are indicated as blue dots. In the~$u$-plane, the wall coincides with the locus where~$K =0$, i.e.~it coincides with the outermost contour plotted in figure \ref{fig:K0contoursu}.
\label{fig:su3ms1}}
\end{figure}

Inspection of the solution given by Corollary \ref{cor:2.3} shows that $Q_{0,1}=Q_{1,1}=0$ when $v=0$, so that $Q(\xi) = \xi^4 Q_{0,0} + \xi^2 Q_{1,0}$. Thus, the periods in the~$u$-plane are given by
\bea
a_1 & = &  (\rho^2 -1) Q_{0,0} +  (\rho-1) Q_{1,0} \,,
\no \\
a_{2} & = &  (\rho^2 -\rho) Q_{0,0} -  (\rho^2 -\rho) Q_{1,0} \,,
\no \\
a_{D1} & = &  (\rho -\rho^2) Q_{0,0} -  (\rho -\rho^2)Q_{1,0}  \,,
\no \\
a_{D2} & = &  (\rho^2 -1) Q_{0,0} +  (\rho-1) Q_{1,0} \,,
\eea
where $\rho = \ep^2 = e^{2 \pi i /3}$. The above expressions for the periods  imply the following relations, independently of the values taken by $Q_{0,0}$ and $Q_{1,0}$, 
\bea
\label{6.v=0a}
a_2 = - a_{D1}\,,\qquad a_{D2} = a_1\,.
\eea
Let us evaluate the K\"ahler potential on this slice. Thanks to~\eqref{6.v=0a}, the two pairs of periods contribute equally to~$K$, which can now be expressed in terms of $a_1$ and $a_{D1}$ only,
\bea
\label{uslK}
K = { i \over 2 \pi} \Big ( a_1 \bar a_{D1} - \bar a_1 a_{D1} \Big )\,.
\eea
This differs from the K\"ahler potential for~$SU(2)$ by an overall factor of~$2$. 

\sm

Now consider two BPS states with (non-vanishing, non-proportional) charge vectors~$\vec \mu = (q_1,q_2; g_1, g_2)$ and $\vec \mu'= ( q_1',q_2' ; g_1', g_2')$. Thanks to (\ref{6.v=0a}) their central charges are given by
\bea
Z[\vec \mu] & = & (q_1 +g_2) a_1 + (g_1 -q_2) a_{D1}~,
\no \\
Z[\vec \mu' ] & = & (q_1' +g_2') a_1 + (g_1' -q_2') a_{D1}\,.
\eea
Requiring these to be related by a real proportionality factor implies that $a_1$ and $a_{D1}$ are also thus related. Thus, as a result of the relations (\ref{6.v=0a}), this case is completely parallel to the case of $SU(2)$ discussed above: there is a candidate wall of marginal stability defined by the curve $\text{Im}(a_{D1}/a_1) =0$ in the $u$-plane, depicted in figure \ref{fig:su3ms1}, and the K\"ahler potential~\eqref{uslK} vanishes there.

\subsubsection{The $u=0$ Slice}
\label{sec:u0su3}

{\it General Discussion}

\medskip

Inspection of the solution given by Corollary \ref{cor:2.3} shows that $Q_{1,0}=Q_{1,1}=0$ for $u=0$, so that $Q(\xi) = \xi^4 Q_{0,0} + \xi Q_{0,1}$. As a result, the periods are given as follows 
\bea
\label{u0su3rels}
a_1 & = & -(2+\rho) Q_{0,0} +  \rho Q_{0,1}\,,
\no \\
a_2 & = & -(1+2 \rho) Q_{0,0} - Q_{0,1}\,,
\no \\
a_{D1} & = & + (1+2\rho) Q_{0,0} -  Q_{0,1}\,,
\no \\
a_{D2} & = & -(2+\rho) Q_{0,0} -  \rho Q_{0,1}\,,
\eea
where $\rho=e^{2 \pi i /3}$. The above expressions for the periods  imply the following inter-relations, independently of the values taken by $Q_{0,0}$ and $Q_{0,1}$, 
\bea
\label{7.rels}
a_2 = - \rho^2 \, a_1\,,  \hskip 1in a_{D2} = \rho \, a_{D1}\,.
\eea
On this slice, the K\"ahler potential takes the form
\bea
\label{6.Kahleru}
K = {i \over 4 \pi} \Big ( (1-\rho) a_1 \bar a_{D1} - (1-\rho^2) \bar a_1 a_{D1} \Big ) 
= {  \sqrt{3} \over 2 \pi} \Big ( |Q_{0,1}|^2 - 3 |Q_{0,0}|^2 \Big )\,.
\eea

As before, we consider two BPS states with (non-vanishing, non-proportional) charge vectors~$\vec \mu = (q_1,q_2; g_1, g_2)$ and $\vec \mu'= ( q_1',q_2' ; g_1', g_2')$. Substituting these charges and the periods~\eqref{u0su3rels} into the central charge formula~\eqref{zdef}, we find
\bea
Z [\mu ] & = & (m_1 + m_2 \rho) Q_{0,0} + (n_1 + n_2 \rho) Q_{0,1}\,,
\no \\
Z[\mu'] & = & (m_1' + m_2' \rho) Q_{0,0} + (n_1' + n_2' \rho) Q_{0,1}\,,
\eea
where,
\begin{align} 
m_1 & =  g_1 - 2 g_2 - 2 q_1 - q_2\,, & n_1 & =  g_1 -  q_2\,,
\no \\
m_2 & = 2 g_1 - g_2 -  q_1 - 2 q_2 \,, & n_2 & = - g_2 +  q_1\,,
\no \\
m_1' & =  g_1' - 2 g_2' - 2 q_1' - q_2' \,, & n_1' & =  g_1' -  q_2' \,,
\no \\
m_2' & = 2 g_1' - g_2' -  q_1' - 2 q_2' \,, & n_2' & = - g_2' +  q_1'\,.
\end{align} 
Note that since the charge vector $(g_1, g_2; q_1, q_2)$ is not identically zero, the same is true for $(m_1, m_2; n_1, n_2)$, and similarly for the primed charges. Thus $Z[\vec \mu], Z[\vec \mu'] \not=0$ for generic~$v$.  

\sm

The condition~\eqref{cwms} for a candidate wall of marginal stability, namely $Z[\vec \mu']=\zeta Z[\vec \mu]$ for $\zeta \in \RR$, may be expressed as follows,
\bea
\label{6.lambda}
\zeta ={ az + b \over cz + d }\,,
\qquad 
z= { Q_{0,1} \over Q_{0,0}}\,,
\eea
with
\begin{align}
a & = n_1'+n_2' \rho \,, & b & = m_1'+m_2'\rho\,,
\no \\
c & = n_1 +n_2 \rho \,,& d & = m_1+m_2 \rho\,.
\end{align}
We now analyze the implications of these equations, recalling from above 
that $c$ and $d$ are not both equal to zero. Let us distinguish the following cases:

\begin{itemize}
\itemsep=0in
\item For the singular case $ad-bc=0$ the numerator $az+b$ is a constant multiple of the denominator $cz+d$ in (\ref{6.lambda}). Since $c$ and $d$ cannot vanish simultaneously, it suffices to analyze the cases  $c \not=0$ and $d\not=0$, for which we have the relations,
\bea
c \not=0 : ~~  \zeta = {a \over c} \,,\hskip 1.2in
d \not=0 : ~~ \zeta = {b \over d} \,.
\eea
Since $a,b,c,d \in \ZZ[\rho]$, the ratios $a/c$ and $b/d$ may be real or complex. If the ratios are not real, then there can be no solution since $\zeta$ must be real. If the ratios are real, then the charges $\vec \mu$, $\vec \mu'$ are proportional to one another, which we assumed was not the case. 

\item For the regular case where $ad-bc \not=0$, the relation between $z$ and $\zeta$  may be inverted to give $z$ as a function of $\zeta$,
\bea
z = {Q_{0,1} \over Q_{0,0}} = { d \zeta -b \over - c \zeta +a} \,.
\eea
For generic electromagnetic charge vectors, the constants $a, b, c, d$ will be complex, and thus~$z$, as a function of $\zeta$, will span an arc of a circle in the complex plane whose center is on the imaginary axis. Below, we will see this explicitly in examples. 

\end{itemize}

\medskip

\noindent{\it Candidate Walls for BPS States that are Stable in the Strong-Coupling Chamber}

\medskip

We will now apply the general discussion above to the six BPS particles that are stable in the strong-coupling chamber of the~$SU(3)$ gauge theory. As we review in appendix \ref{sec:BPSapp}, in our conventions these particles have the following electromagnetic charge vectors~$(q_1, q_2; g_1, g_2)$,
\begin{equation}
\begin{split}
& \vec \mu_{01}  =  (-1, 0; -1, 0)~,\\
& \vec \mu_{02} = (0,1; 0,-1)~,\\
& \vec \mu_{11} = (1,0; -1, -1)~,\\
& \vec \mu_{12} = (-1, 1; 0,1)~,\\
& \vec \mu_{21} = (0,1; 1,1)~,\\
& \vec \mu_{22} = (1,-1; -1,0)~.
\end{split}
\end{equation}
Note that the particle pairs with charges~$\vec \mu_{k1}, \vec \mu_{k2}$ become massless at the three multi-monopole points (lying in the~$v = 0$ slice) corresponding to~$k = 0,1,2$. Combining with~\eqref{u0su3rels}, we see that their central charges in the~$u = 0$ plane take the following form, 
\begin{align}
\label{vplanezs}
 Z[\vec\mu_{01}] & = \thalf (3-i \sqrt{3}) (Q_{0,0} +Q_{0,1}) \,,
\no \\
Z[\vec\mu_{02}] & = \thalf (3-i \sqrt{3}) (Q_{0,0} - Q_{0,1})  \,,
\no \\
Z[\vec\mu_{11}] & =   -i \sqrt{3} (Q_{0,0} - Q_{0,1})  \,,
\no \\
Z[\vec\mu_{12}] & =   -i \sqrt{3}  (Q_{0,0} +Q_{0,1})  \,,
\no \\
Z[\vec\mu_{21}] & = - \thalf(3+i \sqrt{3}) (Q_{0,0} +Q_{0,1})  \,,
\no \\
Z[\vec\mu_{22}] & =    - \thalf (3+i \sqrt{3}) (Q_{0,0} -Q_{0,1})  \,.
\end{align}
Note that the two Argyres-Douglas points lie in the~$u = 0$ plane, at~$v = \pm 1$ (see~\eqref{spp}). Evaluating~\eqref{u0} at these points, we find the following relations between the $Q$-functions, 
\bea
Q_{0,1}\big |_{v=\pm 1} = \pm Q_{0,0} \big |_{v=\pm1}\,.
\eea
We see that the states~$\vec \mu_{01}, \vec\mu_{12}, \vec \mu_{21}$ are massless at the~$v = -1$ Argyres-Douglas point, while the remaining three BPS states~$\vec \mu_{02}, \vec\mu_{11}, \vec \mu_{22}$ are massless at~$v = +1$. 

\sm

Using the central charges in~\eqref{vplanezs}, we can form~$\left(\begin{smallmatrix} 6 \\ 2 \end{smallmatrix}\right) = 15$ different pairwise ratios, of which six are complex constants~$\sim \rho, \rho^2$ so that the corresponding central charges can never align. In order to express the alignment conditions for the remaining nine pairs, we use the variable~$z = Q_{0,1}/Q_{0,0}$. This leads to the following three cases:

\begin{enumerate}
\item The central charges pairwise align as~$Z[\vec \mu_{01}] \sim Z[\vec \mu_{02}]$, $Z[\vec \mu_{12}]  \sim Z[\vec \mu_{11}]$, and~$Z[\vec \mu_{21}]  \sim Z[\vec \mu_{22}]$ (with~$\sim$ indicating real proportionality) if and only if
\begin{equation}
{1 + z \over 1-z} = \zeta \in \RR \quad \Longleftrightarrow \quad z = {\zeta -1 \over \zeta + 1}~.
\end{equation}
This describes a horizontal straight line in the complex~$z$-plane, i.e.~$z=Q_{0,1}/Q_{0,0}$ has to be real, and this can only happen when $v$ is real. 

\item The central charges pairwise align as~$Z[\vec \mu_{01}] \sim Z[\vec \mu_{22}]$, $Z[\vec \mu_{12}]  \sim Z[\vec \mu_{02}]$, and~$Z[\vec \mu_{21}]  \sim Z[\vec \mu_{11}]$ (with~$\sim$ again indicating real proportionality) if and only if
\begin{equation}
\label{circ1}
\rho {1 + z \over 1-z} = \zeta \in \RR \quad \Longleftrightarrow \quad z = {\rho^2 \zeta -1 \over \rho^2 \zeta + 1}~.
\end{equation}
This describes a segment of the circle $|z+i/\sqrt{3}|^2 =4/3$.

\item The central charges pairwise align as~$Z[\vec \mu_{01}] \sim Z[\vec \mu_{11}]$, $Z[\vec \mu_{12}]  \sim Z[\vec \mu_{22}]$, and~$Z[\vec \mu_{21}]  \sim Z[\vec \mu_{02}]$ if and only if
\begin{equation}
\label{circ2}
-\rho^2 {1 + z \over 1-z} = \zeta \in \RR \quad \Longleftrightarrow \quad z = {\rho \zeta +1 \over \rho \zeta - 1}~.
\end{equation}
This describes a segment of the circle $|z-i/\sqrt{3}|^2 =4/3$.

\end{enumerate} 

\begin{figure}[htb]
\begin{center}
\tikzpicture[scale=0.92]
\scope[xshift=0cm,yshift=0cm, scale=0.8]
\draw (-4,0) -- (4,0);
\draw (0,-4) -- (0,4);

\draw  [blue, thick, domain=-90:90] plot ({4/sqrt(3)*cos(\x)},{2/sqrt(3)+4/sqrt(3)*sin(\x)});
\draw  [blue, thick, domain=90:270] plot ({4/sqrt(3)*cos(\x)},{2/sqrt(3)+4/sqrt(3)*sin(\x)});
\draw  [red,  thick, domain=-90:90] plot ({4/sqrt(3)*cos(\x)},{-2/sqrt(3)+4/sqrt(3)*sin(\x)});
\draw  [red,  thick, domain=90:270] plot ({4/sqrt(3)*cos(\x)},{-2/sqrt(3)+4/sqrt(3)*sin(\x)});

\draw  [thick, domain=0:90] plot ({2*sqrt(3)*cos(\x)},{2*sqrt(3)*sin(\x)});
\draw  [thick, domain=90:180] plot ({2*sqrt(3)*cos(\x)},{2*sqrt(3)*sin(\x)});
\draw  [thick, domain=180:270] plot ({2*sqrt(3)*cos(\x)},{2*sqrt(3)*sin(\x)});
\draw  [thick, domain=270:360] plot ({2*sqrt(3)*cos(\x)},{2*sqrt(3)*sin(\x)});

\draw (-0.2,-0.3) node{\small $0$};
\draw (2.5,-0.3) node{\small $1$};
\draw (-2.7,-0.3) node{\small $-1$};
\draw (2,0) node{$\bullet$};
\draw (-2,0) node{$\bullet$};

\draw (3.8,3.4) [fill=white] rectangle (3.2,2.8) ;
\draw (3.5,3.1) node{$z$};
\draw (3.9,-0.5) node{\small $\sqrt{3}$};
\endscope
\scope[xshift=5.8cm,yshift=-2.99cm,scale=1.05]
\begin{axis}[
xmin=-2.2, xmax=2.2,
ymin=-2.2,ymax=2.2]
\addplot +[smooth, thick, color=blue,mark=none] table [x=a, y=b]  {
a b
1 0
1.05506 0.0476384
1.10099 0.0984561 
1.14073 0.152278
1.17521 0.208873
1.20486 0.268
1.22986 0.329405
1.2503 0.392825
1.26619 0.457988 
1.27753 0.524617
1.28432 0.592424
1.28653 0.66112
1.28415 0.730411
1.27719 0.8
1.26565 0.869589
1.24955 0.93888
1.22895 1.00758
1.20388 1.07538
1.17444 1.14201
1.1407 1.20718
1.10277 1.2706
1.06077 1.332
1.01485 1.39113
0.965155 1.44772
0.911856 1.50154
0.855134 1.55236
0.795184 1.59996
0.73221 1.64413
0.666426 1.68468
0.598048 1.72145
0.527286 1.75427
0.454339 1.78301
0.379357 1.80753 
0.302378 1.82775
0.223099 1.84356
0.13986 1.8549
0.0353468 1.86172 
0 1.864
-0.0353468 1.86172
-0.13986 1.8549
-0.223099 1.84356 
-0.302378 1.82775 
-0.379357 1.80753 
-0.454339 1.78301 
-0.527286 1.75427 
-0.598048 1.72145 
-0.666426 1.68468 
-0.73221 1.64413
-0.795184 1.59996
-0.855134 1.55236
-0.911856 1.50154
-0.965155 1.44772 
-1.01485 1.39113
-1.06077 1.332
-1.10277 1.2706 
-1.1407 1.20718 
-1.17444 1.14201
-1.20388 1.07538 
-1.22895 1.00758 
-1.24955 0.93888
-1.26565 0.869589
-1.27719 0.8
-1.28415 0.730411
-1.28653 0.66112
-1.28432 0.592424
-1.27753 0.524617
-1.26619 0.457988 
-1.2503 0.392825
-1.22986 0.329405
-1.20486 0.268
-1.17521 0.208873
-1.14073 0.152278
-1.10099 0.0984561 
-1.05506 0.0476384 
-1 0
};
\addplot +[smooth, thick, color=red,mark=none] table [x=a, y=b]  {
a b
1 0
1.05506 -0.0476384
1.10099 -0.0984561 
1.14073 -0.152278
1.17521 -0.208873
1.20486 -0.268
1.22986 -0.329405
1.2503 -0.392825
1.26619 -0.457988 
1.27753 -0.524617
1.28432 -0.592424
1.28653 -0.66112
1.28415 -0.730411
1.27719 -0.8
1.26565 -0.869589
1.24955 -0.93888
1.22895 -1.00758
1.20388 -1.07538
1.17444 -1.14201
1.1407 -1.20718
1.10277 -1.2706
1.06077 -1.332
1.01485 -1.39113
0.965155 -1.44772
0.911856 -1.50154
0.855134 -1.55236
0.795184 -1.59996
0.73221 -1.64413
0.666426 -1.68468
0.598048 -1.72145
0.527286 -1.75427
0.454339 -1.78301
0.379357 -1.80753 
0.302378 -1.82775
0.223099 -1.84356
0.13986 -1.8549
0.0353468 -1.86172 
0 -1.864
-0.0353468 -1.86172
-0.13986 -1.8549
-0.223099 -1.84356 
-0.302378 -1.82775 
-0.379357 -1.80753 
-0.454339 -1.78301 
-0.527286 -1.75427 
-0.598048 -1.72145 
-0.666426 -1.68468 
-0.73221 -1.64413
-0.795184 -1.59996
-0.855134 -1.55236
-0.911856 -1.50154
-0.965155 -1.44772 
-1.01485 -1.39113
-1.06077 -1.332
-1.10277 -1.2706 
-1.1407 -1.20718 
-1.17444 -1.14201
-1.20388 -1.07538 
-1.22895 -1.00758 
-1.24955 -0.93888
-1.26565 -0.869589
-1.27719 -0.8
-1.28415 -0.730411
-1.28653 -0.66112
-1.28432 -0.592424
-1.27753 -0.524617
-1.26619 -0.457988 
-1.2503 -0.392825
-1.22986 -0.329405
-1.20486 -0.268
-1.17521 -0.208873
-1.14073 -0.152278
-1.10099 -0.0984561 
-1.05506 -0.0476384 
-1 0
};
\addplot +[smooth, thick, color=cyan,mark=none] table [x=a, y=b]  {
a b
-1 0
-0.984555 -0.0404093
-0.965611 -0.0806456
-0.943616 -0.120537
-0.918765 -0.159911
-0.891191 -0.198601
-0.861004 -0.236441
-0.828304 -0.273268
-0.793191 -0.308925
-0.755767 -0.343259
-0.716138 -0.376123
-0.674418 -0.407377
-0.630724 -0.436886
-0.585185 -0.464524
-0.537933 -0.490173
-0.489111 -0.513723
-0.43887 -0.535074
-0.387372 -0.554133
-0.334795 -0.570819
-0.281342 -0.585061
-0.227263 -0.596797
-0.172928 -0.605978
-0.119089 -0.612564
-0.0683537 -0.616527
0 -0.61785
0.068357 -0.616527
0.119089 -0.612564
0.172928 -0.605978
0.227263 -0.596797
0.281342 -0.585061
0.334795 -0.570819
0.387372 -0.554133
0.43887 -0.535074
0.489111 -0.513723
0.537933 -0.490173
0.585185 -0.464524
0.630724 -0.436886
0.674418 -0.407377
0.716138 -0.376123
0.755767 -0.343259
0.793191 -0.308925
0.828304 -0.273268
0.861004 -0.236441
0.891191 -0.198601
0.918765 -0.159911
0.943616 -0.120537
0.965611 -0.0806456
0.984555 -0.0404093
1 0
};
\addplot +[smooth, thick, color=orange,mark=none] table [x=a, y=b]  {
a b
-1 0
-0.984555 0.0404093
-0.965611 0.0806456
-0.943616 0.120537
-0.918765 0.159911
-0.891191 0.198601
-0.861004 0.236441
-0.828304 0.273268
-0.793191 0.308925
-0.755767 0.343259
-0.716138 0.376123
-0.674418 0.407377
-0.630724 0.436886
-0.585185 0.464524
-0.537933 0.490173
-0.489111 0.513723
-0.43887 0.535074
-0.387372 0.554133
-0.334795 0.570819
-0.281342 0.585061
-0.227263 0.596797
-0.172928 0.605978
-0.119089 0.612564
-0.0683537 0.616527
0 0.61785
0.068357 0.616527
0.119089 0.612564
0.172928 0.605978
0.227263 0.596797
0.281342 0.585061
0.334795 0.570819
0.387372 0.554133
0.43887 0.535074
0.489111 0.513723
0.537933 0.490173
0.585185 0.464524
0.630724 0.436886
0.674418 0.407377
0.716138 0.376123
0.755767 0.343259
0.793191 0.308925
0.828304 0.273268
0.861004 0.236441
0.891191 0.198601
0.918765 0.159911
0.943616 0.120537
0.965611 0.0806456
0.984555 0.0404093
1 0
};
\addplot[domain=-pi:pi,samples=200,black,line width=0.7pt]({2.006*sin(deg(x))}, {1.86*cos(deg(x))});
\draw (120,220) node{$\bullet$};
\draw (320,220) node{$\bullet$};
\draw (-100,220) -- (500, 220);
\draw (220,-100) -- (220, 500);
\draw (420,410) [fill=white] rectangle (380,370) ;
\draw (400,390) node{$v$};
\end{axis}
\endscope
\endtikzpicture
\caption{Candidate curves of marginal stability in the $u=0$ slice for $SU(3)$. In the left panel, the curves are circles of radius $\tfrac{2}{\sqrt{3}}$ centered at $\pm \tfrac{i}{\sqrt{3}}$ in the~$z$-plane, and plotted in blue and red. In the right panel, the image of the blue circle is shown in blue and cyan, while the image of the red circle is shown in orange and red. In both panels, the Argyres-Douglas points are indicated by bold black dots, while the~$K = 0$ curve where the K\"ahler potential vanishes is plotted in black. \label{fig:2}}
\end{center}
\end{figure}
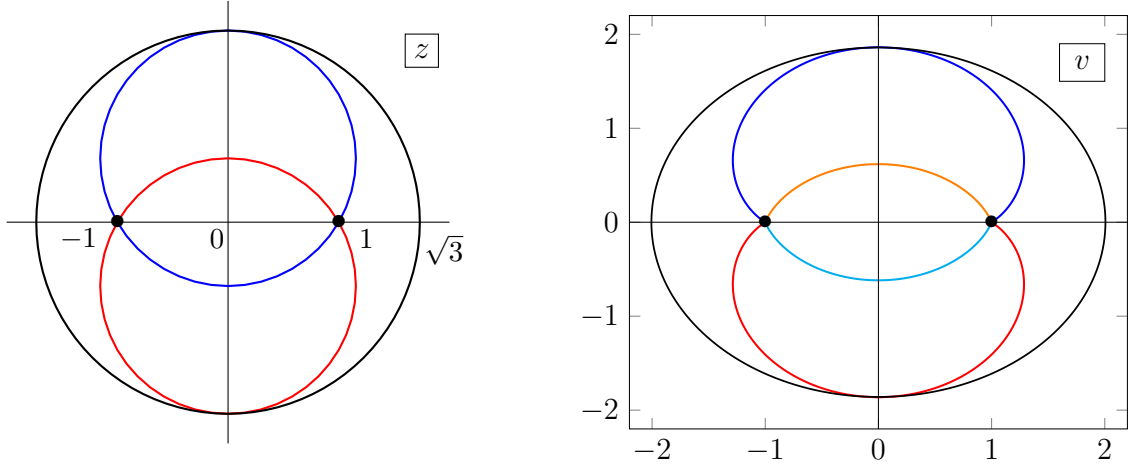

\noindent In the left panel of figure \ref{fig:2} we have plotted the circle described by~\eqref{circ1} in red, and the one described by~\eqref{circ2} in blue. There we also indicate in black the curve corresponding to the vanishing of the K\"ahler potential, $K = 0$. As may be read off from (\ref{6.Kahleru}), this curve is a circle in the $z$-plane of radius $\sqrt{3}$. 

\sm

 In the right panel of figure \ref{fig:2}, the curve of vanishing K\"ahler potential and the candidate curves of marginal stability are plotted in the $v$-plane. The function $z=Q_{0,1}(0,v)/Q_{0,0}(0,v)$ is holomorphic and single-valued away from the Argyres-Douglas branch points.
Thus, the map from $z$ to $v$ is conformal away from the branch points and preserves all angles. Comparison of the curves in the left and right panels of figure \ref{fig:2} clearly shows, however, that the angles between the curves are not preserved at the Argyres-Douglas points, as expected. Using~\eqref{u0}, the precise expression for the map is given by
\bea
z = {Q_{0,1}(0,v) \over Q_{0,0}(0,v)} = { 8 \pi^3 \, v \over 2^{{2 \over 3}} 3^3 \Gamma ( \tfrac{2}{3})^6} 
{ F(\tfrac{1}{3}, \tfrac{1}{3}; \tfrac{3}{2}; v^2) \over F (- \tfrac{1}{6}, - \tfrac{1}{6};  \tfrac{1}{2}; v^2)}\,.
\eea
The lowest order approximation, where the hypergeometric functions are set to 1, gives the approximation $z \approx 0.93875 \, v$ and translates the circle $|z|^2 =3$ into the circle $|v|^2 \approx 3.4043$, which provides  a reasonable approximation to the curve of vanishing K\"ahler potential in the $v$-plane. The behavior near the Argyres-Douglas points may be obtained by using the analytic continuation formulas for the hypergeometric functions, which leads to
\bea
z = v \Big ( 1- 0.47310 (1-v^2)^{5 \over 6} +\cO(1-v^2) \Big )\,.
\eea
The points $v = \pm 1 $ are clearly mapped to the points $z=\pm 1$, but the map is not conformal at those points, which explains the widening of the angles in the $v$ plane.

\subsection{Generalization to the~$u_0$ Slice for~$SU(N)$} 

The approach adopted above for the $u=0$ slice in $SU(3)$ extends almost verbatim to the 1-complex-dimensional slice $u_0\not=0$ and $u_n=0$ with $n =1,\ldots, N-2$ for~$SU(N)$ gauge group. Recall from Corollary \ref{cor:4} that in this case, 
\bea
a_{DI} & = & Q(\ep^{2I}) - Q(\ep^{2I-1})\,,
\no \\
a_I & = &  \sum_{J=1}^I \Big \{ Q(\ep^{2J-1}) - Q(\ep^{2J-2}) \Big \}\,,
\eea
where $\ep=e^{2 \pi i /2N}$ and the function $Q(\xi)$ for any $\xi^{2N}=1$  is given by,
\bea
Q(\xi) = \xi Q_1 + \xi^{N+1} Q_{N+1}\,,
\eea
where $Q_1$ and $Q_{N+1}$ are functions of $u_0$ only. Substituting the form of these functions into the periods, we obtain, 
\bea
a_{DI} & = &  \ep^{2I-1} \Big \{ (\ep-1) Q_1 + (\ep+1) Q_{N+1} \Big \}\,,
\no \\
a_I & = &  \sum_{J=1}^I \ep^{2J-2} \Big \{ (\ep-1) Q_1 - (\ep+1) Q_{N+1} \Big \}\,.
\eea
The central charge of a BPS state with electromagnetic charge vector\\ $\vec\mu = (q_1, \ldots, q_{N-1}; g_1, \ldots, g_{N-1})$ is given by
\bea
Z[\vec \mu] = m Q_1 + n Q_{N+1}\,,
\eea
where $m,n \in \ZZ[\ep]$ are given in terms of the charge vector,
\bea
m & = &  \sum_{I=1}^{N-1} g_I \, \ep^{2I-1} (\ep-1) +   \sum_{I=1}^{N-1} q_I \, \sum_{J=1}^{I}  \ep^{2J-2} (\ep-1)\,,
\no \\
n & = &  \sum_{I=1}^{N-1} g_I \, \ep^{2I-1} (\ep+1) -   \sum_{I=1}^{N-1} q_I \, \sum_{J=1}^{I}  \ep^{2J-2} (\ep+1) \,.
\eea
Now we simply repeat the argument used to determine candidate curves of marginal stability in the $u=0$ slice for $SU(3)$ gauge group (see section~\ref{sec:u0su3} above). 
Consider two charge vectors~$\vec \mu, \vec \mu'$ with corresponding central charges, 
\bea
Z[\vec \mu] & = & m Q_1 + n Q_{N+1}\,,
\no \\
Z[\vec \mu'] & = & m' Q_1 + n' Q_{N+1}\,.
\eea
Marginal stability requires $Z[\vec\mu']=\zeta Z[\vec\mu]$ for $\zeta \in \RR$, or equivalently
\bea
\zeta = { n' z + m'   \over nz+ m   } \,, \hskip 0.9in z = { Q_{N+1} \over Q_1}\,.
\eea
For the singular case $n'm-m'n=0$ and, say, $m \not=0$, we have $\zeta = m'/m$. There are then two possibilities: if $m'/m$ is not real, then there are no solutions; while if $m'/m=n'/n$ are real, then the charge vectors are proportional to one another. 
For the regular case $n'm-m'n\not =0$, the relation between $z$ and $\zeta$ may be inverted and we have,
\bea
z = { Q_{N+1} \over Q_1} = {m \zeta - m' \over -n\zeta +n'}\,.
\eea
For generic charge vectors, $z$ traces an arc of a circle in the complex plane.

\appendix

\addtocontents{toc}{\protect\setcounter{tocdepth}{1}}

\newpage

\section{Proof of Theorem \ref{thm:1}}
\setcounter{equation}{0}
\label{sec:A}

In this appendix, we shall provide a complete proof of Theorem \ref{thm:1}. 

\subsection{Taylor series expansion of $\lambda$} 

To obtain a Taylor expansion for  the periods $a_I$ and $a_{DI}$ at the $\ZZ_{2N}$ symmetric curve, we begin by Taylor expanding the Seiberg-Witten differential  in the moduli $u_n$ by setting,  
\bea
A(x) = x^N - U(x)\,,
\hskip 0.9in 
U(x) = \sum _{\ell=0}^{N-2} u_\ell \, x^\ell\,,
\eea
and expanding in powers of the polynomial $U(x)$, 
\bea
\lambda
= \sum _{m,n=0}^\infty { \Gamma (m+\half) \Gamma (n+\half) \over \Gamma (\half)^2 \, m! \, n!} 
{ \big (N x^N - x U'(x) \big ) \, U(x)^{m+n} \over (x^N- 1)^{\half +m} (x^N+ 1)^{\half +n}} \, dx\,.
\eea
To obtain this formula, it is convenient to derive the expansions arising from the factors $A(x) \pm 1$ separately and then multiply both series together. Furthermore, to arrive at an integrand whose integrals are easily computed, it will be convenient to multiply numerator and denominator by the factor $(x^N+1)^m (x^N-1)^n$, so that all contributions have a common denominator in the form of a power of $(x^{2N}-1)$. Changing summation variables from $m,n$ to $m$ and $M=m+n$, the result may be expressed as follows, 
\bea
\lambda
= \sum _{M=0}^\infty P_M(x^N)  \, {\big (N x^N - x U'(x) \big ) \, U(x)^M \over (x^{2N}- 1)^{\half +M} } \, dx~.
\eea
Here we have introduced a family of polynomials $P_M(z)$, defined by
\bea
P_M(z) = \sum_{m=0}^M { \Gamma (m+\half) \Gamma (M-m+\half) \over \Gamma (\half)^2 \, m! \, (M-m)!} 
 (z+1)^m (z-1)^{M-m}\,.
\eea
Alternatively, one may define these polynomials by their generating function, 
\bea
\sum_{M=0}^\infty x^M P_M(z) = { 1 \over \sqrt{(1-xz)^2 - x^2}} \,.
\eea
The polynomial $P_M(z)$ is of degree $M$ in $z$, satisfies the parity relation $P_M(-z) = (-)^M P_M(z)$, and belongs to a class of polynomials  that generalizes  Jacobi polynomials. Its expansion in powers of $z$ defines the coefficients $p_M(\ell)$ as follows, 
\bea
P_M(z) = \sum_{m=0}^M p_M(m) \, z^m\,.
\eea
The parity relation for $P_M(z)$ implies that the coefficients $p_M(m)$ vanish unless $M$ and $m$ are either both even or both odd, in which case we have the following expression for $p_M(m)$ obtained using {\tt Mathematica},
\bea
\label{A.pMm}
p_M(m) = { 2^{m-M} \, M! \over m! \, \Gamma (\thalf (M-m) +1)^2}~, \qquad 0\leq m \leq M, \quad M \equiv m ~ (\rm{mod} ~ 2)\,.
\eea
 We shall also use the multinomial expansion of $U(x)^M$ in powers of the moduli,
\bea
U(x)^M = \sum _{\ell_0, \ldots , \ell_{N-2}=0 }^\infty \delta \left ( M - \sum_{j=0}^{N-2} \ell_j \right ) 
\binom{M}{\ell_0, \ldots, \ell_{N-2}} \,
u_0^{\ell_0} \cdots u_{N-2}^{\ell_{N-2}} \, x^L\,,
\eea
where $L$ and $M$ are related to the exponents $\ell_j$ by
\bea
\label{5.dept}
L = \sum_{j=0}^{N-2} j \ell_j \,, \hskip 0.9in M = \sum _{j=0}^{N-2} \ell_j\,.
\eea
Putting all together, we obtain the following expansion for the SW differential,
\bea
\label{5.SWexp}
\lambda = \sum _{\ell_0, \ldots , \ell_{N-2}=0 }^\infty 
\binom{M}{\ell_0, \ldots, \ell_{N-2}} \,
u_0^{\ell_0} \cdots u_{N-2}^{\ell_{N-2}} \, \sum_{m=0}^M p_M(m) \, \lambda _M (Nm+L)\,,
\quad
\eea
where $L$ and $M$ are given in terms of the exponents $\ell_j$ by the relations of (\ref{5.dept}) and 
the differential $(1,0)$-form $\lambda _M (k)$ is given as follows, 
\bea
\label{5.lamM}
\lambda _M (k)
= {(Nx^N - x U'(x)) \, x^k \,  dx \over (x^{2N}- 1)^{\half +M} }~.
\eea
To derive the expansion of Theorem \ref{thm:1} we need to obtain the period integrals of the differential $\lambda_M(k)$ and to carry out the sum over $m$.

\subsection{The basic integrals}

The period integrals of $\lambda$ may be expressed in terms of those of $\lambda_M(k)$ through (\ref{5.SWexp}), which in turn may be obtained as finite linear combinations of integrals of the  type, 
\bea
\int _0 ^y  { x^{\g-1} \, dx \over \left ( x^{2N} - 1 \right )^{{P \over 2}}}
= i^P \, { y^\g \over \g} \, F \left ( {P \over 2} , {\g \over 2N} ; 1+{\g \over 2N}; y^{2N} \right )\,,
\eea
for $\Re(\g) >0$ and odd integer $P\geq 1$. The branch cut has been chosen so that $\sqrt{x^{2N}-1} = -i \sqrt{1-x^{2N}}$ when  $|x|<1$, for which the square root $\sqrt{1-x^{2N}}$ is positive for $x$ real. Actually, in view of (\ref{2.aQ}) and (\ref{2.Q}), evaluating the period integrals  requires the above integrals only at the points $y = \xi$ where $\xi^{2N}=1$. To obtain these, we choose the integration path to be the straight line from 0 to $\xi$, as illustrated  for $N=3$ by the green lines in figure \ref{fig:1}.  Since the hypergeometric function then has argument 1, it may be simplified using Gauss's formula, 
\bea
F(a,b;c;1)= { \Gamma (c) \Gamma (c-a-b) \over \Gamma (c-a) \Gamma (c-b)}\,,
\eea
and we obtain, 
\bea
\label{5.int}
\int ^\xi  _0 { x^{\g-1} \, dx \over \left ( x^{2N} - 1 \right )^{{P \over 2}}}
= i^P \, { \xi^\g \over \g} \, 
 { \Gamma (1+{\g \over 2N}) \Gamma (1-{P \over 2}) \over \Gamma (1+{\g \over 2N}-{P \over 2}) }\,.
 \eea
Using the decomposition of the differential $(1,0)$-form $\lambda _M (k)$ in terms of the above integrands, 
\bea
\lambda _M (k)
={ N x^{N+k} \,  dx \over (x^{2N}- 1)^{\half +M} }
- \sum _{j=1}^{N-2} j u_j { x^{k+j} \,  dx \over (x^{2N}- 1)^{\half +M} }\,,
\eea
its integral is readily obtained with the help of (\ref{5.int}), 
\bea
\label{5.intl}
{ 1 \over  \pi i} \int ^\xi _0 \lambda _M (k) 
& = & 
{\xi ^{k+N+1}  \,  \Gamma ({k+1 \over 2N}+\half ) 
\over  2 \Gamma ({k+1 \over 2N}+1-M) \Gamma (M+\half)}\,,
\no \\ &&
- \sum _{j=1}^{N-2} j \, u_j \,  
{ \xi^{k+j+1} \, \Gamma ({k+j+1\over 2N} ) 
\over 2N \, \Gamma ({k+j+1  \over 2N}+ \half -M) \Gamma (M+\half )}~,
\eea
where we used $\Gamma (\half -M) \Gamma (\half +M) = \pi (-)^M$ for integer $M$ to simplify the result. 

\sm

The integral of the SW differential from 0 to an arbitrary $2N$-th root of unity $\xi$,
\bea
\label{5.Qxi}
Q(\xi) = { 1 \over  \pi i} \int ^\xi_0 \lambda \,,
\eea
may then be expressed in terms of the integrals of the differentials $\lambda_M(k)$, 
\bea
\label{5.W}
W_M(\xi, L) = M! \sum_{m=0}^M p_M(m) \, {1 \over  \pi i} \int ^\xi _0 \lambda _M (Nm+L)\,,
\eea
as follows, 
\bea
\label{5.Qfin}
Q(\xi) = \sum _{\ell_0, \ldots , \ell_{N-2}=0 }^\infty 
{  u_0^{\ell_0} \cdots u_{N-2}^{\ell_{N-2}} \over \ell_0! \cdots \ell_{N-2}!}  \, W_M(\xi, L)\,,
\eea
where in both formulas $L$ and $M$ are given in terms of the $\ell_j$ by (\ref{5.dept}).

\subsection{Carrying out the sum over $m$}

Substituting the result (\ref{5.intl}) for the integral of $\lambda_M(k)$ into the expression for $W_M(\xi,L)$ in~(\ref{5.W}), we obtain 
\bea
W_M(\xi, L) 
& = & 
{1  \over 2} \, \xi ^{NM+L+N+1}  \sum_{m=0}^M {p_M(m)  \,  \Gamma ({ m \over 2} + \half +{L +1 \over 2N} ) \, M!
\over   \Gamma ({m \over 2} + {L+1 \over 2N}+1-M) \Gamma (M+\half)}
\no \\ &&
- {1 \over 2N}  \xi^{NM+L+j+1} \, \sum _{j=1}^{N-2} j \, u_j \,  \sum_{m=0}^M
{  p_M(m) \, \Gamma ({m \over 2} + {L+j+1\over 2N} ) \, M!
\over \Gamma ({m \over 2} + {L+j+1  \over 2N}+ \half -M) \Gamma (M+\half )}\,.
\eea
We have used the fact that $p_M(m)$ vanishes unless $M$ and $m$ are both even or both odd to set $\xi^{Nm}=\xi^{NM}$, thereby allowing us to extract this factor from under the summation symbol.   Both sums over $m$ are of the following form for an arbitrary $\gamma \in \CC$, 
\bea
S_M(\g) = \sum_{m=0}^M  {  p_M(m) \, \Gamma ({m \over 2} + \g ) \, M!
\over \Gamma ({m \over 2} + \g + \half -M) \Gamma (M+\half )}\,,
\eea
in terms of which $W_M$ can be expressed as follows,
\bea
\label{5.Wl}
W_M(\xi, L) 
& = & 
{1 \over 2} \, \xi ^{NM+L+N+1}  S_M(\tfrac{N+L +1}{2N}) 
- {1 \over 2N}   \, \xi^{NM+L+1} \sum _{j=1}^{N-2} j \, u_j \,  \xi^j S_M(\tfrac{L+j+1}{2N})\,.
\eea
To evaluate the functions $S_M(\gamma)$ we need the following lemma:

{\lem
\label{lem:1}
The function $S_M(\g)$ evaluates to the following expression, 
\bea
S_M (\g) = { 2^{M+1-2\g}  \over \pi^2 }  \Gamma (2 \g) \Gamma( \tfrac{M+1}{2}  -\g)^2 \sin^2 \big ( \pi ( \tfrac{M+1}{2} -\g) \big )~,
\eea
for integer $M\geq 0$.}

\sm

The formula was obtained by induction from the form of $S_M(\g)$ for low values of $M$, and then verified using {\tt Maple} for all values of $M$ up to 200. It may be proven analytically by appealing to the hypergeometric function ${}_3 F_2$ as follows: using the explicit expression for $p_M(m)$ given in (\ref{A.pMm}), the sum over $m$ in $S_M(\g)$ may be carried out to obtain
\bea
S_M(\gamma) = 
{ 2^{M+2\sigma}  \Gamma (\tfrac{M}{2} + \thalf + \sigma)^2 \Gamma (\g+\sigma)
 \over \pi \Gamma (M +\thalf) \Gamma (\thalf + \sigma -M +\g)} \, 
 {}_3 F_2 \left ( \bma \sigma -\tfrac{M}{2},\sigma -\tfrac{M}{2},\g+\sigma \cr \thalf+ 2 \sigma , \thalf +\sigma -M +\g \ema ;  1 \right )\,,
\eea
where $\sigma = \tfrac{M}{2} - \left [ \tfrac{M}{2} \right ]$ takes the value $0$ when $M$ is even and $\thalf$ when $M$ is odd. Next, we use the analogue of Gauss's formula for ${}_3F_2$,
\bea
{}_3 F_2 \left ( \bma a,b,-n \cr  c, 1+a+b-c-n \ema ;1 \right ) 
= { \Gamma(c-a+n) \Gamma (c-b+n) \Gamma (c) \Gamma (c-a-b) 
\over \Gamma (c-a) \Gamma (c-b) \Gamma (c+n) \Gamma (c-a-b+n)}\,,
\eea
with $a=\sigma -\tfrac{M}{2}$, $b=\gamma+\sigma $, $c=\thalf+2 \sigma$, and $n = \tfrac{M}{2}-\sigma$.  After some simplifications, this expression combines to give the formula of Lemma \ref{lem:1} and completes its proof.

\sm

Substituting~Lemma \ref{lem:1} into~\eqref{5.Wl}, we find the following formula for $W_M(\xi,L)$, 
\bea
\label{5.WW}
W_M(\xi, L) 
& = & 
{  2^{M-(L +1)/N}   \over 2 \pi^2 } \, \xi ^{NM+L+N+1}  
\Gamma ( \tfrac{N+L+1}{N}) \Gamma( \tfrac{NM-L-1}{2N}  )^2 \sin^2 \big ( \pi  \tfrac{NM-L-1}{2N}  \big )
\no \\ &&
- {2^{M-(L+1)/N}  \over \pi^2 N }   \sum _{j=1}^{N-2} j \, u_j \,  \xi^{NM+j+L+1} \,  2^{- \tfrac{j}{N}}  
 \Gamma ( \tfrac{j+L+1}{N})  \Gamma( \tfrac{NM+N-j-L-1}{2N} )^2 
\no \\ && \hskip 1.5in 
\times \sin^2 \big ( \pi  \tfrac{NM+N-j-L-1}{2N}  \big )\,.
\eea

\subsection{Final simplification}

We now substitute the formula for $W_M(\xi,L)$ in~\eqref{5.WW} above into the expression for $Q(\xi)$ in~(\ref{5.Qfin}) to obtain
\bea
Q(\xi) & = & 
 \sum _{\ell_0, \ldots , \ell_{N-2}=0 }^\infty 
{ u_0^{\ell_0} \cdots u_{N-2}^{\ell_{N-2}} \over \ell_0! \, \cdots \ell_{N-2}! } 
{  2^{M-(L +1)/N}   \over 2 \pi^2 } \, \xi ^{NM+L+N+1}  
\Gamma ( \tfrac{N+L+1}{N}) \Gamma( \tfrac{NM-L-1}{2N}  )^2 \sin^2 \big ( \pi  \tfrac{NM-L-1}{2N}  \big )
\no \\ &&
-  \sum _{\ell_0, \ldots , \ell_{N-2}=0 }^\infty 
{ u_0^{\ell_0} \cdots u_{N-2}^{\ell_{N-2}} \over \ell_0! \, \cdots \ell_{N-2}! } 
{2^{M-(L+1)/N}  \over \pi^2 N }   \sum _{j=1}^{N-2} {j \, u_j \,  \xi^{NM+j+L+1} \over 
 2^{\tfrac{j}{N}}  } \Gamma ( \tfrac{j+L+1}{N}) 
 \no \\ && \hskip 1in \times \,
\Gamma( \tfrac{NM+N-j-L-1}{2N} )^2 \sin^2 \big ( \pi  \tfrac{NM+N-j-L-1}{2N}  \big )\,,
\eea
where on both lines we use the expressions for $L$ and $M$ given in (\ref{5.dept}). 
In the sum over $j$ on the second and third lines, we combine the lone factor of $u_j$ with the monomial
$u^{\ell_j} \to u_j^{\ell_j+1}$ and change variables $\ell_j +1 \to \ell_j$. The net effect  is to bring out a factor of $\ell_j$ and to decrease the values of $M$ and $\ell$ as follows, $M \to M-1$ and $L \to L - j$. Carrying out these three changes at once, and factorizing common parts, gives
\bea
Q(\xi) & = & 
 \sum _{\ell_0, \ldots , \ell_{N-2}=0 }^\infty 
{ u_0^{\ell_0} \cdots u_{N-2}^{\ell_{N-2}} \over \ell_0! \, \cdots \ell_{N-2}! } 
{  2^{M-(L +1)/N}   \over 2 \pi^2 } \, \xi ^{NM+L+N+1}  
 \Gamma( \tfrac{NM-L-1}{2N}  )^2 
\no \\ && \qquad
\times \sin^2 \big ( \pi  \tfrac{NM-L-1}{2N}  \big )
\left [ \Gamma (1+ \tfrac{L +1}{N}) - \Gamma ( \tfrac{L+1}{N}) \sum _{j=0}^{N-2}  { j \, \ell_j \over N} \right ] \,.
\eea
The expression inside the brackets simplifies to $\Gamma ( \tfrac{L+1}{N})/N$, which leads to our final result, 
\bea
Q(\xi) & = & 
 \sum _{\ell_0, \ldots , \ell_{N-2}=0 }^\infty 
{ u_0^{\ell_0} \cdots u_{N-2}^{\ell_{N-2}} \over \ell_0! \, \cdots \ell_{N-2}! } 
{  2^{M-(L +1)/N}   \over 2 \pi^2 N } \, \xi ^{NM+L+N+1}  
\no \\ && \hskip 0.8in \times 
\Gamma ( \tfrac{L +1}{N}) \Gamma( \tfrac{NM-L-1}{2N}  )^2 \sin^2 \big ( \pi  \tfrac{NM-L-1}{2N}  \big )\,.
\no
\eea
This expression  may be recast in the form of Theorem \ref{thm:1}, thereby completing its proof.

\section{The $SU(3)$ Solution in Terms of  Appell Functions}
\setcounter{equation}{0}
\label{sec:B}

In this appendix, we prove Corollary \ref{cor:2.3} and thereby show that the results obtained in Theorem \ref{thm:1} for arbitrary $N$ reproduce the solution in terms of Appell functions obtained in~\cite{Klemm:1995wp}.

\sm

The starting point for the proof  is the expression for $Q(\xi)$ for the case $N=3$. We shall use the simplified notation $v=u_0$ and $u=u_1$, and express the sum in terms of $\ell=\ell_0$ and $k=\ell_1$ so that $M=k+\ell$ and $L=k$. In terms of these variables, the result of Theorem \ref{thm:1} reduces to the following expression for $Q(\xi)$,  
\bea
Q(\xi) = 
 \sum _{k,\ell=0 }^\infty 
{  2^{(3\ell+2k-1)/3}   \over 6 \pi^2 \, k! \, \ell! } \, \xi ^{3\ell+4k+4}  \, u^k \, v^\ell  \, 
\Gamma ( \tfrac{k+1}{3}) \Gamma( \tfrac{3\ell+2k-1}{6}  )^2 \sin^2 \big ( \pi  \tfrac{3\ell+2k-1}{6}  \big )\,.
\eea
To decompose the function $Q(\xi)$ into powers of $\xi$, we decompose the summation variables $k$ and $\ell$ modulo 3 and 2 respectively, 
\begin{align}
k & = 3m + \mu & m & \geq 0 & \mu &= 0,1,2
\no \\
\ell & = 2n + \nu & n & \geq 0 & \nu & =0,1
\end{align}
so that the $\xi$-dependence of $Q(\xi)$ is contained entirely in $\mu, \nu$ and independent of $m,n$.
The function $Q(\xi)$ then decomposes as follows,
\bea
Q(\xi) & = & \sum _{\mu=0,1,2} \, \sum _{\nu=0,1} \xi^{4 \mu + 3 \nu +4} \, Q_{\mu, \nu}
\no \\ & = & 
\xi^4 Q_{0,0} + \xi^2 Q_{1,0} +\xi^0 Q_{2,0} +\xi  ^1Q_{0,1} +\xi^5 Q_{1,1} +\xi^3 Q_{2,1} \,,
\eea
where the coefficient functions are given by,
\bea
Q_{\mu, \nu} & = &
{ 1 \over 6 \pi^2} \sin^2 \big ( \pi  \tfrac{3\nu+2\mu-1}{6}  \big ) \sum_{m,n=0}^\infty {  2^{2m+2n +\nu +(2\mu -1)/3}   \over  (3m+\mu )! \, (2n+\nu)!  }  \, u^{3m+\mu}  \, v^{2n+\nu}  
\no \\ && \hskip 1.2in \times 
\Gamma ( m + \tfrac{\mu+1}{3})  \Gamma( m+n + \tfrac{3\nu +2\mu-1}{6}  )^2 \,.
\eea
Using the duplication and triplication formulas for the factorials in the denominators,
\bea
\Gamma (2n+\nu+1) & = & 
{ 2^{2n+\nu} \over \sqrt{\pi}} \Gamma(n+\tfrac{\nu}{2} +\thalf) \Gamma(n+\tfrac{\nu}{2} +1)\,,
\no \\
\Gamma (3m +\mu +1) & = & 
{ 3^{3m+\mu+\thalf} \over 2 \pi} \Gamma(m+\tfrac{\mu}{3} +\tfrac{1}{3}) \Gamma(m+\tfrac{\mu}{3} +\tfrac{2}{3})
\Gamma(m+\tfrac{\mu}{3} +1)\,,
\eea
we obtain, 
\bea
Q_{\mu, \nu} & = &
{ 2^{-1/3} \over 3 \sqrt{3 \pi}  } 
\sum_{m,n=0}^\infty  {  \sin^2 \big ( \pi  \tfrac{3\nu+2\mu-1}{6}  \big ) \,  \Gamma( m+n + \tfrac{3\nu +2\mu-1}{6}  )^2 \, \left ( { 4u^3 \over 27} \right )^{m+{\mu \over 3}}  \, v^{2n+\nu}  
 \over 
  \Gamma(n+\tfrac{\nu}{2} +\thalf) \Gamma(n+\tfrac{\nu}{2} +1) \,  
  \Gamma(m+\tfrac{\mu}{3} +\tfrac{2}{3})
\Gamma(m+\tfrac{\mu}{3} +1) }  \, .
\eea
Of the six inequivalent representations of $\ZZ_6$, $Q_{2,0}$ multiplies the trivial representation of $\ZZ_6$ and cancels in the differences giving the periods. Also, the sine-factor vanishes identically for $\mu=2$ and $\nu=1$, so that we have
\bea
Q_{2,1}=0~.
\eea
The remaining four functions correspond to $\mu, \nu =0,1$ and evaluate as follows, 
\bea
Q_{0, 0} & = &
{ 2^{-1/3} \over 12 \sqrt{3 \pi}  } 
\sum_{m,n=0}^\infty  {    \Gamma( m+n - \tfrac{1}{6}  )^2 \, 
 \over 
 \Gamma(m +\tfrac{2}{3})   \Gamma(n +\thalf)   \, m! \, n! }  \, \left ( { 4u^3 \over 27} \right )^{m}  \, v^{2n}  \,,
\no \\
Q_{0, 1} & = &
{ 2^{-1/3} \over 4 \sqrt{3 \pi}  } \, 
v \sum_{m,n=0}^\infty  {   \Gamma( m+n + \tfrac{1}{3}  )^2 \, 
 \over 
  \Gamma(m +\tfrac{2}{3})   \Gamma(n+\tfrac{3}{2})  \, m! \, n!} \, \left ( { 4u^3 \over 27} \right )^{m}  \, v^{2n}  \,,
\no \\
Q_{1, 0} & = &
{ 2^{1/3} \over 36 \sqrt{3 \pi}  } \, u
\sum_{m,n=0}^\infty  {    \Gamma( m+n + \tfrac{1}{6}  )^2  
 \over     \Gamma(m+\tfrac{4}{3} ) \Gamma(n +\thalf)   \, m! \, n! }  \, 
 \left ( { 4u^3 \over 27} \right )^m  \, v^{2n} \,,
 \no \\
Q_{1, 1} & = &
{ 2^{1/3} \over 12 \sqrt{3 \pi}  } \, uv
\sum_{m,n=0}^\infty  {    \Gamma( m+n + \tfrac{2}{3}  )^2 \, 
 \over 
 \Gamma(m+\tfrac{4}{3} ) \Gamma(n+\tfrac{3}{2} ) \, m! \, n!}  \, \left ( { 4u^3 \over 27} \right )^m \, v^{2n}  \,.
\eea
Using the definition of the Appell function $F_4$ in the variables $x=4u^3/27$ and $y=v^2$, we easily convert these expressions into those stated in Corollary \ref{cor:2.3}.

\newpage

\section{Aspects of Elliptic Functions and Modular Forms}
\setcounter{equation}{0}
\label{sec:C}

In this appendix, we provide a brief review of elliptic functions and modular forms as needed here.  A standard and useful reference is \cite{BatemanII}, whose notations we follow.

\subsection{Weierstrass Elliptic Functions}

Given a lattice in $\CC$ with periods $2 \om, 2 \om'$ the Weierstrass $\wp$ function $\wp(\upsilon | 2\om, 2 \om')$ satisfies the following differential equation,
\bea
\left ( { \p \wp (\upsilon | 2\om, 2 \om') \over \p \upsilon} \right )^2= 4 \, \wp(\upsilon | 2\om, 2 \om')^3 - g_2(2\om, 2 \om') \, \wp(\upsilon | 2\om, 2 \om') -g_3(2\om, 2 \om')\,,
\eea
where each of these quantities is given by the following lattice sums,
\bea
\wp(\upsilon | 2\om, 2\om') & = & { 1 \over \upsilon^2} + \sum _{(m,n) \not= (0,0)} \left ( 
{1 \over (\upsilon + 2 m \om + 2 n \om')^2} - {1 \over (2 m \om + 2 n \om')^2} \right )\,,
\no \\
g_2 (2\om, 2 \om') & = & 60 \sum _{{m,n}\not= (0,0)} { 1 \over (2m \om  +  2n \om' )^4}\,,
\no \\
g_3 (2 \om , 2 \om') & = & 140 \sum _{{m,n}\not= (0,0)} { 1 \over (2m \om  +  2n \om' )^6}\,.
\eea 
It will be convenient to use canonical normalizations instead  in which the periods are normalized to $1, \tau=\om'/\om$, and the argument  is normalized accordingly to $z=\upsilon /2 \om$. The relation between these two normalizations amounts to a scaling factor by powers of the period $2\om$ (the functions with both normalizations are denoted by the same symbol),
\bea
(2 \om)^{-2} \, \wp (z|\tau) & = &  \wp (\upsilon | 2\om,2\om')\,,
\no \\
(2 \om)^{-4} \, g_2(\tau) & = &  g_2(2\om,2\om')\,,
\no \\
(2 \om)^{-6} \, g_3(\tau)  & = &  \, g_3(2\om,2\om')\,,
\eea 
or equivalently $g_2(\tau)=g_2(1,\tau)$ and $g_3(\tau)=g_3(1,\tau)$. In the sequel, the argument $\tau $ will not be exhibited if its dependence is clear from the context.  The differential equation for the Weierstrass function  is homogeneous under this scaling, and the canonical Weierstrass function $\wp(z)$ satisfies, 
\bea
 \wp' (z) ^2= 4 \, \wp(z)^3 - g_2 \wp(z) -g_3\,.
\eea
Henceforth, we shall use this canonical form which is the derivative of the Weierstrass $\zeta (z)$ function
(not to be confused with the Riemann $\zeta$-function which will not enter here), 
\bea
\wp(z) = - \zeta '(z)\,, \qquad \zeta (-z) = -\zeta (z)\,.
\eea
The $\zeta(z)$ function has the following monodromy relations,
\bea
\label{C.zeta}
\zeta (z+1)-\zeta (z) & = & 4 \om \eta\,,  \hskip 0.9in 2 \om \, \eta =  \zeta (\tfrac{1}{2}) \,,
\no \\
\zeta (z+\tau)-\zeta (z) & = & 4  \om\eta' \,, \hskip 0.9in 2 \om \, \eta' =  \zeta (\tfrac{\tau}{2}) \,.
\eea
Note that $\zeta (\tfrac{1}{2}) $ and $\zeta (\tfrac{\tau}{2}) $ are functions of $\tau$ only but, because of the extra factor of $\om$ in their definition, the parameters $\eta$ and $\eta '$ depend on both $\om$ and $\tau$. 
The relations are readily established by using the fact that $\zeta (-z)=\zeta (z)$ and setting $z$ equal to $-\tfrac{1}{2}$ and $-\tfrac{\tau}{2}$, respectively. The periods $2 \om$ and $2 \om'$ and the parameters $\eta$ and $\eta'$ satisfy the following relation, 
\bea
\label{square}
2 \eta \, \om' - 2 \eta ' \om = \tau \zeta (\tfrac{1}{2})  - \zeta (\tfrac{\tau}{2})  & = & i \pi\,.
\eea
One may use this relation to obtain $\eta'$ and $\zeta( \tfrac{\tau}{2})$ in terms of the other data.
A useful formula for $\zeta (\tfrac{1}{2})$  and thus $\eta$ in terms of  the discriminant $\Delta = g_2^3 - 27 g_3^2 $ is as follows,
\bea
\label{zetaDelta}
\zeta (\tfrac{1}{2}) = -{ i \pi \over 12} \, \p_\tau \ln \Delta = { \pi^2 \over 6} E_2(\tau)\,.
\eea
The combinations $g_2, g_3$ and $\Delta$ are holomorphic modular forms.

\subsection{Modular Transformations and Modular Forms}

Modular transformations form the group $SL(2,\ZZ)$ and act on $\tau$ by M\"obius transformations,   
\bea
\tau \to \tilde \tau  = { a \tau + b \over c \tau +d} \,,
\hskip 0.9in
\left ( \bma a & b \cr c & d \ema \right ) \in SL(2, \ZZ)\,.
\eea
They are generated by the transformations $S$ and $T$, under which $S:\tau\to -1/\tau$ and $T: \tau \to \tau+1$, and whose matrix form is as follows,
\bea
S = \left ( \bma 0 & -1 \cr 1 & 0 \ema \right )\,,
\hskip 0.9in 
T = \left ( \bma 1 & 1 \cr 0 & 1 \ema \right ) \,.
\eea
Under an arbitrary modular transformation $\tau \to \tilde \tau$, the half-periods $ \om$ and $ \om'$ and the parameters $\eta$ and $\eta'$ transform linearly, 
\begin{align}
\label{omzetamod}
\tilde \om' & =  a \om' + b \om\,, & \tilde \eta' & =  a \eta ' + b \eta \,,
\no \\
\tilde \om & =  c \om' + d \om \,,& \tilde \eta & =  c \eta ' + d \eta \,,
\end{align}
while the combinations $g_2$, $g_3$, $\Delta = g_2^3 - 27 g_3^2$, and $j=1728 \, g_2^3/\Delta$  transform as follows, 
\begin{align}
g_2(\tilde \tau) & =  (c \tau+d)^4 \, g_2(\tau) \,, & \Delta (\tilde \tau) & =  (c \tau +d)^{12} \, \Delta (\tau)\,,
\no \\
g_3(\tilde \tau) & =  (c \tau+d)^6 \, g_3(\tau) \,, & j (\tilde \tau) & =  j (\tau)\,.
\end{align}
The combinations $g_2, g_3$, and $\Delta$ are referred to as holomorphic modular forms of weight $(4,0)$, $(6,0)$, and $(12,0)$ respectively, while $j$ is a meromorphic modular function, since it is invariant under $SL(2,\ZZ)$.  We note that $\zeta (\tfrac{1}{2})$ and $\zeta (\tfrac{\tau}{2})$ are not modular forms. Indeed, by combining the expression for $E_2$ in terms of $\Delta$ with the transformation law for $\Delta$, we obtain, 
\bea
\label{C.E2}
E_2(\tilde \tau) = (c \tau +d)^2 E_2(\tau) + {12\over 2 \pi i}  c (c \tau +d) \,,
\eea
and $E_2$ is referred to as a quasi-modular form.

\sm

The standard fundamental domain for $SL(2,\ZZ)$ is given by,
\bea
\cF = \{ \tau  \in \CC \hbox{ such that } 0 < \tau_2 , \, 1 \leq  |\tau|, \, |\tau_1|\leq \thalf \}\,.
\eea
It contains the orbifold points $i$, $\rho=e^{2 \pi i /3}$, and $\rho' = \rho+1$, which are fixed points under the transformations $S:i\to i$, $ST: \rho \to \rho$ and $TS: \rho' \to \rho'$. The $j$-function provides a holomorphic bijection from $\cF$ to the Riemann sphere $\hat \CC$. 

\sm

Convergent Taylor series expansions in terms of the variable $e^{2 \pi i \tau}$ for $\zeta(\thalf)$, $g_2$ and $g_3$ with~$\tau$ in the standard fundamental domain $\cF$ are given by
\begin{align}
\zeta (\thalf |\tau) & = { \pi^2 \over 6} E_2 (\tau) \,, & E_2 (\tau) & = 1 -24 \sum _{n=1}^\infty \sigma _1(n) \, e^{ 2 \pi i n \tau} \,,
\no \\
g_2(\tau) & =  {4  \pi ^4 \over 3} E_4(\tau) \,, &  E_4(\tau) & = 1 + 240 \sum _{n=1}^\infty \sigma _3(n) \, e^{ 2 \pi i n \tau} \,,
\no \\
g_3(\tau) & =  {8  \pi ^6 \over 27} E_6(\tau) \,,& E_6(\tau) & = 1 - 504 \sum _{n=1}^\infty \sigma _5(n) \, e^{ 2 \pi i n \tau} \,,
\end{align}
where $\sigma _k (n) = \sum _{d |n} d^k$ with $d >0$ are the standard sum-of-divisor functions. The discriminant takes the form,
\bea
\label{Delta}
\Delta (\tau) 
= {( 2 \pi)^{12} \over 1728} \Big ( E_4(\tau)^3 - E_6(\tau) ^2 \Big ) 
= (2 \pi)^{12} \, e^{2 \pi i \tau}  \prod _{n=1}^\infty ( 1 -e^{2 \pi i n \tau} )^{24}\,.
\eea
The $j$-function is  normalized so that its pole in $q=e^{2 \pi i \tau}$ at the cusp has unit residue, 
\bea
j(\tau) =  e^{-2 \pi i \tau} + 744 + \cO(e^{2 \pi i \tau}) \,.
\eea

\subsection{Special Values}

In this final subsection, we review the reality conditions  for $g_2, g_3, \Delta, E_2$ and $\zeta (\tfrac{1}{2}) $ and $\zeta (\tfrac{\tau}{2})$, and obtain their special values at the orbifold points $i$ and $ \rho$ and at the cusp $i \infty$. 
The modular function $j$, the modular forms $g_2, g_3, \Delta$, as well as $E_2$ and thus $\zeta (\thalf)$ are real  for $\tau \in i \RR$ and $ \tau \in \pm \thalf + i \RR$. 

\sm

At the orbifold points $i , \, \rho$ and at the cusp $ i \infty$, the functions  $E_4(\tau)$, $E_6(\tau)$ , $j(\tau)$, $\zeta (\thalf |\tau)$ and $\zeta (\tfrac{\tau}{2} |\tau)$ take the following values, 
\begin{align} 
\label{table1}
E_2(i \infty) & = 1 & E_2(i) & = \tfrac{3}{\pi} & E_2(\rho) & = \tfrac{ 2 \sqrt{3}}{ \pi}
\no \\
E_4(i \infty) & = 1 & E_4(i) & = 48 \Gamma(\tfrac{5}{4})^4/ (\pi^2 \Gamma(\tfrac{3}{4})^4) & E_4(\rho) & = 0
\no \\
E_6(i \infty) & = 1 & E_6(i) & = 0 & E_6(\rho) & =729 \Gamma(\tfrac{4}{3})^6/(2 \pi^3 \Gamma(\tfrac{5}{6})^6)
\no \\
j(i \infty) & = \infty & j(i) & = 1728 & j(\rho) & =0
\no \\
\zeta (\thalf | i \infty ) & = \tfrac{\pi^2}{6}  & \zeta (\thalf | i ) & = \tfrac{\pi}{2} & \zeta (\thalf | \rho ) & = \tfrac{\pi}{ \sqrt{3}}
\no \\
\zeta (\tfrac{\tau}{2} | \tau )& \approx \tfrac{\pi^2}{6} \tau - i \pi  & \zeta (\tfrac{i}{2} | i ) & = - \tfrac{i \pi } {2} & \zeta (\tfrac{\rho}{2} | \rho ) & = - \tfrac{\pi}{2 \sqrt{3}} - \tfrac{i \pi}{2}  
\end{align}
The values for the cusp $\tau=i \infty$ follow from the series expansions of $E_2, E_4$ and $E_6$, and the relation between $\zeta(\thalf )$ and $E_2$ and (\ref{square}). 
The cancellations of $E_4(\rho)$  and $E_6(i)$  follow from the modular transformations \cite{apostol}: the relation   $Si=i$ implies $E_6(Si) =  - E_6(i)$ so that $E_6(i)=0$ and $j(i) = 1728$. Similarly, the relation $ST\rho = \rho$ implies  $E_4(ST \rho)  = \rho^2 E_4(\rho)$ so that $E_4(\rho)=0$ and $j(\rho)=0$. The values of $E_4(i)$  and $E_6(\rho)$ may be found on page 7 in \cite{DS}. 

\sm

The values of $\zeta (\tfrac{1}{2}|\tau)$ and $\zeta (\tfrac{\tau}{2}|\tau )$ at the fixed points may be obtained from the values of $E_2$ at the fixed points combined with the relations (\ref{square}). Applying the modular transformation rule for $E_2$ for the values $\tau=\tilde \tau =i$ and $\tau=\tilde \tau =i$ with the modular transformations $S$ and $ST$ respectively, we find, 
\bea
E_2(i) = - E_2(i) + { 6 \over \pi}\,,
\hskip 0.9in 
E_2(\rho) = \rho E_2 (\rho) + { 6 \rho^2 \over \pi}\,.
\eea
Solving these equations gives the entries in the first line of (\ref{table1}).

\newpage

\section{Numerical Methods for $SU(3)$}
\setcounter{equation}{0}
\label{sec:appnum}

In this appendix, we describe two numerical techniques to evaluate the K\"ahler potential for the case of~$SU(3)$ gauge group. 

\subsection{Numerically Evaluating  $F_4$}
\label{sec:F4}

The Appell function $F_4(a,b,c_1,c_2;x,y)$ may be evaluated by summing its Taylor series at $x=y=0$ in the domain of convergence $\sqrt{|x|}+\sqrt{|y|} <1$. Beyond this domain, the function enjoys an inversion formula, given in (\ref{analF4}), which allows one to extend the domain to $\sqrt{|x|} + 1 < \sqrt{|y|}$ and $\sqrt{|y|} + 1 < \sqrt{|x|}$. Still, the union of all these domains does not cover all of $x,y \in \CC$, and in particular excludes the domain that is of greatest interest to us when $|x|\sim |y|\sim1$. {\tt Maple} sports a preprogrammed function for $F_4$, which has difficulties precisely in this physically interesting region as well.  To this end, we now present a numerical approach that circumvents these problems. It is based on numerically integrating a first order ODE along a ray $x(t) = t x, y(t)=t y$ for a given point $(x,y) \in \CC^2$ and $t \in [0,1]$. We shall now present the essential components of the method.

\subsubsection{Conversion to a First Order System}

Following Appell and Kamp\'e de F\'eriet \cite{Appell}, we transform the system of two second order differential equations for $F_4$,
\bea
x(1-x) \p_x^2 f - y ^2 \p_y ^2 f - 2 x y \p_x \p_y f 
+\big (c_1- c_0 x \big )  \p_x f - c_0 y \p_y f - ab f  & = & 0
\no \\
y(1-y) \p_y^2 f - x ^2 \p_x ^2 f - 2 xy \p_x \p_y f 
+\big (c_2 -c_0 y \big )  \p_y f -c_0 x \p_x f  - ab f  & = & 0
\eea
with $c_0=a+b+1$,  into a system of 4 first order differential equations. Clearly, the dimension of the first order system must be 4 since we know that there must be 4 independent solutions (for generic parameters). Using the notation, 
\begin{align}
\label{D.pqs}
p&={ \p f \over \p x} \,, & q & = { \p f \over \p y} \,,& s & = { \p^2 f \over \p x \, \p y}\,,
\end{align}
the system of first order differential equations is conveniently expressed in terms of matrix-valued differential form notation, 
\bea
\label{6.dif1}
d \Phi 
= ( M_x dx + M_y dy) \Phi \,,
\eea
where,
\bea
\Phi = \left ( \bma f  \cr p \cr q \cr s \ema \right )\,,
\hskip 0.3in
M_x = \left ( \bma 0 & 1 & 0 & 0 \cr A_1 & A_2 & A_3 & A_4 \cr 0 & 0 & 0 & 1 \cr C_1 & C_2 & C_3 & C_4 \ema \right ) \,,
\hskip 0.3in
M_y = \left ( \bma 0 & 0 & 1 & 0 \cr 0 & 0 & 0 & 1 \cr B_1 & B_2 & B_3 & B_4 \cr D_1 & D_2 & D_3 & D_4 \ema \right ) \,.
\eea
The rows that involve only 0 and 1 in $M_x$ and $M_y$ readily result from the definitions of $p,q,s$ in (\ref{D.pqs}). The  entries $A_1, \ldots, A_4$ and $B_1, \ldots, B_4$  may be obtained by transforming the system of second order equations into an equivalent system in which one equation involves $\p_x p$ but not $\p_y q$ and vice-versa. This system is given as follows, 
\bea
\label{D.App}
x(1-x-y)\p_x p - 2 xy s +(c_1-c_1y-c_0x) p + (c_2-c_0) y q - abf & = & 0\,,
\no \\
y(1-x-y)\p_y q - 2 xy s +(c_2-c_2x-c_0y) q + (c_1-c_0) x p - abf & = & 0\,.
\eea
As a result, we have,
\begin{align}
A_1 & = {ab \over x(1-x-y)} & 
B_1 & = {ab \over y(1-x-y)} 
\no \\
A_2 & = { c_0x+c_1y-c_1 \over x (1-x-y)} & 
B_2 & = {(c_0-c_1)x \over y(1-x-y)}
\no \\
A_3 & = {(c_0-c_2)y \over x(1-x-y)} & 
B_3 & = { c_2x+c_0y-c_2 \over y (1-x-y)} 
\no \\
A_4 & = { 2 y \over 1-x-y} & 
B_4 & = { 2 x \over 1-x-y} 
\end{align}
To obtain the entries $C_1, \ldots, C_4$ and $D_1, \ldots, D_4$ we begin by taking the $\p_y$ derivative of the first equation in (\ref{D.App}) and the $\p_x$ derivative of the second equation.   Taking linear combinations that produce one equation involving only $\p_x s$ and another involving only $\p_y s$  and eliminating the derivatives $\p_x p$ and $\p_y q$ using (\ref{D.App}), one obtains,
\bea
C_i = { \cC_i \over x (1-x-y)N}\,, \hskip 0.5in 
D_i = { \cD_i \over y (1-x-y)N}\,,
\eea
where $N = (1-x-y)^2 -4xy$. The coefficients $\cC_i$ are given by,
\bea
\cC_1 & = & ab (c_0-2c_1+c_2+1) x +ab (c_0-c_2+1)(1-y) \,,
\no \\
\cC_2 & = & (-2ab+c_0^2-c_0c_1+c_0c_2-c_1c_2-c_0+c_1)x^2 
\no \\ &&
+ (-2ab-c_0^2+3c_0c_1+c_0c_2-2c_1^2-c_1c_2-3c_0+3c_1) xy 
\no \\ &&
+ (2ab+c_0^2-3c_0c_1-c_0c_2+2c_1^2+c_1c_2+3c_0-3c_1)x\,,
\no \\
\cC_3 & = & (ab-c_0c_2+c_2^2+c_0 -c_2 ) (1-x)^2 + (ab-c_0^2+c_0c_2) y^2 
\no \\ &&
+(2ab+c_0^2-2c_0c_1-2c_0c_2+2c_1c_2+c_2^2+3c_0-3c_2) xy
\no \\ &&
+ (-2ab+c_0^2-c_2^2-c_0+c_2) y\,,
\no \\
\cC_4 & = & (c_0-2c_2+2)x^3 + c_1 y^3 +(2c_0-3c_1+2)x^2y +(-3c_0+2c_1+2c_2-4)xy^2
\no \\ &&
+(-2c_0- c_1+4c_2-4)x^2 - 3 c_1 y^2 +(2c_0-4c_1+2) xy
\no \\ &&
+ (c_0+2c_1-2c_2+2) x + 3c_1 y -c_1 \,,
\eea
and the coefficients $\cD_i$ are given by, 
 \bea
\cD_1 & = & ab (c_0-c_1+1) (1-x) +ab (c_0+c_1 -2 c_2+1)y\,,
\no \\
\cD_2 & = & (ab-c_0^2+c_0c_1)x^2  + (ab-c_0c_1+c_1^2+c_0-c_1)(1-y)^2
\no \\ &&
+ (2ab+c_0^2-2c_0c_1-2c_0c_2+c_1^2+2c_1c_2+3c_0-3c_1)xy
\no \\ &&
+ (-2ab+c_0^2-c_1^2-c_0+c_1) x \,,
\no \\
\cD_3 & = & (-2ab+c_0^2+c_0c_1-c_0c_2-c_1c_2-c_0+c_2) y^2 
\no \\ &&
+(-2ab-c_0^2+c_0c_1+3c_0c_2-c_1c_2-2c_2^2 -3 c_0 + 3c_2) xy
\no \\ &&
+ (2ab+c_0^2-c_0c_1-3c_0c_2+c_1c_2+2c_2^2 +3c_0-3c_2) y\,,
\no \\
\cD_4 & = & c_2 x^3 + (c_0-2c_1+2)y^3 +(-3c_0+2c_1+2c_2-4)x^2y + (2c_0-3c_2+2)xy^2 
\no \\ &&
- 3 c_2 x^2 +(-2c_0+4c_1-c_2-4)y^2  +(2c_0-4c_2+2) xy
\no \\ &&
 + 3c_2 x + (c_0-2c_1+2c_2+2) y -c_2 \,.
\eea
Swapping $x \leftrightarrow y$ and simultaneously $c_1 \leftrightarrow c_2$, we verify 
$(A_1, A_2, A_3, A_4) \leftrightarrow (B_1, B_3, B_2, B_4)$ and 
$(\cC_1, \cC_2, \cC_3, \cC_4) \leftrightarrow (\cD_1, \cD_3, \cD_2, \cD_4)$.

\subsubsection{Solution on a Ray via ODE}

Fix a point $(x,y) \in \CC^2$ and parametrize a ray from the $\ZZ_6$ symmetric point to $(x,y)$ by
\begin{align}
x(t) & = t \a^2 \,,& x & = \a^2\,,
\no \\ 
y(t) & = t \b^2 \,,& y & =\b^2\,.
\end{align}
In terms of this parametrization, the denominators $(1-x-y)$ and $N$ have simple zeros,
\bea
1-x(t) -y(t)  & = & 1-t(\a^2+\b^2) \,,
\no \\
N(x(t) , y(t)) & = & \big (1-t (\a - \b) ^2 \big ) \big (1-t (\a + \b) ^2 \big ) \,.
\eea
On this ray, the system of first order equations in two variables collapses to a system in just one variable $t$,
\bea
\label{6.dif2}
{ d \over dt}  \Phi = (\a^2 M_x + \b^2 M_y) \Phi \,,
\eea
where $\Phi$, $M_x$, and $M_y$ are all evaluated at $x(t), y(t)$. The initial values at the $\ZZ_6$-symmetric point may be obtained from the Taylor expansion of $F_4$ in powers of $x,y$. The integration of this ODE for each one of the $F_4$ functions appearing in the $SU(3)$ solution is now standard, and thankfully proves to be numerically fast.

\subsection{Numerically Evaluating the Derivatives of $K$}
\label{sec:numint}

We now describe a method for numerically evaluating the $SU(3)$ K{\"a}hler potential by integrating its derivatives with respect to the moduli~$u, v$. 
Our starting point is the expression~\eqref{5.Ka} for the derivatives of $K$ as two-dimensional integrals, which we repeat here
\bea
\label{dkdu2}
{ \p K \over \p \bar u_n}  
=
{1 \over 8 \pi^3} \lim _{R \to \infty} \int _{|x|<R} d^2x  \, {  xA'(x) \,  \bar x^{n} \over |A(x)^2 - 1|}\,,
\eea
and likewise for the complex conjugate derivatives.
For $N=3$ the two complex moduli are $u_1=u$ and $u_0=v$, with $A(x) = x^3- u x - v$.

Straightforward numerical integration of \eqref{dkdu2} fails, due to the fact that the integrand has poles at $x_i$ satisfying $A(x_i)^2= 1$, as well as at infinity (for the $v$-derivative).
We proceed by explicitly subtracting the residues of these poles from the integrand, numerically integrating, and then adding back in the subtracted contributions.  For instance, $dK/d\bar{u}$ is computed as 
\bea
{ \p K \over \p \bar u} 
&=
{2 i \over 8 \pi^3} \lim_{\epsilon\to0} \left[ \int _{-\infty}^\infty \int _{-\infty}^\infty  d\text{Re}(x) \, d\text{Im}(x)    \, I(x,\epsilon) + \sum_i  J(x_i,\epsilon)\right]\,,
\label{dderiv}
\eea
where $I(x,\epsilon)$ and $J(x_i,\epsilon)$  are given as,
\bea
\label{Ix}
I(x,\epsilon) & = & {  |x|^2A'(x)    \over |A(x)^2 - 1|}   - \frac{3 x^2}{ |x|^4 }e^{- \frac{1}{|x|^2}}
- \sum_{i}   \frac{ |x_i|^2 A'(x_i) e^{- \epsilon |x-x_i|^2}}{2 |A'(x_i)||x-x_i|} \,,  \nonumber \\
 J(x_i,\epsilon)  &= &   \frac{ x_i A'(x_i) }{2 |A'(x_i)|} ( 2\pi)\int_{0}^\infty\, dr\,  e^{-\epsilon r^2}\,.
\eea
When evaluating these formulas numerically we take $\epsilon$ to be a small number (e.g.~$\epsilon = 0.01$) whose precise value demonstrably does not affect the results of the integration. The expression \eqref{dderiv} for $\partial K/\partial\bar{u}$, along with the analogous 
formulae for $\partial K/\partial\bar{v}$ and their complex conjugates,
can then be straightforwardly numerically integrated from an initial point (which we take to be the multi-monopole point with real $u$ for which $K=0$) to a generic point $(u,v)$. In this way, $K$ can be evaluated on a grid of complex $(u,v)$ values.

We have subjected the numerical evaluation method described above to various consistency checks: 

\begin{itemize}

\item Applying the same method to the case of $SU(2)$ gauge group, we have verified that the resulting $K(u)$ matches the known hypergeometric function representation depicted in figure \ref{fig:Ksu2better}.

\item We have verified that on the $u=0$ and $v=0$ slices of moduli space for which the Appell functions reduce to hypergeometric functions,  the numeric evaluation of $K(u,v)$ reproduces the correct values. This includes matching to the analytic values of $K$ at the origin (computed in~\eqref{symm}) and at the Argyres-Douglas points~(computed in~\eqref{KADpt}).

\item Within the region of convergence of \textsc{Maple}'s predefined Appell $F_4$ function, we have verified for a variety of $(u,v)$ that the numerical evaluation of~$K(u, v)$ matches the numerical evaluation of the exact formula of~$K$ given in terms of Appell functions. 

\item We have, in some regimes, checked the two numerical methods described in appendices~\ref{sec:F4} and~\ref{sec:numint} against each other and found them to match with high precision. 
 
\end{itemize}

\section{The Strong-Coupling Spectrum for $SU(N)$}
\setcounter{equation}{0}
\label{sec:BPSapp}

In this appendix, we enumerate the stable BPS particles in the strong-coupling chamber of four-dimensional $\mathcal{N}=2$ pure supersymmetric gauge theory with gauge group~$SU(N)$, though our primary interest is the case~$N = 3$. 

\sm

These BPS states were determined, for all~$SU(N)$ gauge groups, in~\cite{Lerche:2000uy,Alim:2011kw,Chuang:2013wt}. We follow the discussion in~\cite{Alim:2011kw}, where a basis different from ours is used. It is convenient to work at the origin of moduli space, i.e.~at the~$\ZZ_{2N}$-symmetric point. The left panel of figure \ref{fig:cyclesorigin} shows the branch cut conventions and various cycles used in~\cite{Alim:2011kw}, while the right panel shows our choice of branch cuts and cycles, both for the case of~$SU(3)$ gauge group. We see that the branch cuts, as well as the~$\hat{\mA}_I$ (and hence the~${\mA}_I$) cycles and the intersection pairing  $\#( \mA_I,  \mB_J) = \delta_{IJ}$, are the same in both figures. However, the figures differ in the definition of the~$\mB_I$ cycles. 

\sm

In order to translate their results into our conventions, we denote their $\mB_I$ cycles -- referring to the left panel of figure \ref{fig:cyclesorigin} -- by $\mB_I^\text{them}$, while our $\mB_I$ cycles -- referring to the right panel of figure \ref{fig:cyclesorigin} -- are denoted by $\mB_I^{\text{us}}$. By examining these figures, we see that the cycles are related as follows,
\bea
\label{bcycrel}
\mB_1^{\text{us}} = \mB_1^\text{them} - \mA_1\, , 
\hskip 0.9in
 \mB_2^{\text{us}} =  \mB_2^\text{them} + \mA_2\, .
\eea

\tikzset{
  branch cut/.style={
    decorate,decoration=snake,
    to path={
      (\tikztostart) -- (\tikztotarget) \tikztonodes
    }
    }
}
\tikzset{middlearrow/.style={
        decoration={markings,
            mark= at position 0.5 with {\arrow[scale=1.5]{#1}} ,
        },
        postaction={decorate}
    }
}
\tikzset{quarterarrow/.style={
        decoration={markings,
            mark= at position 0.25 with {\arrow[scale=1.5]{#1}} ,
        },
        postaction={decorate}
    }
}

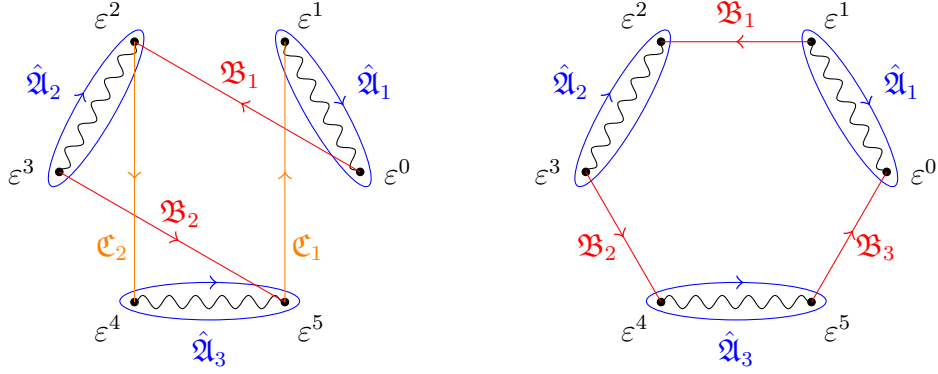
\begin{figure}[t!]
\centering
 \begin{tikzpicture}
 \begin{scope}[shift={(-3.5,0)}]
 
       \filldraw[color=black] (2,0) circle (0.05);
   \filldraw[color=black] (1, 1.732) circle (0.05);
     \filldraw[color=black] (-1, 1.732) circle (0.05);
      \filldraw[color=black] (-2,0) circle (0.05);
       \filldraw[color=black] (-1, -1.732) circle (0.05);
         \filldraw[color=black] (1, -1.732) circle (0.05);
 
    \draw[branch cut] (2,0) to (1, 1.732);
    \draw[branch cut] (-1, 1.732)  to (-2,0);
    \draw[branch cut] (-1, -1.732)  to (1, -1.732) ;

   \node at  (2.5, 0)  (e0) {$ \ep^0$};
      \node at  (1+0.35, 1.732+0.35)  (e1) {$ \ep^1$};
         \node at (-1-0.35, 1.732+0.35)  (e2) {$ \ep^2$};
            \node at  (-2.5,0)  (e3) {$ \ep^3$};
               \node at  (-1-0.35, -1.732-0.35)  (e4) {$ \ep^4$};
                  \node at  (1+0.35, -1.732-0.35) (e5) {$ \ep^5$};
   
       \node at (2.2,1.2)  {$\textcolor{blue}{\hat{\mA}_1}$};
       \node at (-2.2,1.2)  {$\textcolor{blue}{\hat{\mA}_2}$};
      \node at (0,-2.35)  {$\textcolor{blue}{\hat{\mA}_3}$};
      
         \node at (0.4,1.3)  {$\textcolor{red}{{\mB}_1}$};
          \node at (-0.4,-0.55)  {$\textcolor{red}{{\mB}_2}$};
       
        \node at (1.3,-1)  {$\textcolor{orange}{{\mC}_1}$};
        \node at (-1.3, -1)  {$\textcolor{orange}{{\mC}_2}$};

   \draw[rotate around={-60:(1.73,0.95)},blue,quarterarrow={<}](1.71,0.708) ellipse (34pt and 7.1pt);
     \draw[rotate around={60:(-1.73,0.95)},blue,quarterarrow={<}](-1.71,0.708) ellipse (34pt and 7.1pt);
         \draw[blue,quarterarrow={<}](0,-1.72) ellipse (34pt and 7.1pt);

   \draw[red,middlearrow={<}] (-1, 1.732) -- (2, 0);
     \draw[red,middlearrow={<}] (1, -1.732) -- (-2, 0);
     
     \draw[orange,middlearrow={>}] (1, -1.732) -- (1, 1.732);
        \draw[orange,middlearrow={<}] (-1, -1.732) -- (-1, 1.732);

    \end{scope}
     \begin{scope}[shift={(3.5,0)}]
     
       \filldraw[color=black] (2,0) circle (0.05);
   \filldraw[color=black] (1, 1.732) circle (0.05);
     \filldraw[color=black] (-1, 1.732) circle (0.05);
      \filldraw[color=black] (-2,0) circle (0.05);
       \filldraw[color=black] (-1, -1.732) circle (0.05);
         \filldraw[color=black] (1, -1.732) circle (0.05);
 
    \draw[branch cut] (2,0) to (1, 1.732);
    \draw[branch cut] (-1, 1.732)  to (-2,0);
    \draw[branch cut] (-1, -1.732)  to (1, -1.732) ;

   \node at  (2.5, 0)  (e0) {$ \ep^0$};
      \node at  (1+0.35, 1.732+0.35)  (e1) {$ \ep^1$};
         \node at (-1-0.35, 1.732+0.35)  (e2) {$ \ep^2$};
            \node at  (-2.5,0)  (e3) {$ \ep^3$};
               \node at  (-1-0.35, -1.732-0.35)  (e4) {$ \ep^4$};
                  \node at  (1+0.35, -1.732-0.35) (e5) {$ \ep^5$};
   
       \node at (2.2,1.2)  {$\textcolor{blue}{\hat{\mA}_1}$};
       \node at (-2.2,1.2)  {$\textcolor{blue}{\hat{\mA}_2}$};
      \node at (0,-2.35)  {$\textcolor{blue}{\hat{\mA}_3}$};
      
         \node at (0,2.1)  {$\textcolor{red}{{\mB}_1}$};
          \node at (-1.85,-1)  {$\textcolor{red}{{\mB}_2}$};
                   \node at (1.85,-1)  {$\textcolor{red}{{\mB}_3}$};

   \draw[rotate around={-60:(1.73,0.95)},blue,quarterarrow={<}](1.71,0.708) ellipse (34pt and 7.1pt);
     \draw[rotate around={60:(-1.73,0.95)},blue,quarterarrow={<}](-1.71,0.708) ellipse (34pt and 7.1pt);
         \draw[blue,quarterarrow={<}](0,-1.72) ellipse (34pt and 7.1pt);

   \draw[red,middlearrow={>}] (1, 1.732) -- (-1, 1.732);
      \draw[red,middlearrow={>}] (-2, 0) -- (-1, -1.732);
           \draw[red,middlearrow={<}] (2, 0) -- (1, -1.732);

    \end{scope}
    \end{tikzpicture}
     \caption{ The left panel shows the conventions for the $\hat{\mA}_I$, $\mB_I$, and $\mC_I$ cycles that are used in \cite{Alim:2011kw}. The right panel shows our conventions for the $\hat{\mA}_I,\mB_I$ cycles. In this figure we have specialized to $N=3$, with $\ep=e^{\frac{2\pi i}{6}}$. \label{fig:cyclesorigin}}
\end{figure}

In the conventions of~\cite{Alim:2011kw}, the $N(N-1)$ BPS particles that exist at the origin of~$SU(N)$ gauge theory have charge vectors~$\vec \mu_{kI}$ that correspond to the following cycles~$\mu_{kI}$,\footnote{~As is common in the literature on BPS particles, we do not list their corresponding anti-particles.} 
\bea
	\mu_{0I} = - \mB_I^\text{them}\,,
	\hskip 0.4in
	 \mu_{1I} = \mC_I\,,
	 \hskip 0.4in
	 \mu_{kI} = \mu_{k-1,I-1} + \mu_{k-1,I+1} - \mu_{k-2,I}\,.
\label{eq:cyclee}
\eea
Here the labels run over  $k=2,\ldots,N-1$ and $I=1,\ldots,N-1$, so that we set $\mu_{kI}=0$ for $I<1$ and $I > N-1$. The $\mC_I$ cycle is obtained from $\mB^\text{them}_I$ by a $-\frac{\pi}{N}$ rotation, as indicated in figure \ref{fig:cyclesorigin} for the case $N=3$. It can be expressed in terms of the $\mA_I$ and $\mB^\text{them}_I$ cycles:
\bea
\mC_I = \left\{ \begin{array}{cc} - \mA_{I-1} + 2 \mA_I- \mA_{I+1} + \mB^\text{them}_I & I\ \text{even} \\   - \mA_{I-1} + 2 \mA_I - \mA_{I+1} - \mB^\text{them}_{I-1}-\mB^\text{them}_I-\mB^\text{them}_{I+1} & I \ \text{odd} \end{array} \right.  \,.
\label{eq:ci}
\eea
The tower of $N-1$ mutually local dyons that become massless at the $k$'th multi-monopole point corresponds to the set of $\mu_{kI}$ cycles, with $I=1,\cdots,N-1$. 

\begin{table}[h!]
\centering
\begin{tabular}{|c||c|c|c|}
\hline
$\mu_{kI}$ &  Cycles &   $(\vec{q}; \vec{g})^\text{them}$ & $(\vec{q}; \vec{g})^\text{us}$  \\ \hline \hline
$\mu_{01}$ & $-\mB_1^\text{them}$   &   $(0,0;-1,0)$ & $(-1,0;-1,0)$ \\ \hline
$\mu_{02}$ &  $-\mB^\text{them}_2$  &  $(0,0;0,-1)$  & $ (0,1;0,-1)$  \\ \hline \hline
$\mu_{11}$ & $\mC_1=2\mA_1 - \mA_2 - \mB^\text{them}_1-\mB^\text{them}_2$   &  $(2,-1;-1,-1)$ &  $ (1,0;-1,-1)$  \\ \hline
$\mu_{12}$ & $\mC_2 = - \mA_1+ 2 \mA_2 + \mB^\text{them}_2$  &  $(-1,2;0,1)$  & $(-1,1;0,1)$  \\ \hline \hline
$\mu_{21}$ &  $\mC_2+\mB^\text{them}_1$ & $(-1,2;1,1)$  & $ (0,1;1,1)$  \\ \hline
$\mu_{22}$ &  $\mC_1+\mB^\text{them}_2$  & $( 2,-1;-1,0)$   &$ (1,-1;-1,0)$   \\ \hline 
\end{tabular}
\caption{The six stable BPS particles in the strong-coupling chamber of $SU(3)$ gauge theory, together with the corresponding cycles and charges in the conventions of~\cite{Alim:2011kw}, as well as the translation of the charges into our conventions.\label{tab:charges}}
\end{table}

Specializing to $N=3$, there are six BPS particles in the strong-coupling chamber, corresponding to cycles $\mu_{kI}$ with $k=0,1,2$ and $I=1,2$. The corresponding charges $\vec \mu^\text{them} = (\vec q; \vec g)^\text{them}$ can be read off immediately from the specialization of~\eqref{eq:cyclee} and~\eqref{eq:ci} to $N = 3$, while the charges~$\vec \mu^\text{us} = (\vec q; \vec g)^\text{us}$ in our basis can be determined from~\eqref{bcycrel},
\begin{equation}
\begin{split}
\vec g^\text{us} & = \vec g^\text{them}~,\\
q_1^\text{us} & = q_1^\text{them} + g_1^\text{them}~,\\
q_2^\text{us} & = q_2^\text{them} - g_2~.
\end{split}
\end{equation}
The results are summarized in~table \ref{tab:charges}. The states in the table are grouped into three pairs $\mu_{kI}$, labeled by $k=0,1,2$, of mutually local dyons, with one pair becoming massless at each of the three multi-monopole points. At each of the two Argyres-Douglas points, one dyon from each pair becomes massless. At the origin of moduli space, all six states are massive and degenerate.

\newpage


\bibliographystyle{utphys}
\bibliography{biblio}

\end{document}